\documentclass[%
 reprint,
superscriptaddress,
 amsmath,amssymb,
 aps,
]{revtex4-2}
\usepackage[dvipsnames]{xcolor}
\definecolor{DarkNavy}{RGB}{0,0,150}
\usepackage{array}
\usepackage{algorithm}
\usepackage{algorithmic}

\usepackage{graphicx}
\usepackage{dcolumn}
\usepackage{bm}
\usepackage[colorlinks=true,linkcolor=blue,urlcolor=blue,citecolor=blue]{hyperref}
\usepackage{comment}

\usepackage[normalem]{ulem}



\begin{document}

\preprint{APS/123-QED}

\title{Localising entropy production along non-equilibrium trajectories }
\author{Biswajit Das}
\email{bd18ip005@iiserkol.ac.in}
\affiliation{Department of Physical Sciences, Indian Institute of Science Education and Research Kolkata, Mohanpur Campus, Mohanpur, West Bengal 741246, India}
\author{Sreekanth K Manikandan}
\email{sreekanth.manikandan@physics.gu.se}
\affiliation{Department of Chemistry, Stanford University, Stanford, CA, USA 94305}
\affiliation{Department of Physics, Gothenburg University, Gothenburg, Sweden}

\date{\today}

\begin{abstract}

{\color{black} Entropy production is a universal measure of irreversibility and energy dissipation in physical, chemical, and biological systems operating far from equilibrium. However, quantifying and spatiotemporally localising it in complex processes directly from experimental data remains a major open challenge.}
Here we address this issue through a data-driven approach that combines the recently developed short-time thermodynamic uncertainty relation based inference scheme with machine learning techniques. Our approach leverages the flexible function representation provided by deep neural networks to achieve accurate reconstruction of high-dimensional, potentially time-dependent dissipative force fields as well as the localization of fluctuating entropy production in both space and time along nonequilibrium trajectories. We demonstrate the versatility of the framework through applications to diverse systems of fundamental interest and experimental significance, where it successfully addresses distinct challenges in localising entropy production.

\end{abstract}

\maketitle


\begin{figure*}[htb]
    \centering
    \includegraphics[width=0.99\linewidth]{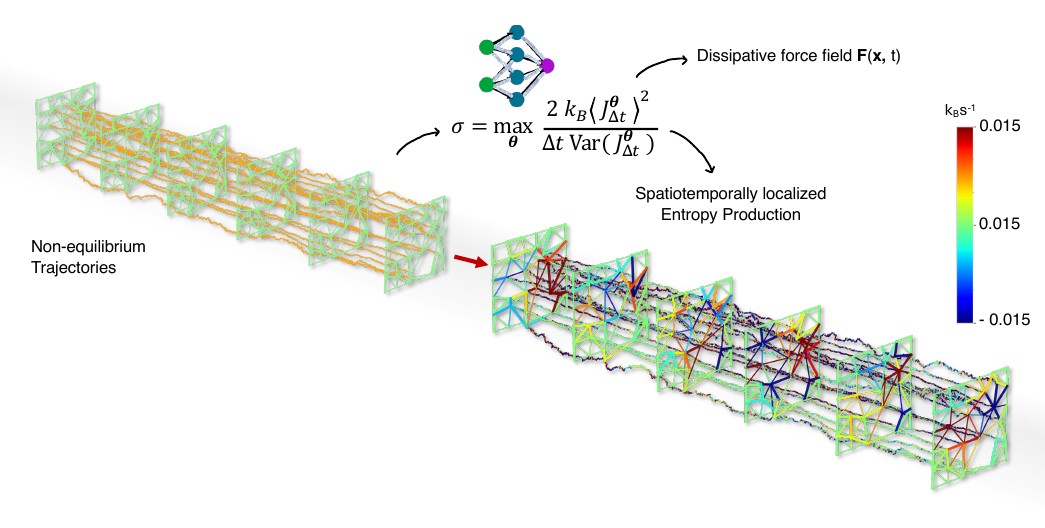}
    \caption{\textbf{Schematic of entropy production inference in an active biological network model.}
\textbf{Input}: The method processes experimentally measurable trajectory data  without requiring prior knowledge of system parameters. \textbf{Outputs}: Using short-time Thermodynamic Uncertainty Relations and neural networks \textcolor{black}{(schematically illustrated at the top with a cartoon)}, we simultaneously infer (i) the dissipative (thermodynamic) force field ${\bm F}({\bm x},t)$ driving the nonequilibrium dynamics and (ii) the corresponding fluctuating entropy production  (color scale: $\pm 0.015 k_B/s$), localized in both space and time. \textcolor{black}{Here, 
$\sigma$ denotes the local entropy production rate, and 
$J_{\Delta t}$ represents a generalized current in phase space.}}
    \label{fig:scematics}
\end{figure*}

\section{Introduction}
Processes that operate far from thermodynamic equilibrium are ubiquitous in nature~\cite{bustamante_phystoday_2005}. They are characterized by non-vanishing phase space fluxes and the continuous dissipation of energy into their surroundings~\cite{battle2016broken,gnesotto2018broken}. Extracting the underlying dissipative force fields and the associated energetic costs from experimentally tractable observations, such as single-molecule trajectories, fluorescence imaging, or spectroscopic data is a challenging task and a vastly developing field of research \cite{roldan2010estimating, noninvasive,harada2005equality,frishman:sfi,gnesotto:lnb, manikandan2020inferring,manikandan2024estimate,boffi2024deep,roldan2021quantifying, di2024variance, degunther2024fluctuating, degunther2024general, majhi2025decoding}. Recent advances in this field, primarily driven by the developments in stochastic thermodynamics~\cite{seifert2008stochastic,seifert2019stochastic,Ciliberto:2017est}, combine far-from-equilibrium statistical physics with experiments/simulations and data-driven approaches \cite{li2019quantifying,otsubo2022estimating,manikandan2021quantitative,kim2020learning,boffi2024deep}. These methods have, for example,  successfully quantified entropy production in cellular processes where dissipation rates could be orders of magnitude below the detection limits of conventional calorimetry \cite{manikandan2024estimate, roldan2021quantifying,oocites,ariga2018nonequilibrium,di2024variance}.

One of the main contributions of this field is the rigorous formulation of entropy production along individual realizations of  nonequilibrium processes  where thermal fluctuations are significant \cite{seifert2005entropy, seifert2019stochastic}. The rate of entropy production ($\sigma$) quantifies, simultaneously, the breaking of time-reversal symmetry of the process, as well as the continuous dissipation of energy to the surroundings \cite{seifert2005entropy}. Furthermore, a non-vanishing conjugate force field, referred to as the thermodynamic force field \( {\bm F}({\bm x}, t) \), can always be associated with nonequilibrium processes. When contracted with the infinitesimal increment of the phase space trajectory \( d{\bm x}_t \), this force field yields the local entropy production along the trajectory~\cite{li2019quantifying}. The thermodynamic force field vanishes everywhere in space and time when the system is at equilibrium; thus, it governs the local dissipative dynamics, providing a direct link between microscopic irreversibility and energy dissipation.

Despite their fundamental importance, directly measuring the entropy production or dissipative force field from experimental data remains a significant challenge. Conventional approaches rely on the knowledge of the underlying dynamical equations, such as Fokker-Planck and Master equations, and their solutions, which are often unknown in realistic settings. 
In a number of such cases, in principle, a global estimate of the entropy production rate of a process can be directly estimated from experimental data, particularly for systems with Markovian dynamics, using a variety of methods. These include the application of the Harada-Sasa equality \cite{harada2005equality}, which involves perturbing systems from the non-equilibrium steady state and analyzing their response; methods that depend on the estimation of the average steady-state currents and the steady-state probability distribution \cite{noninvasive}; and time-irreversibility measures, which quantify the breaking of time-reversal symmetry in the dynamics \cite{sekimoto1997kinetic,sekimoto1998langevin,seifert2005entropy,Seifert:2012stf,maes2003time,gaspard2004time}. Path probability estimators, which compute the relative probabilities for forward and time-reversed trajectories directly from data, have also been developed \cite{roldan2010estimating,andrieux2008thermodynamic,EpTA}. For overdamped diffusive processes, another approach involves inferring the drift and diffusion terms from experimental data \cite{frishman:sfi, gnesotto:lnb}, from which $\sigma$ can subsequently be deduced. A more recent method, called the variance sum rule (VSR), derives a formula for the entropy production rate $\sigma$ by incorporating the second derivative of the position correlation function and the covariances of microscopic forces~\cite{di2024variance}. While its direct application was initially limited by experimental challenges in measuring mobility and forces, a recently proposed modification to VSR has overcome this limitation~\cite{diterlizzi2025forcefree}.

In addition to these approaches, a number of variational optimization schemes have been developed to quantify the entropy production rate. A notable method is based on the thermodynamic uncertainty relation (TUR) \cite{barato2015thermodynamic, gingrich2017, Gingrich2016, manikandan2020inferring, van2020entropy, Shun:eem, manikandan2021quantitative, das2022inferring,otsubo2022estimating,li2019quantifying}, which translates the task of identifying entropy production into an optimization problem over the space of a single projected fluctuating current in the system. Unlike trajectory-based methods that require estimating probability distributions over the phase space, these inference schemes typically rely only on the means and variances of measured currents, making them more effective for higher-dimensional systems \cite{li2019quantifying}. For overdamped diffusive processes, it has since been shown that such optimization problems yield the exact value of the entropy production rate, as well as the exact thermodynamic force field, provided that short-time currents are used \cite{manikandan2020inferring, Shun:eem, van2020entropy} -- hereafter we refer to this method as the \textit{the short-time inference scheme}. Notably, this method works in both stationary regimes~\cite{manikandan2021quantitative} and in time-dependently driven cases~\cite{otsubo2022estimating}, and obtains the average entropy production rate as well the dissipative force field. A similar variational approach using neural networks has also been proposed \cite{kim2020learning}, where the estimator includes an exponential average over current fluctuations. This method has the additional advantage of being applicable to both overdamped diffusive processes and Markov jump processes.

 The aforementioned studies have primarily focused on obtaining global estimates of the average entropy production rate. However, there are a few notable exceptions~\cite{li2019quantifying,kim2020learning,degunther2024fluctuating,degunther2024general,boffi2024deep}. In Ref.~\cite{li2019quantifying}, the authors developed a brute-force statistical binning approach to estimate the thermodynamic force field. However, this method scales poorly to higher dimensional systems. The method proposed in Ref.~\cite{kim2020learning} directly learns entropy production along trajectories; however, it does not study the structure of the underlying dissipative force fields, even though this information is, in principle, available. In Ref.~\cite{degunther2024fluctuating,degunther2024general}, the authors introduced a notion of ‘Markovian events’ which formally defines fluctuating entropy production for coarse-grained processes in a thermodynamically consistent manner. Most recently, Ref.~\cite{boffi2023probability,boffi2024deep} showed how part of a dissipative force field can be inferred directly from data, assuming the remaining components are known in advance. They also provided clear demonstrations of spatially localized, ensemble averaged entropy production in the stationary states of high-dimensional, interacting active particle systems~\cite{boffi2024deep}. This approach has since been extended to the underdamped regime, and to cases where the equations of motion are unknown~\cite{boffi2024modelfree}, enabling the model-free learning of probability flows and spatial maps of average entropy production.

 In this work, we build on the short time inference scheme and Machine Learning techniques and present a systematic, data driven study of fluctuating, trajectory resolved local entropy production (see Figure~\ref{fig:scematics} for a schematic overview) and it's statistics. We apply this framework to a broad class of systems, encompassing stationary and non-stationary dynamics as well as linear and non-linear models, and show that entropy production can be robustly inferred at the single trajectory level. This enables us to identify when and where second-law–violating events (negative entropy production) and near-reversible fluctuations occur, even in globally irreversible systems. Our method relies on a single neural-network architecture that performs robustly across all examples without extensive hyperparameter tuning, making it broadly applicable in practice. We further characterize the local fluctuations of entropy production and show that they reveal detailed information about the underlying nonequilibrium dynamics, while exhibiting robust, system independent properties. For instance, when entropy production fluctuations are spatially localized to specific regions, the log probability ratio of positive to negative fluctuations in each region follows the form predicted by the fluctuation theorem, despite substantial heterogeneity in the qualitative nature of the fluctuations across regions. Similarly we verify that the skewness of cumulative entropy production obeys theoretically predicted universal properties. We also show that the inferred fluctuating entropy production remains statistically consistent with theoretical predictions under temporal coarse graining and in the presence of hidden degrees of freedom, two features that are intrinsic to essentially all experimental settings.  Taken together, these results establish local fluctuating entropy production as a physically interpretable and statistically consistent quantity that characterize non-equilibrium states beyond average entropy production rates and can be inferred robustly from experimental data.

\section{Methods}
\label{sec:methods}
The methods discussed in this work are applicable to experimental data from overdamped diffusive processes but do not require the knowledge of their underlying equations. To establish notations and definitions, we introduce a generic overdamped Langevin model in  a $d$ -dimensional space, described by:
\begin{align}
    \dot{{\bm x}}(t)={\bm A}({\bm x}(t),t)+{\bm B}({\bm x}(t), t)\cdot {\bm \xi}(t),\label{eq: Langevin}
\end{align}
where ${\bm A}({\bm x}, t)$ is the drift vector, and ${\bm B}({\bm x}, t)$ is a $d \times d$ matrix, and ${\bm \xi}(t)$ represents a Gaussian white noise satisfying $\langle \xi_i(t) \xi_j(t')\rangle = \delta_{ij}\delta(t-t')$. Both ${\bm A}({\bm x}, t)$ and ${\bm D}({\bm x}, t)$ could be non-linear functions in ${\bm x}$ and $t$. 

The Fokker-Planck equation that describes the time evolution of the probability density function $p({\bm x}, t)$ is given by,
\begin{align}
\label{eq:generic}
    \partial_t p({\bm x}, t) &= -{\bm \nabla}\cdot{\bm j}({\bm x}, t),\\
    j_i({\bm x}, t) &= A_i({\bm x}, t) p({\bm x}, t) - \sum_j\nabla_j\left[D_{ij}({\bm x}, t)p({\bm x}, t) \right],\label{eq:prob_cur}
\end{align}
where $\bm D$ is the diffusion matrix defined as
\begin{align}
    {\bm D}({\bm x}, t)=\frac{1}{2}{\bm B}({\bm x}, t){\bm B}({\bm x}, t)^T
\end{align}
and ${\bm j}({\bm x},t)$ is the probability current. 

A nonvanishing probability current --- ${\bm j}({\bm x},t)\neq 0$ is then a characteristic property of a process that is out of equilibrium. The extent of the non-equilibrium character can then be quantified using the average rate of the entropy production at a given instant $\sigma(t)$, defined as \cite{Spinney2012} 
\begin{align}
\label{eq:sigmatime}
    \sigma(t)=\int {\rm d}{\bm x}\; {\bm F}({\bm x},t) \cdot {\bm j}({\bm x},t),
\end{align}
 where ${\bm F}({\bm x},t)$ is the conjugate thermodynamic force field or dissipative force field defined as
 \begin{align}
 \label{eq:Ffield}
 {\bm F}({\bm x},t) &= \frac{{\bm j}^T({\bm x},t){\bm D}({\bm x}, t)^{-1}}{p({\bm x},t)}.
 \end{align}
Here we have set Boltzmann's constant $k_B = 1$. Further, when contracted with the infinitesimal increment of the phase space trajectory \( d{\bm x}_t \), this force field yields the local entropy production along the trajectory at time $t$, as
\begin{align}
    {\rm d}S(t) = {\bm F}\!\left({\bm x}(t), t\right)\circ {\rm d}{\bm x}(t),
    \label{eq: single_ep}
\end{align}
where $\circ$ denotes the Stratonovich product.
 In a steady state, where the probability density and the associated drift and diffusion terms are time independent, the quantity defined in Eq.~\eqref{eq: single_ep} coincides exactly with the total stochastic entropy production ${\rm d}S_{\rm tot}(t)$ across an infinitesimal transition $\bm{x}_t \to \bm{x}_{t+\mathrm dt}$~\cite{seifert2005entropy}.
In contrast, for explicitly time-dependent processes, ${\rm d}S(t)$ defined above differs from the total entropy production by an additional term arising from the explicit time dependence of the probability density, namely $-\partial_t \ln p(\bm x,t)\,\mathrm dt$.
Since this additional contribution vanishes upon ensemble averaging, as a direct consequence of probability normalization, ${\rm d}S(t)$ as defined above captures all irreversible contributions to the average entropy production rate $\sigma(t)$ through the integral in Eq.~\eqref{eq:sigmatime}.
It also isolates the fluctuation contributions arising from dynamics against the irreversible thermodynamic force field, independently of contributions associated with the explicit time dependence of $p(\bm x,t)$.

At equilibrium, detailed balance implies $\bm j(\bm x,t)=0$, such that both the force field $\bm F(\bm x,t)$ and the local entropy production $\mathrm dS(t)$ vanish identically. Away from equilibrium, their spatiotemporal structure quantifies where and when entropy production occurs, providing a localised characterization of nonequilibrium dynamics. 

Indeed, if the explicit form of the overdamped Langevin equation in Eq.~\eqref{eq: Langevin}, together with the corresponding Fokker--Planck equation in Eq.~\eqref{eq:generic} and its time-dependent solution, are known, one can straightforwardly determine the functional form of the thermodynamic force field in Eq.~\eqref{eq:Ffield} and use it to evaluate entropy production along individual realizations of the process. However, this is only possible in a limited number of analytically tractable cases.
In the following, we discuss the TUR based \textit{short-time inference scheme} that can estimate $\sigma(t)$, ${\bm F}({\bm x},t)$ and $dS(t)$ from overdamped non-equilibrium trajectories, without requiring the prior knowledge of the dynamical equations or their solution. \\

\subsection{Inferring local entropy production using the short-time TUR} Our approach is based on the thermodynamic uncertainty relation (TUR), which provided a universal constraint on non-equilibrium current fluctuations in terms of the entropy production rate \cite{barato2015thermodynamic}. In the original form, TUR provides a lower bound for the entropy production rate in terms of the first two cumulants of non-equilibrium current fluctuations in a steady state as,
\begin{eqnarray}
\label{eq:TUR}
\sigma \geq \frac{2\left<J_{\bm d}\right>^2}{\tau {\rm Var}(J_{\bm d})},
\end{eqnarray}
where $J_{\bm d}$ is a generalized current of length $\tau$ given by $J_{\bm d}:= \int_{{\bm x}(0)}^{{\bm x}(\tau)}{\bm d}({\bm x}(t), t)\circ {\rm d}{\bm x}(t)$ defined with some coefficient field ${\bm d}({\bm x})$. The originally known limit in which the lower bound saturates is the equilibrium limit where $\sigma \rightarrow 0$.
It was then shown that, for overdamped Langevin dynamics, TUR also saturates in the $\tau \rightarrow 0$ limit of current fluctuations\cite{manikandan2020inferring,Shun:eem,van2020entropy}. Crucially, the proof in Ref.~\cite{Shun:eem} is also valid for non-stationary dynamics.\\ \indent
This gives a variational representation of the entropy production rate, as
\begin{eqnarray}
\label{eq:TURinf}
\sigma_{\rm TUR}(t) := \frac{1}{{\rm d}t}\max_{{\bm d}}\frac{2\left<J_{\bm d}\right>^2}{{\rm Var}(J_{\bm d})},
\end{eqnarray}
where $J_{\bm d}$ is the (single-step) generalized current given by $J_{\bm d}:= {\bm d}({\bm x}(t))\circ {\rm d}{\bm x}(t)$. The expectation and the variance are taken with respect to the joint probability density $p({\bm x}(t), {\bm x}(t+{\rm d}t))$.
In the ideal short-time limit ${\rm d}t\to 0$, the estimator gives the exact value, i.e., 
$\sigma_{\rm TUR}(t) = \sigma(t)$ holds for the optimal current $J^*_{\bm d}$ that maximizes Eq.~\eqref{eq:TURinf}.
Further, the optimal coefficient field is known to be proportional to the thermodynamic force field as
${\bm d}^*({\bm x}) = c\,{\bm F}({\bm x}, t)$. See Refs. \cite{van2020entropy, Shun:eem} for proofs. 
The constant of proportionality arises because the TUR objective function
$2\langle J_{\bm d}\rangle^2/{\rm Var}(J_{\bm d})$ is invariant under a global rescaling of the current.
The constant factor $c$ can be fixed by using the short-time property of the total entropy production
for overdamped diffusive processes, namely
${\rm Var}(\Delta S_{\rm tot})/\langle \Delta S_{\rm tot}\rangle = 2$ as ${\rm d}t\to 0$,
which implies $2\langle J^*_{\bm d}\rangle/{\rm Var}(J^*_{\bm d}) = 1/c$~\cite{manikandan2020inferring,otsubo2022estimating, Manikandan:2018erf}. 

From an experimental point of view, d$t$, however is set by the sampling interval $\Delta t$, which is often fixed by practical constraints. When its choice is flexible, a natural dimensionless criterion is $\Delta t / \tau_{\min} \ll 1$, where $\tau_{\min}$ denotes the shortest relevant correlation or relaxation time of the system. 
\subsection{Machine learning the dissipative force field and entropy production}
The variational representation of the $\sigma(t)$ given in Eq.~\eqref{eq:TURinf} takes a form that is well-established in control theory. A common strategy in such settings is to approximate the optimal solution, the thermodynamic force field,  using a parameterized function,
\begin{equation}
\label{eq:nn}
 {\bm d}({\bm x}, t) = \hat{\bm d}({\bm x}, t; {\bm \theta}).
\end{equation}
where ${\bm \theta}$ represents a set of tunable parameters. However, the complexity of high-dimensional, many-body interactions poses significant challenges for traditional optimization methods. To overcome these difficulties, we leverage gradient-based techniques enhanced by deep neural networks, which excel at approximating complex, high-dimensional functions~\cite{rotskoff2018advances,chizat2018advances,mei2018mean,sirignano2020mean,barron1993universal,cybenko1989approximation}. This combination enables efficient exploration of the parameter space, even in scenarios where analytical solutions are intractable. The detailed steps of our approach are outlined in \textbf{Algorithm 1}, as well as in the accompanying Python implementation~\cite{git_repo}. 

We first focus on stationary non-equilibrium processes, for which the explicit
time dependence in Eq.~\eqref{eq:nn} can be omitted. Our approach builds on the
deep-Ritz neural-network architecture \cite{yu2018deep} used in Ref.~\cite{yan2022learning}, where
optimal-control techniques were combined with the expressive power of deep
neural networks to efficiently sample rare events in interacting many-body
systems. Motivated by this approach, we approximate $\hat{\bm d}(\bm x;\theta)$ using a multi-layer neural network that maps the input 
$\bm x \in \mathbb R^{d}$ to an output in the same physical dimension. The input is first mapped to a hidden space of dimension $h$ via $\bm X = P(\bm x)$, where $P$ denotes either zero--padding or a linear projection, depending on the relative sizes of $d$ and $h$. The network consists of a sequence of hidden layers, each of which applies two linear transformations with nonlinear activations and adds the result back to the input,
\begin{equation}
L_i(\bm X)
= \bm \phi\!\left(
W_{i,2}\,\bm \phi\!\left(W_{i,1}\bm X + \bm b_{i,1}\right) + \bm b_{i,2}
\right)
+ \bm X ,
\end{equation}
where $W_{i,j} \in \mathbb{R}^{h \times h}$ and $\bm b_{i,j} \in \mathbb{R}^{ h}$ are trainable parameters and $\phi$ is the activation function. This additive structure helps preserve information across layers and stabilizes training. The hidden layers are then composed to produce the hidden representation
\begin{equation}
\bm z({\bm X};{\bm \theta})
= L_n \circ \dots \circ L_1\!\left({\bm X}\right)
\in \mathbb R^{h},
\end{equation}
which is then mapped back to the physical dimension through a linear output layer to obtain $\hat{\bm d}({\bm x};{\bm \theta})$ as,
\begin{equation}
\hat{\bm d}({\bm x};{\bm \theta})
= W_{\mathrm{out}}\,\bm z({\bm X};{\bm \theta}) + \bm b_{\mathrm{out}},
\end{equation}
where $W_{\mathrm{out}} \in \mathbb{R}^{d \times h}$ and 
$\mathbf{b}_{\mathrm{out}} \in \mathbb{R}^{d}$. We use the bounded activation function $\tanh(\cdot)$ throughout; a comparison with other activations is provided in \textbf{Supplementary Note 1}.

\begin{figure}
\label{algo:1}
    \centering
\includegraphics[width=0.99\linewidth]{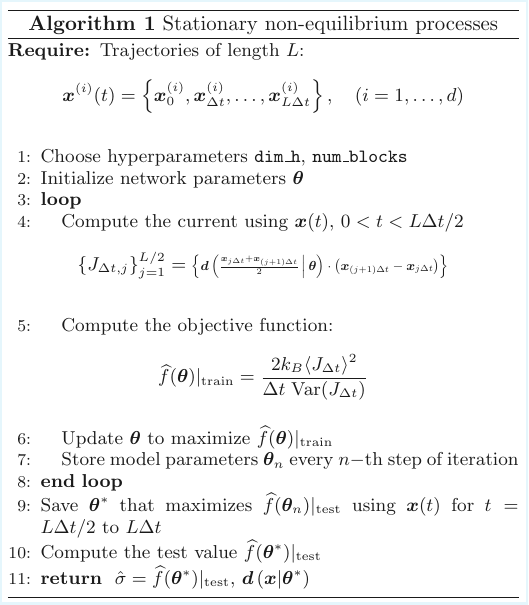}
\end{figure}

Note that, this framework redefines the task of measuring entropy production in nonequilibrium systems as a machine learning problem centered on approximating the associated dissipative force field. This strategy eliminates the reliance on explicit dynamical descriptions of the process, and offers a scalable solution for analyzing entropy production in systems with nontrivial interactions between different degrees of freedom. Once trained, the machine-learned force field can be used to visualize entropy production along individual non-equilibrium trajectories. 

\subsection{Algorithm}
 For processes in a stationary state, our inference scheme proceeds as follows (see Algorithm 1). 
From discretely sampled trajectories $\{\boldsymbol{x}^{(i)}(t)\}$, of length $\tau = L \Delta t$ where $i= 1 \dots d$, where $d$ is the dimensionality of the system, and $\Delta t$ is the sampling interval/ step size, and $L$ is the length of the time series, we aim to infer the thermodynamic force field and an estimate of the total entropy production rate $\hat{\sigma}$. 
We parameterize the force field using a fully connected neural network 
$\hat{\bm{d}}(\bm{x}\mid\boldsymbol{\theta})$, whose parameters $\boldsymbol{\theta}$ are optimized by maximizing the variational objective in Eq.\ \eqref{eq:TUR}.  
In practice, each trajectory is split into two halves: the first half is used for training, while the second half is reserved for unbiased validation.

Training proceeds iteratively. At each step, we use the stratanovich convention and compute a generalized current over short trajectory segments from the training half,
\begin{equation}
\label{eq:discrete_current}
J_{\Delta t}
=
\hat{\boldsymbol{d}}\!\left(
\frac{\boldsymbol{x}_{t} + \boldsymbol{x}_{t+\Delta t}}{2}
\,\middle|\,
\boldsymbol{\theta}
\right)
\cdot
\left(
\boldsymbol{x}_{t+\Delta t} - \boldsymbol{x}_{t}
\right).
\end{equation}
From these currents, we evaluate the training objective,
\begin{equation}
\label{eq:training_objective}
\widehat{f}(\boldsymbol{\theta})\big|_{\mathrm{train}}
=
\frac{2 k_B \langle J_{\Delta t} \rangle^2}
{\Delta t \, \mathrm{Var}(J_{\Delta t})}.
\end{equation}
The parameters $\boldsymbol{\theta}$ are then updated via gradient ascent to maximize 
$\widehat{f}(\boldsymbol{\theta})\big|_{\mathrm{train}}$.
During training, we additionally store intermediate parameter values 
$\{\boldsymbol{\theta}_n\}$ along the optimization trajectory.

After training, all stored models are evaluated on the held-out validation data, and the final model parameters $\boldsymbol{\theta}^*$ are chosen as those that maximize the  objective function on the validation data.
For relatively short data sets, it is often observed that an intermediate iterate outperforms the final converged model on the validation set. This is indicative of  `overfitting' at late training times.
In contrast, when sufficient data are available, overfitting becomes negligible and the model that maximizes the training objective also maximizes the validation objective.
For the low-dimensional systems studied here, we empirically find that data sets of size $L \sim 10^{4}\text{--}10^{5}$ time steps typically fall into this large-data regime.

The final estimate of the entropy production rate is therefore computed as
\begin{equation}
\label{eq:final_sigma_estimate}
\hat{\sigma}
=
\widehat{f}(\boldsymbol{\theta}^*)\big|_{\mathrm{test}},
\end{equation}
ensuring an estimate that generalizes beyond the training data.
The optimized force field $\boldsymbol{d}(\boldsymbol{x}\mid\boldsymbol{\theta}^*)$ can then be directly used for trajectory-resolved analyses of fluctuating entropy production.

For time-dependent dynamics (See Algorithm 2), the network architecture is extended so that the neural net approximation to the force field takes time $t$ as an additional input.
Crucially, the training objective is now evaluated and aggregated over a mini-batch of sampled time points 
$\{t_k\}$,
\begin{equation}
\label{eq:time_dependent_objective}
\widehat{f}(\boldsymbol{\theta})
=
\sum_{k=1}^{\mathrm{batch\_size}}
\frac{2 \langle J_{\Delta t,k} \rangle^2}
{\Delta t \, \mathrm{Var}(J_{\Delta t,k})},
\end{equation}
where $J_{\Delta t,k}$ denotes the generalized current computed at time $t_k$. This formulation offers several advantages over training separate models at each discrete time point: 
it enforces temporal smoothness through shared network parameters, enables generalization to unobserved times, 
and efficiently leverages data across the entire time domain. 
The resulting optimized force field $\hat{\boldsymbol{d}}(\boldsymbol{x}, t \vert \boldsymbol{\theta}^*)$ is then expected to be a good approximation of the thermodynamic force field ${\bm F}({\bm x}, t)$, and can be used to obtain associated local contributions to the total entropy production, $dS(t)$.
\begin{figure*}[htb]
    \centering
\includegraphics[width=0.99\linewidth]{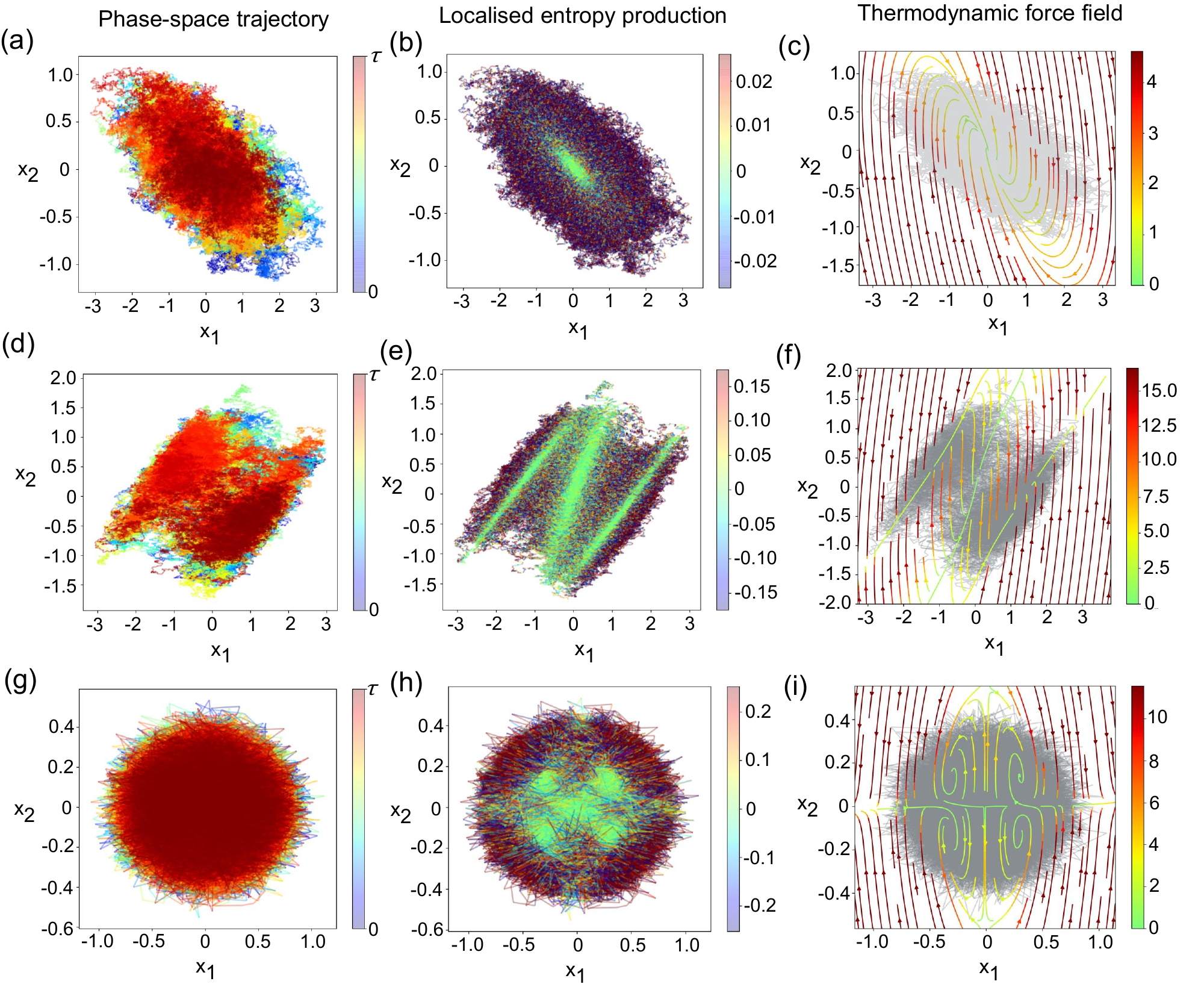}
    \caption{\textbf{Local entropy production in Brownian gyrator models.} (a) 2D-dimensional trajectories of a Brownian gyrator system with harmonic confining potential. [Parameters: $k_1 = 1$, $k_2 = 2$, $\gamma = 1$, $\theta = \pi/4$, $D_1 = 1$, $D_2 = 0.1$].  (b) local entropy production rate and (c) thermodynamic force field for the system with harmonic confinement -  estimated using the neural network representation. 
    (d) 2D-dimensional trajectories of a Brownian gyrator system with a bi-stable confining potential. [Parameters: $k = 1$, $b = 1$, $\gamma = 1$, $\theta = \pi/4$, $D_1 = 1$, $D_2 = 0.1$] (e) local entropy production rate and (f) thermodynamic force field estimated using the neural network representation.  (g) 2D-dimensional trajectories of a Brownian gyrator system with a quartic confining potential. [Parameters: $k_1 = k_2 = 10$,  $\gamma = 1$, $\theta = \pi/4$, $D_1 = 10$, $D_2 = 1$] (h) local entropy production rate and (i) thermodynamic force field estimated using the neural network representation.
    The colours corresponding to the local entropy production rate (in units of $k_B/s$) of the gyrators are thresholded between $[-\alpha\,\mathrm{median},\, \alpha\,\mathrm{median}]$
, where $\alpha$ (typically $20 -50$) multiplies the median of the corresponding local entropy production dataset. Values outside these ranges are clipped for visualisation purposes to prevent rare large fluctuations from dominating the colour mapping. 
Similarly, the thermodynamic force field values for the gyrators are thresholded within $[0,\, \mathrm{median}]$. 
The numerical trajectories are usually generated for $2000s$ with a sampling rate of $1\ kHz$ - from which trajectory traces of $500 s$ are shown in the plots. The colorbars in panels
(a), (d), and (e) indicate the progression along the trajectory. }
    \label{fig:br_gyrator}
\end{figure*}

\begin{figure*}[htb]
    \centering
    \includegraphics[width=0.9\linewidth]{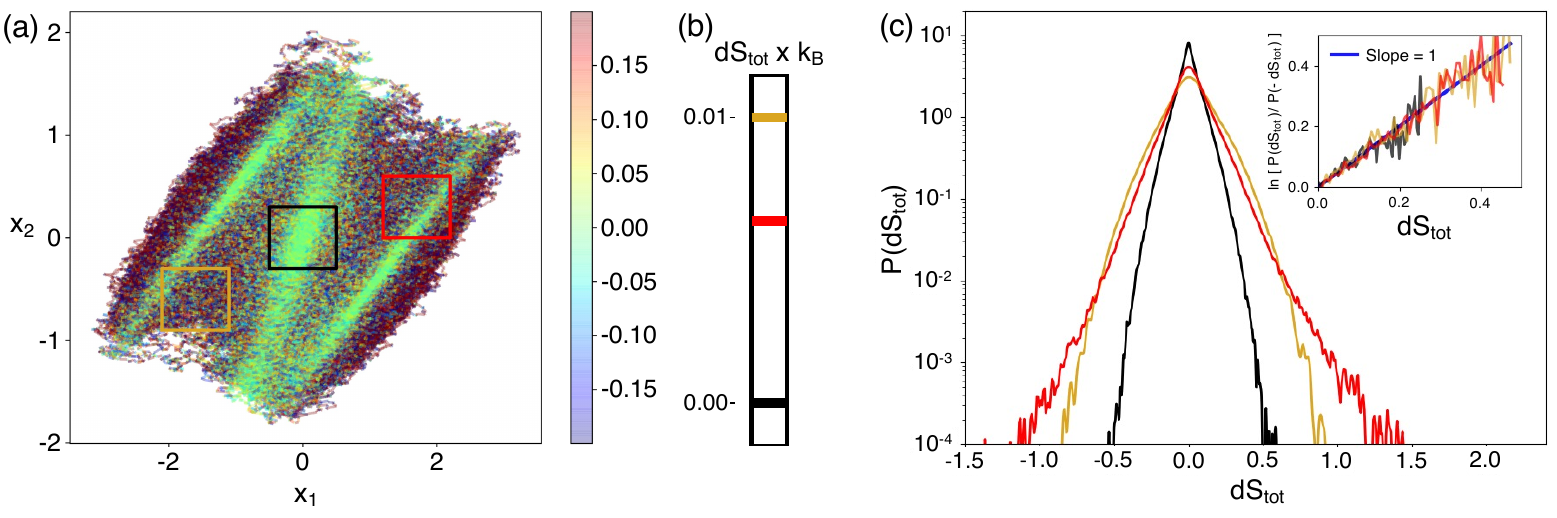}
    \caption{\textbf{Local tests of fluctuation theorem.}  (a) The boxed regions highlight three representative areas characterized by low (black), intermediate (red), and high (goldenrod) local dissipation (average values shown in (b)), chosen to probe distinct dynamical environments. (c) Probability distributions $P(dS_{tot})$ conditioned on these regions, illustrating pronounced region-dependent differences in the statistics of entropy production. The low-dissipation region exhibit narrow, nearly symmetric distributions, while higher-dissipation regions display broader, strongly skewed distributions with extended tails. The inset shows the corresponding fluctuation ratios $\ln[P(dS_{tot})/P(-dS_{tot})]$ as a function of $dS_{tot}$, demonstrating that each region independently satisfies a local fluctuation theorem with unit slope.
 }
    \label{fig:FT}
\end{figure*}

\begin{figure*}[htb]
    \centering
    \includegraphics[width=0.99\linewidth]{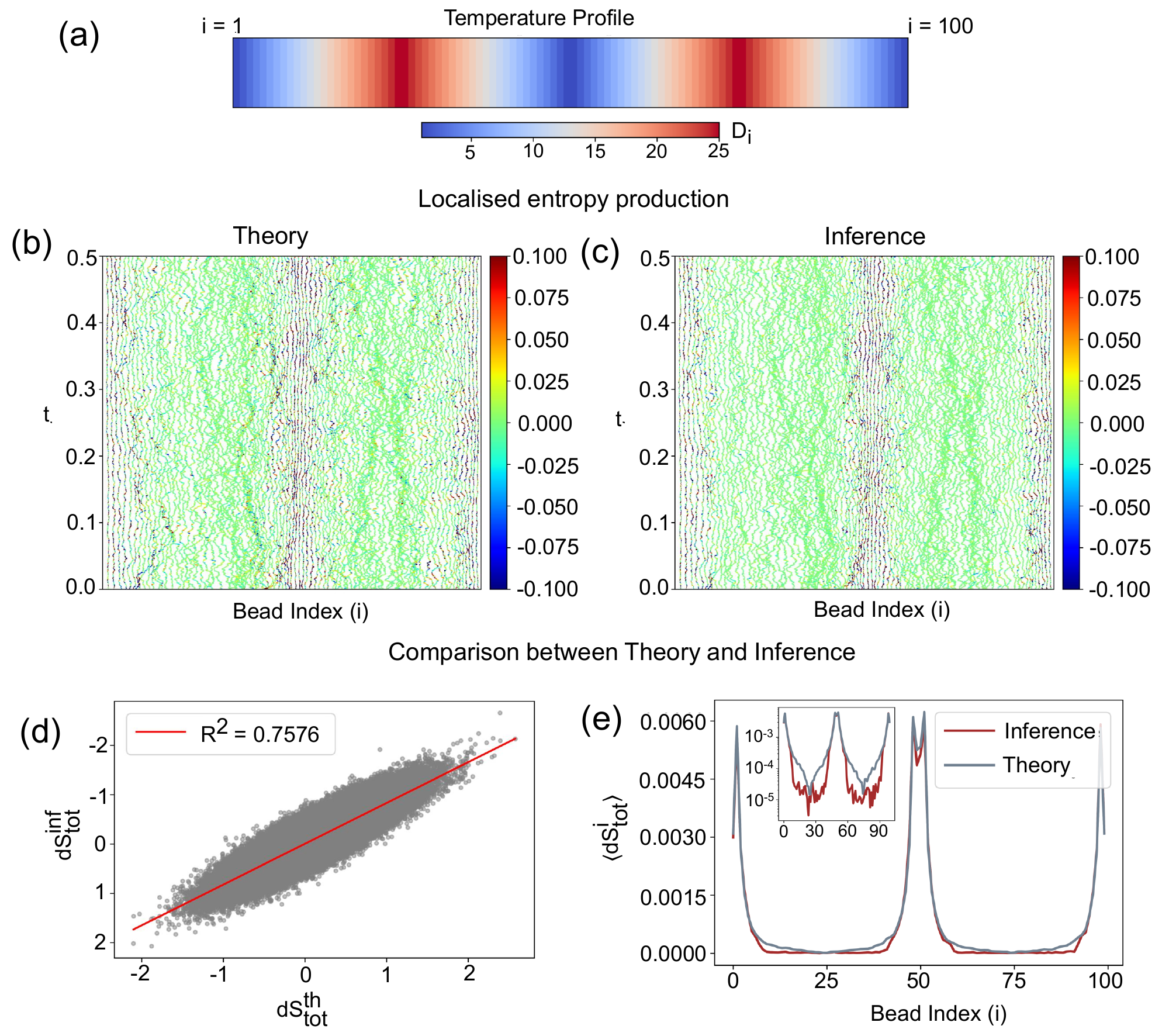}
    \caption{\textbf{Bead wise local entropy production for N-dimensional brownian gyrator model. }  (a) Temperature profile of the $N$-dimensional gyrator setup. \textcolor{black}{$D_i$ denotes the diffusion coefficient of $i$-th bead as $k_B = \gamma = 1$.} (b) Analytically estimated local entropy production rate (in units of $k_B/s$) of the system. (c) Local entropy production rate (in units of $k_B/s$) inferred from the numerical trajectories using a neural network representation. The colors do not indicate the true values of the fluctuating entropy current, but they are thresholded for better visualisation. (d) Convergence test ($R^2$ test) of the neural network–based estimation of the fluctuating entropy production rate for an $N$-dimensional Brownian gyrator with $N=100$. The inferred and analytical local entropy production rates, averaged over all beads, exhibit a finite spread around the linear fit. (e) Comparison of the inferred average entropy production for each bead with the corresponding theoretical estimate. (Inset) The same data shown on a logarithmic (y-) scale reveals that dissipation of beads associated with low irreversible signature (entropy production) are challenging for the neural network to capture, resulting in a mismatch with the theoretical prediction. }
    \label{fig:nBgr}
\end{figure*}

\section{Results and Discussion}
\label{sec:applications}
In the following, we demonstrate that the methodology described in the previous section is versatile and applicable to a wide range of processes of experimental significance, each presenting unique challenges in localizing entropy production. We further show that these inferred fluctuations obey robust and system-independent properties which can be directly verified.
\subsection{Stationary non-equilibrium systems: Brownian Gyrators}
Brownian gyrator is a simplistic model of an autonomous heat engine~\cite{filliger2007brownian,argun2017experimental}. In its most basic implementation, a mesoscopic particle is confined in a two-dimensional conservative potential in the presence of two heat baths at different temperatures. The dynamics of the system can be expressed as a $2D$ overdamped \textit{Langevin} equation as,
\begin{equation}
\label{eq:gyrator_eq}
    \gamma \dot{x}_i(t) = -\ \partial_{x_i} U(x_i, x_j) + \sqrt{2D_i\gamma^2} \xi_i(t),
\end{equation}
with $\langle \xi_i(t)\rangle = 0$, $\langle \xi_i(t)\xi_j(t^\prime)\rangle = \delta_{ij}\delta(t-t^\prime)$. Here, the viscous drag of the medium is denoted as $\gamma$, and $D_i = k_B T_i/\gamma$ denotes the diffusion constants along two directions, which are proportional to the temperatures ($T_i$) of the corresponding baths. In the absence of any coupling through the potential between the degrees of freedom of the system, each degree reaches an equilibrium steady state characterised by the temperatures of the corresponding heat bath. However, if the degrees of freedom are coupled, the system will in general reach a non-equilibrium stationary state, with non-vanishing phase space currents. The total entropy production rate then captures the average dissipation of such systems and their dependence on the system parameters~\cite{das2022inferring}.  
However, how the fluctuations in total entropy production is distributed across the phase space of such systems having different forms of confinements has not been explored before.

We consider four different examples of Brownian gyrators: (1) the harmonic gyrator, where a particle is confined in a 2D harmonic potential \( U_{\text{har}}(x_1^\prime, x_2^\prime) = \frac{1}{2}(k_1 x_1^{\prime 2} + k_2 x_2^{\prime 2}) \) with a rotated coordinate system such that $(x_1^\prime, x_2^\prime)^T = \mathbf{R}(\theta)\cdot (x_1,x_2)^T$, $\mathbf{R} = \begin{pmatrix}
    \cos \theta & -\sin \theta\\
    \sin \theta &\cos\theta
\end{pmatrix}$,
(2) the anharmonic gyrator with a bistable potential \( U_{\text{dw}}(x_1^\prime, x_2^\prime) = x_1^{\prime 4} - 2b x_1^{\prime 2} + 0.5 k x_2^{\prime 2} \) and (3) a quartic potential \( U_{\text{qu}}(x_1^\prime, x_2^\prime) = 0.5 (k_1 x_1^{\prime 2} + k_2 x_2^{\prime 2})^2 \), both having prominent non-linear dynamics and finally (4) a multidimensional, linear gyrator which exemplifies a high-dimensional system. For examples (1) and (4), since the dynamics is governed by linear Langevin equation, exact analytic solutions are available for the thermodynamic force field. This is, however, not the case for the anharmonic gyrators~\cite{chang2021autonomous,das2022inferring}.

Figure~\ref{fig:br_gyrator} summarizes the results for 2D Brownian gyrator systems. The top panel corresponds to the harmonic confining potential case, showing: (\textit{a}) the characteristic phase-space trajectory, (\textit{b}) the entropy production along trajectories (with positive, negative, and near-zero values appearing prominently near the origin), and (\textit{c}) the inferred dissipative force field color-coded by magnitude. Notably, the force field magnitude reaches its minimum in the neighborhood of the potential minimum at (0,0), which coincides with the region of near-zero entropy production visible in panel (\textit{b}). In these plots, the colour bars are clipped to threshold values to ensure good contrast between different regimes, and to prevent large fluctuations from dominating the color mapping. For the local entropy production plots, the color bar is symmetrically clipped to the interval 
$[-\alpha\,\mathrm{median},\, \alpha\,\mathrm{median}]$, where $\alpha$ (typically $10 -50$) multiplies the median of the corresponding local entropy production dataset. For the force field, we color it according to the magnitude, and the colour scale is clipped to the interval $[0,\mathrm{median}]$, where the median is computed from the corresponding dataset.

The second row of Figure~\ref{fig:br_gyrator}(d-f) corresponds to the anharmonic Brownian gyrator system with a bistable confining potential. In this case, both the dissipative force field and local entropy production exhibit a highly counterintuitive contour of near-zero entropy production, which would be very difficult to predict \textit{a priori}. Similarly, in the case of the quartic Brownian gyrator too, we observe a highly nontrivial spatiotemporal structure for both local entropy production and the dissipative force field, as illustrated in the third row, Figure~\ref{fig:br_gyrator}(g-i).  Notably, the thermodynamic force field is found to feature four vortices that cancel out at the origin. These vortices are found in the low-entropy regions of phase space.

In all the trajectory entropy production plots (second column), we note that while the majority of the trajectory shows positive entropy production values (consistent with the Second Law), the regions near zero entropy production are accompanied by clearly contrasted regimes of negative entropy production. This is an expected consequence of low-entropy production regimes, where fluctuations should follow near-Gaussian statistics, making negative entropy production values more probable as compared to the rest of the phase space, where fluctuations can be highly non-Gaussian and positively skewed~\cite{manikandan2022nonmonotonic}. 

With detailed knowledge of the fluctuating local entropy production, these qualitative features can be further quantified by conditioning the statistics on distinct regions of phase space. To this end, we consider three representative regions, highlighted in Fig.~\ref{fig:FT}(a), chosen to sample qualitatively different dynamical environments characterized by low, intermediate, and high levels of local dissipation $dS_{tot} (t)$. Their average values are shown in Fig.~\ref{fig:FT}(a).  This allows us to probe how the statistics of the local entropy production depend on the underlying phase-space region. The corresponding probability distributions, shown in Fig.~\ref{fig:FT}(c), reveal pronounced region-dependent differences: low-dissipation regions (black) exhibit narrow, nearly symmetric distributions, whereas higher-dissipation regions (red and goldenrod) display broader, skewed distributions with extended tails. Despite these marked differences in the statistics of $dS_{tot}$, the inset of Fig.~\ref{fig:FT}(c) demonstrates that each region independently satisfies a local fluctuation theorem, as evidenced by the linear relation $\ln[P(dS_{tot})/P(-dS_{tot})] = dS_{tot}$ with unit slope. These results demonstrate that fluctuation-theorem symmetries persist locally in phase space, providing a stringent, data-driven consistency check on the inferred entropy production.

Next, in Figure~\ref{fig:nBgr}, we show results for $N$-dimensional Brownian gyrator system considered here as a controlled testbed to assess how the proposed inference method scales with system dimensionality and to identify challenges that arise in high-dimensional force and entropy-production inference. The dynamics is described by:
\begin{align}
    \dot{{\bm x}}(t)={\bm A}{\bm x}(t)+{\bm \xi}(t),\label{eq: BGLangevin}
\end{align}
where $\xi_i(t)$ represents Gaussian white noise with correlations: $\langle \xi_i(t) \xi_j(t') \rangle = 2D_{ij}\delta(t-t')$, where the diffusion matrix is diagonal: $D_{ij} = D_i \delta_{ij}$. The stationary state of the system is non-equilibrium whenever $\mathbf{A}^\top \mathbf{D}^{-1} \neq \mathbf{D}^{-1} \mathbf{A}$, with a corresponding entropy production rate $\sigma =
\frac{1}{2} \, \mathrm{Tr}\left( \mathbf{A} \mathbf{C} \left( \mathbf{A}^\top \mathbf{D}^{-1} - \mathbf{D}^{-1} \mathbf{A} \right) \right)$.  Here ${\bm C}$ is the steady state covariance matrix of the system, which solves the Lyapunov equation $\mathbf{A}\mathbf{C} + \mathbf{C}\mathbf{A}^\top + 2\mathbf{D} = 0$. The corresponding thermodynamic force field is given by 
\begin{equation}
\label{eq:th_ffield}
   {\bm F}({\bm x}) = \mathbf{D}^{-1} \left( \mathbf{A} + \mathbf{D} \mathbf{C}^{-1} \right) {\bm x} \ .
\end{equation}
Here, we consider a high-dimensional realization of this system with $N = 100$, having the drift matrix
\begin{equation}
    A_{ij} = \kappa \begin{cases}
        -2 & \text{if } i = j \\
        1 & \text{if } |i-j| = 1 \\
        0 & \text{otherwise.}
    \end{cases}
    \label{eq:AMatrix}
\end{equation}
Further we choose
the temperatures $D_i$ to follow a periodic sawtooth profile (Figure~\ref{fig:nBgr}(a)), with maximum values at $i = N/4$ and
$i = 3N/4$.  This parameter choice, obtained by trial and error, induces strong spatial heterogeneity in the local entropy production, with bead-resolved contributions spanning several orders of magnitude. As a result, the system simultaneously probes strongly irreversible and near-equilibrium regimes within a single, analytically tractable model. This allows us to assess how the inference method performs across widely separated dissipation scales.

In order to apply the inference algorithm, we generate stationary trajectories of length $2\times 10^6$, with a timestep of 0.001s. In Figure~\ref{fig:nBgr}(b-c), we show the results of inferring local dissipation. For better visualization, we color individual trajectories $x_i(t)$ with the local entropy production of each bead ($i$-th bead) defines as 
\begin{equation}
\label{eq:th_localepr}
    dS_{tot}^i(t) = \bm F_{i}({\bm x}(t)) \circ dx_i(t) .
\end{equation}
Figure~\ref{fig:nBgr}(b) depicts the local entropy production using the theoretically known form of ${\bm F}({\bm x})$, while  Figure~\ref{fig:nBgr}(c) shows the same obtained from solving the inference algorithm. As we see, there is good visual agreement between the theory and the results obtained from the inference algorithm.

To quantify the agreement between theory and inference, Figure~\ref{fig:nBgr}(d) shows a scatter plot comparing the analytically computed and learned entropy production, summed over all beads. The data follow a clear linear trend with an $R^2$ value of 0.7576, with noticeably better agreement in the high-dissipation regime. To investigate the origin of the remaining spread, Figure~\ref{fig:nBgr}(e) shows the time-averaged entropy production of each bead. This reveals a separation of roughly two to three orders of magnitude between beads with high and low entropy production, and shows that the discrepancy between theory and inference is noticeably high at the low-dissipation beads. This can be explained by noting that, in these weakly dissipative regions, irreversible signatures are small compared to fluctuations. As a result, the effective signal-to-noise ratio available for training the model is strongly reduced, limiting its ability to generalize uniformly across degrees of freedom with widely separated entropy production scales.
\textcolor{black}{Note that for low-dimensional systems, however, the learning algorithm performs significantly better, as demonstrated in \textbf{Supplementary Note 2}. }

\begin{figure*}[htb]
    \centering
    \includegraphics[width=0.99\linewidth]
{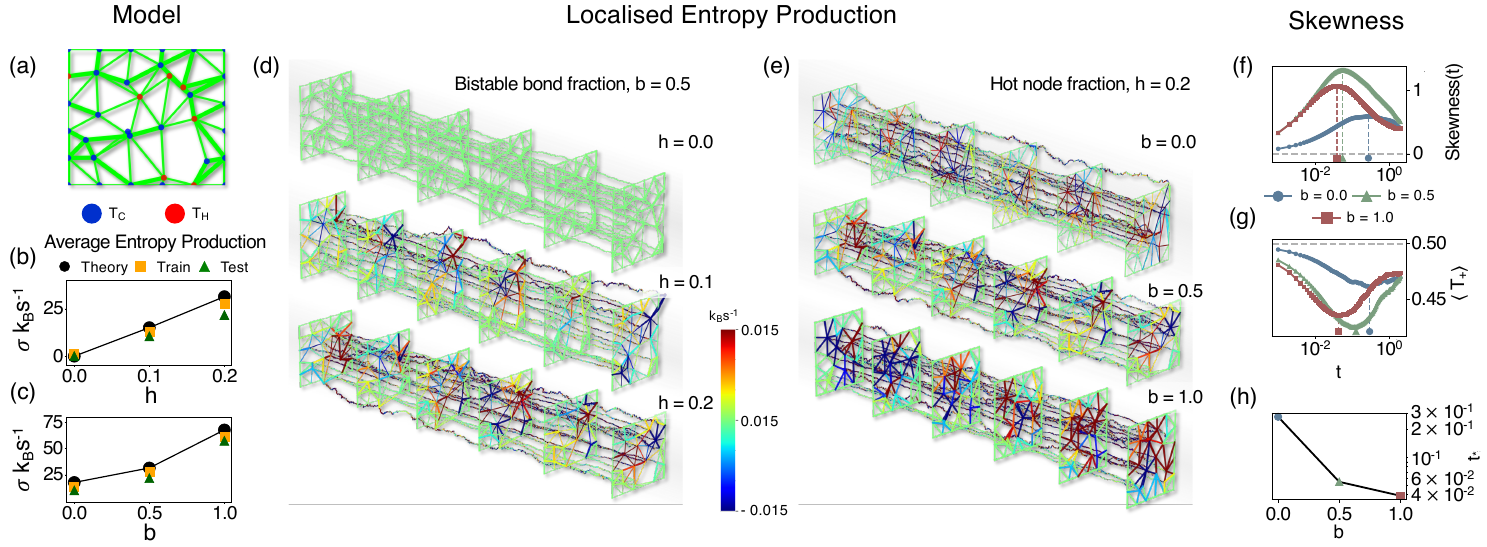}
   \caption{
\textbf{Entropy production and finite-time fluctuations in active--bistable mechanical networks.}
(a) Example disordered two-dimensional spring network with a fraction of hot nodes (red, temperature $T_{\mathrm{hot}}$) and cold nodes (blue, $T_{\mathrm{cold}}$). Thin green bonds denote linear springs, while thick green bonds indicate bistable springs.  
(b,c) Average entropy production rate as a function of the fraction of hot nodes $h$ at fixed bistable bond fraction $b=0.5$ (b), and as a function of the bistable bond fraction $b$ at fixed $h=0.2$ (c). Symbols show inferred values from training and test data, while solid lines indicate theoretical predictions for the total entropy production rate.  
(d) Spatial maps of the inferred local entropy production rate for increasing hot-node fraction $h$ at fixed $b=0.5$. 
(e) Spatial maps of the inferred local entropy production rate for increasing bistable bond fraction $b$ at fixed $h=0.2$.  In (d) and (e), the network bonds are colored by the mean value of dissipation of the two nodes in the bond. Additionally, the time-series corresponds to 1000 consecutive steady state configurations.
(f) Skewness of the time-integrated entropy production $\Delta S_{\rm tot}$ as a function of the integration time $t$ for different bistable bond fractions $b$, showing a pronounced nonmonotonic dependence.  
(g) Fraction of time-integrated entropy production fluctuations lying above the mean, $\langle T_+(t)\rangle = \mathbb{P}(\Delta S_{\rm tot}>\langle \Delta S_{\rm tot}\rangle)$, demonstrating a finite-time bias with $\langle T_+(t)\rangle<1/2$.  
(h) Characteristic integration time $t_\ast$ at which the skewness is maximal, as a function of the bistable bond fraction $b$.
Increasing the fraction of bistable bonds systematically amplifies finite-time asymmetries in cumulative entropy production and shifts the characteristic timescale to shorter values, demonstrating that mechanical nonlinearity enhances emergent non-Gaussian entropy production statistics at experimentally relevant finite times.
}
    \label{fig:bionet}
\end{figure*}

\subsection{Biological model I: Active-bistable mechanical network}

Building on earlier studies of two-temperature systems, we now examine the dynamics of a disordered 2D network with random connectivity and a mix of linear and bistable bonds. This setup is designed to capture essential features of biological networks, from cytoskeletal structures to extracellular matrices, and the goal is to see how well the inferred fluctuating entropy production can quantitatively distinguish different non-equilibrium states of this system. We obtain the computational model by extending the model previously introduced in Ref.\ \cite{gnesotto:lnb} to include bistable bonds, which may represent molecular bonds or cross-links that switch between metastable states under load. We further impose a temperature gradient across nodes which drives nonequilibrium activity.

The system consists of a triangular lattice of $N \times N$ nodes, where particles are connected by two types of springs: (1) linear springs with stiffness $k=1.0$ and rest length $l_0=1.0$, and (2) bistable springs (the fraction of such bonds $b$ can be varied) with a double-well potential characterized by minima at $a_1=0.75l_0$ and $a_2=0.25l_0$ and stiffness $k_{\text{bistable}}=1.0$. The bistable force follows
\begin{equation}
F = -k_{\text{bistable}}(r-a_1)(r-a_2)(2r-a_1-a_2),
\end{equation}
enabling cooperative transitions between extended and contracted states under strain. Connections are initialized via Delaunay triangulation \cite{delaunay1934, SciPy, 2020SciPy-NMeth}, with 10\% of bonds randomly removed to introduce structural disorder, mimicking the heterogeneity of biological networks. One such configuration is shown in Figure~\ref{fig:bionet}(a) with $N = 6$.

The nodes evolve under overdamped Langevin dynamics with a friction coefficient $\gamma=1.0$, subject to forces from their neighbors and thermal noise. A fraction of randomly chosen nodes (that we can vary) are coupled to a ``hot'' thermal bath ($T_{\text{hot}}=1$), while the remainder are held at a ``cold'' temperature ($T_{\text{cold}}=0.1$), creating a sustained temperature gradient. Fixed boundary conditions are imposed mimicking anchored structures (e.g., adherent cell membranes or tissue boundaries). For the analysis, we simulate a stationary trajectory of length $2\times 10^6$ points with timestep $dt = 0.001s$.

 Using our short-time inference scheme, we systematically probe how temperature gradients and mechanical nonlinearities govern spatio-temporal entropy production in disordered elastic networks. Figure~\ref{fig:bionet} shows two complementary sweeps: (A) fixing the fraction of bistable bonds at 50\%  ($b = 0.5$) while varying the fraction of nodes at higher temperature ($h$), and (B) fixing the fraction of hot nodes ($h = 0.2$) while varying the degree of nonlinearity. In both cases, the temperature ratio \( T_{\text{hot}}/T_{\text{cold}} = 10 \) is held constant. As expected, increasing the number of hot nodes leads to higher entropy production due to stronger thermal driving (Figure~\ref{fig:bionet}(b)). More surprisingly, we find that increasing the fraction of bistable bonds---i.e., introducing more mechanical nonlinearity---also systematically enhances entropy production (Figure~\ref{fig:bionet}(c)). This dependence on nonlinearity is nontrivial and highlights the subtle role of local mechanical instabilities in shaping the global non-equilibrium response. The inferred mean entropy production rates are further consistent with the true average total entropy production rate, which we can independently obtain in this case in terms of the medium entropy production~\cite{seifert2005entropy}.

Figure~\ref{fig:bionet}(d) and  Figure~\ref{fig:bionet}(e) demonstrate how dissipation is spatiotemporally organized across the network as the parameters $h$ and $b$ are varied. The inferred local entropy production highlights highly heterogeneous patterns within a single frame across the network, across frames, and across the chosen parameter values.  Access to this information further enables analysis of how they shape finite-time integrated observables. They are of significant practical interest since function in cellular nonequilibrium processes is often encoded in cumulative outputs accumulated over finite operational times. Relatedly, recent theoretical work identified a counterintuitive finite-time effect in nonequilibrium steady states: time-integrated current measurements are more likely to lie below their steady-state average for most of the measurement duration \cite{manikandan2022nonmonotonic}. This finite-time bias was shown to be associated with a \emph{positive skewness} of entropy production fluctuations and a \emph{nonmonotonic dependence} of the skewness on the integration time $t$, implying an optimal timescale at which the discrepancy between above- and below-average fluctuations is maximal. These predictions were previously derived and tested in low-dimensional overdamped systems.

Our results enable a data-driven test of this finite-time property in this experimentally relevant model. From the inferred instantaneous entropy production rate, we construct the time-integrated entropy production $\Delta S_{\rm tot}$ over a sliding window of duration $t$, and quantify (i) its skewness
${\rm Skewness}(t) = \frac{\langle (\Delta S_{\rm tot}-\langle \Delta S_{\rm tot}\rangle)^3\rangle}{\langle (\Delta S_{\rm tot}-\langle \Delta S_{\rm tot}\rangle)^2\rangle^{3/2}}$,
and (ii) the fraction of fluctuations lying above the mean,
$
\langle T_+(t)\rangle = \mathbb{P}\!\left(\Delta S_{\rm tot} > \langle \Delta S_{\rm tot}\rangle\right)$.
As shown in Fig.~\ref{fig:bionet}(f), we observe a pronounced nonmonotonic skewness and find that typical finite-time measurements satisfy $\langle T_+(t)\rangle < 1/2$ (Fig.~\ref{fig:bionet}(g)). Consistent with this behavior, the discrepancy between above- and below-average fluctuations is maximal at a characteristic integration time. Crucially, increasing the fraction of bistable bonds systematically amplifies this finite-time asymmetry and shifts the characteristic timescale to shorter values (Fig.~\ref{fig:bionet}(h)). This implies that mechanical heterogeneity can significantly affect the emergent non-Gaussian statistics of entropy production at experimentally relevant finite times.

\begin{figure*}
    \centering
    \includegraphics[width=0.95\linewidth]{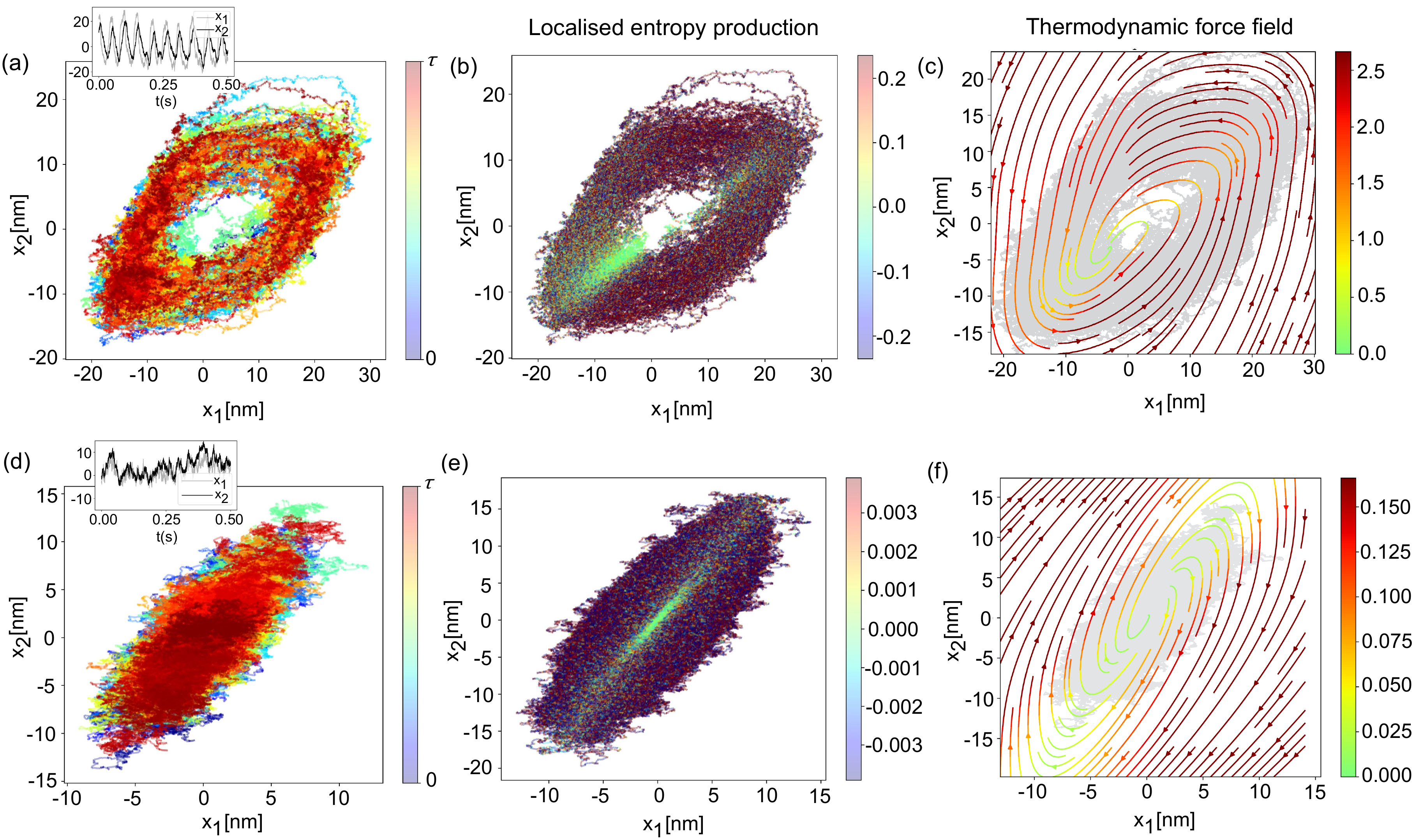}
    \caption{\textbf{Hair-cell bundle oscillation.} (a) Representation of numerical 2D-trajectories in the $(x_1,x_2)$ phase space corresponding to $5s$ simulation sampled at $100\ kHz$. The colorbar indicates progression along the trajectory. The inset plot indicates the oscillatory nature of dynamics for $F_{max} = 57.14\ pN$ and $S = 0.94$. (b) The local entropy production rate (in units of $k_B/s$) is computed using the neural network representation. It captures the active state of dynamics. The colours are for comparative visualisation and do not represent the true value. (c) Thermodynamic force field of the oscillatory state of the dynamics.   (d) Representation of numerical 2D-trajectories in the $(x_1,x_2)$ phase space corresponding to $5s$ simulation sampled at $100\ kHz$. The colorbar indicates progression along the trajectory. The inset plot indicates the quiescent (non-oscillatory) state of dynamics for $F_{max} = 40\ pN$ and $S = 1$. (e) Entropy production rate (in units of $k_B/s$) is locally computed along such trajectories. The colours are for comparative visualisation and do not represent the true value. (f) Thermodynamic force field of the oscillatory state of the dynamics. \textcolor{black}{The colour scale of local entropy production is thresholded symmetrically between $[-10\times\mathrm{median},\, 10\times\mathrm{median}]$ for the oscillatory case and $[-500\times\mathrm{median},\, 500\times\mathrm{median}]$ for the quiescent case, while the colour scale for the force field of both cases is thresholded between  $[0, \mathrm{median}]$}. The other system parameters remain the same as mentioned in Fig.(4) of Ref.~\cite{roldan2021quantifying}: $\gamma_1 = 2.8\ \mu N\ s/m$,  $\gamma_2 = 10\ \mu N\ s/m$, $k_{\text{gs}} = 0.75\ pN/nm$, $k_{\text{sp}} = 0.6\ pN/nm$, $D = 61\ nm$, $N = 50$, $\Delta G = 10k_BT$, $k_BT = 4.143\ pNnm$ and $T_{eff} = 1.5T$.  }
    \label{fig:hair_cell}
\end{figure*}

\subsection{Biological model II: Spontaneous hair-cell bundle oscillations }

Next, we examine another biologically relevant model developed to explain the spontaneous oscillations of hair cells in the bullfrog's inner ear~\cite{roldan2021quantifying}. This highly nonlinear model features two dynamical degrees of freedom: the instantaneous position of the hair bundle tip (\(x_1\)) and the center of mass of the molecular motor (\(x_2\)). Remarkably, the model captures the interplay between mechanosensitive ion channels, molecular motor activity, and calcium feedback, while also elucidating key thermodynamic features of spontaneous hair-bundle oscillations.

Following Ref.~\cite{roldan2021quantifying}, the dynamics of the system are governed by:
\begin{align}
    \begin{split}
        \gamma_1 \dot{x}_1(t) &= -\partial_{x_1} U(x_1, x_2) + \sqrt{2D_1 \gamma_1^2}\, \xi_1(t), \\
        \gamma_2 \dot{x}_2(t) &= -\partial_{x_2} U(x_1, x_2) - f_{\text{act}}(t) + \sqrt{2D_{\text{eff}} \gamma_2^2}\, \xi_2(t),
    \end{split}
    \label{eq:hair_cell}
\end{align}
where \(\gamma_1\) and \(\gamma_2\) are friction coefficients, related to the diffusion terms via \(D_1 = k_B T_1 / \gamma_1\) and \(D_{\text{eff}} = k_B T_{\text{eff}} / \gamma_2\). The nonlinear potential \(U(x_1, x_2)\) (Eq.~(14) in Ref.~\cite{roldan2021quantifying}) encodes conservative forces from elastic elements and the gating dynamics of mechanosensitive ion channels. Its explicit form is:
\begin{align}
    \begin{split}
        U(x_1, x_2) =& \frac{k_{\text{gs}}(x_1 - x_2)^2 + k_{\text{sp}}x_1^2}{2} \\
        &- N k_B T_1 \ln\left[\exp\left(\frac{k_{\text{gs}} D (x_1 - x_2)}{N k_B T_1}\right) + A\right],
    \end{split}
    \label{eq:hair_cell_pot}
\end{align}
with stiffness coefficients \(k_{\text{gs}}\) and \(k_{\text{sp}}\), gating swing \(D\) of a transduction channel, and parameter \(A = \exp[(\Delta G + k_{\text{gs}} D^2 / (2 N)) / (k_B T_1)]\), where \(\Delta G\) is the energy difference between open and closed channels and \(N\) is the number of transduction elements.

The nonequilibrium features originate from the activity of the motor characterised through the higher effective temperature $T_{eff}$ and the active force,
\begin{align}
    F_{act}(t) = F_{max}(1-SP_0(x_1(t),x_2(t))).
\end{align}
Here, the maximum motor force is denoted as $F_{max}$, $S$ controls the
strength of calcium-mediated feedback, and $P_0(x_1(t), x_2(t)))$ [Eq.(16) of Ref.~\cite{roldan2021quantifying}] represents
the probability of the ion channel opening~\cite{nadrowski2004active}, given as, 
\begin{align}
    P_0(x_1,x_2) = \frac{1}{1+ A\ \exp(-k_{\text{gs}}D(x_1 - x_2)/Nk_BT)}.
\end{align}
Note that there remains considerable interest in the estimation of the entropy production rate from the experimental traces of hair-cell oscillation to understand the active feature of the underlying dynamics~\cite{roldan2021quantifying,tucci2022modelling, diterlizzi2025forcefree}. The steady-state mean entropy production rate of the system within this model can be analytically estimated using trajectory energetics~\cite{sekimoto1998langevin,roldan2021quantifying}. However, it would be significantly challenging to obtain the fluctuating entropy production along the phase space trajectory due to the inherent nonlinearity and complexity of the model.
  
In this context, our technique of learning the optimal force field through a neural network representation should be able to resolve entropy production in both space and time. To estimate local entropy production in $(x_1,x_2)$ phase space, Eq.~\eqref{eq:hair_cell} are numerically solved and $2D$ trajectory data is generated.  
As reported earlier in Ref.~\cite{roldan2021quantifying}, the dynamics can be oscillatory or quiescent depending on the value of maximum motor force $F_{max}$ and the feedback strength $S$, and we consider these two cases separately.  We first look into the oscillatory dynamics with $F_{max} = 57.14\ pN$ and $S = 0.94$ - which correspond to the active state of the system (Figure~\ref{fig:hair_cell}(a)). \textcolor{black}{In this case we found that a 5s long trajectory with a time step $\Delta t = 10^{-5}$  was sufficient to ensure convergence of the inference algorithm}. The corresponding localised entropy production is mostly high and homogeneous over the 2D phase space, as shown in Figure~\ref{fig:hair_cell}(b). The computed thermodynamic force field is visualised to be a strongly circulating flow, indicating a strongly irreversible state of the dynamics. On the contrary, the dynamics with $F_{max} = 40 \ pN$ and $S = 1$ shows non-oscillatory features (Figure~\ref{fig:hair_cell}(d)).  \textcolor{black}{In this case we simulated 100s long trajectories with a time step $\Delta t = 10^{-5}$ to ensure convergence of the inference algorithm, and to avoid overfitting}. The quiescent state of the system is characterised by a significantly lower entropy production rate as can be visualised both in the local entropy production (Figure~\ref{fig:hair_cell}(e)) and in corresponding thermodynamic force field  (Figure~\ref{fig:hair_cell}(f)). This shows that in these regimes, a large part of phase space operate close to thermodynamic equilibrium. Together, these results show that local entropy production resolves how irreversibility is distributed across phase space rather than being uniformly associated with a dynamical state. 

\begin{figure}[h]
    \centering
\includegraphics[width=0.99\linewidth]{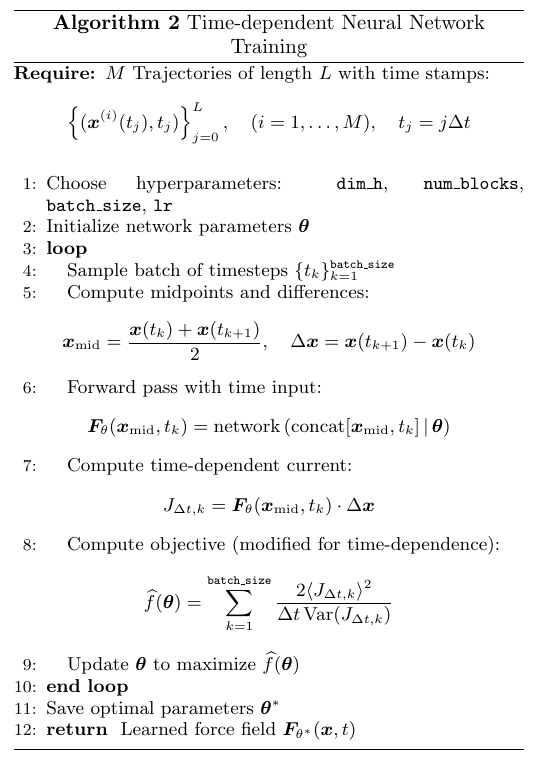}
\end{figure}

\subsection{Time-dependent nonequilibrium process: bit-erasure protocol}

\begin{figure*}
    \centering
\includegraphics[width=0.99\linewidth]{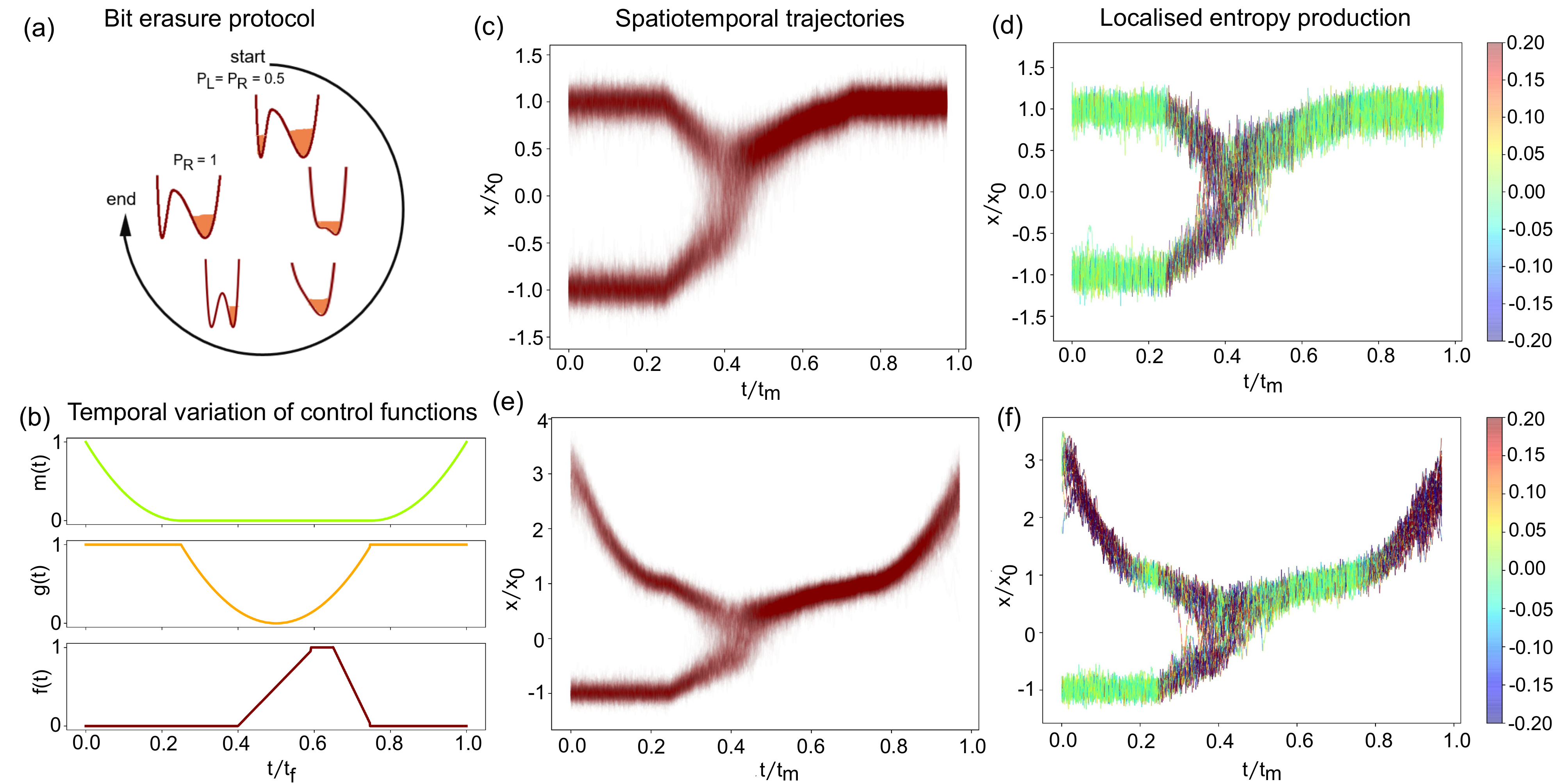}
    \caption{\textbf{Bit-erasure protocol.} (a) Schematic of a bit-erasure protocol. The probability of finding the particle in one of the wells is equal ($P_L = P_R = 0.5$) at the onset and at the end of the protocol, the particle ends up in the right well with $P_R = 1$.  (b) Plot of the time-dependent control functions: $m(t)$, $g(t)$ and $f(t)$. We used the functional form of these functions as provided in the \textit{supplementary} of Ref.~\cite{singh2024inferring}. (c) Representative trajectories undergoing erasing protocol in a symmetric potential ($\eta = 1$). (d) Trajectories with the localised entropy production rate (in units of $k_B/s$) of the particle during the erasing process with a symmetric potential. (e) Representative trajectories undergoing erasing protocol in an asymmetric potential ($\eta = 3$) (f) Trajectories with the localised entropy production rate (in units of $k_B/s$) of the particle during the erasing process with an asymmetric potential ($\eta = 3$). {\color{black} In both cases, the local entropy production obtained here differs from the total entropy production by an additional term arising from the explicit time dependence of the probability density, namely $-\partial_t \ln p(\bm x,t)\,\mathrm dt$ as discussed in the methods section}.  
    [Parameters: $E_b = 13k_BT$, $k_BT = 4.14\times 10^{-3} pN\mu m$, $A = 0.3$, $x_m = 0.77\mu m$, $D = 0.23 \mu m^2/s$, $t_m \equiv (2x_0)^2/D = 10s$.]}
    \label{fig:bit_erasure}
\end{figure*}

Thus far, we have considered nonequilibrium systems in stationary states, where the thermodynamic force field is time-independent. In contrast, for explicitly time-dependent processes, both the thermodynamic force field and the entropy production rate become time-varying quantities. These temporal dependencies introduce significant computational challenges, making direct estimation of these quantities highly nontrivial. Interestingly, a modified version of the inference algorithm can be utilised to still determine the time-dependent thermodynamic force field. As described in the Methods section, this force field can then be used isolate the associated irreversible contributions to total entropy production ~\cite{otsubo2022estimating}. We provide a pseudocode in \textbf{Algorithm 2}.

For a demonstration, here we examine the well-studied process: finite-time bit erasure, where a single bit of information (in a coarse-grained phase space) is erased using a finite-duration protocol. Recent experimental studies have implemented various bit-erasure protocols in colloidal systems with feedback-driven optical traps~\cite{berut2012experimental,jun2014high,gavrilov2016erasure}. In such systems, a bit is typically represented by the coarse-grained position fluctuations of a colloidal particle in a double-well potential. The erasure protocol begins with the particle in either the left well (L, state 0) or the right well (R, state 1), and modulates the potential to ensure the particle always ends up in the right well (R, state 1). 

We numerically implement a protocol using the time-dependent asymmetric double-well potential experimentally realized by Gavrilov~\textit{et al.}~\cite{gavrilov2016erasure}, given by:
\begin{align}
    U(x,t) = 4E_b \left[-\frac{g(t)}{2}\tilde{x}^2 + \frac{\tilde{x}^4}{4} - Af(t)\tilde{x}\right],
\end{align}
where
\begin{align}
\tilde{x} =
\begin{cases} 
\frac{x}{x_0}, & \text{if } x < 0, \\[10pt]
\frac{x}{x_0} \left[1 + m(t)(\eta - 1)\right]^{-1}, & \text{if } x \geq 0.
\end{cases}
\end{align}
Here, $E_b$ is the barrier height between the potential minima, and $g(t)$, $f(t)$, $m(t)$ are time-dependent control functions. Initially, the potential minima are located at $-x_0$ and $+\eta x_0$, with $\eta$ as the asymmetry parameter. The protocol manipulates these control functions to drive the particle into the right well (for $A > 0$), thereby erasing the 1-bit information about its initial state (Figure~\ref{fig:bit_erasure}(a)). Here we numerically realize a \textit{fast} erasing process (cycle time ($t_f$) $\sim$ mean diffusion time ($t_m$)) numerically by solving the corresponding Langevin equation of the particle, $\gamma \dot{x}(t) = -\ \partial_x U(x,t) + \sqrt{2D \gamma^2}\xi(t)$, with the first-order Euler-Maryama integration algorithm.  The time-dependent control functions of the potential that are used to carry out the erasing operation are shown in Figure~\ref{fig:bit_erasure}(b). Note that, we used the time-dependent functional form of these functions as provided in the supplementary material of Ref.~\cite{singh2024inferring}. Representative trajectories following this time-dependent protocol for both symmetric and asymmetric potential are shown in  Figure~\ref{fig:bit_erasure}(c) and  Figure~\ref{fig:bit_erasure}(e), respectively.   We then apply the inference technique on ensembles of trajectories starting from both left and right wells to map out the spatio-temporal dissipation of the process.

Figure~\ref{fig:bit_erasure}(d) and Figure~\ref{fig:bit_erasure}(f) illustrate the local entropy production along individual bit-erasure trajectories.  We stress that these are contributions to total entropy production  arising solely from the thermodynamic force field. It allows one to directly identify entropy production generated by irreversible forces acting along individual trajectories. As a result, the observed bursts and suppressions of entropy production can be unambiguously attributed to genuinely irreversible dynamics at the trajectory level.
Figure~\ref{fig:bit_erasure}(d) demonstrates that for a symmetric potential ($\eta = 1$), trajectories starting from both wells exhibit low entropy production until $t = 0.2 t_m$, followed by an increase in the interval $0.2 t_m < t < 0.5 t_m$. In contrast, the asymmetric case ($\eta = 3$) in Figure~\ref{fig:bit_erasure}(f) shows trajectories originating from the right well with high entropy production throughout most of the interval $0 < t < 0.5 t_m$, reflecting sustained dissipation during erasure. {\color{black} We note that, in either case, the observed negative fluctuations do not necessarily correspond to second-law--violating events, since determining such events requires including the additional contribution to total entropy production arising from the time dependence of the probability density.}
\begin{figure}
    \centering
    \includegraphics[width=0.95\linewidth]{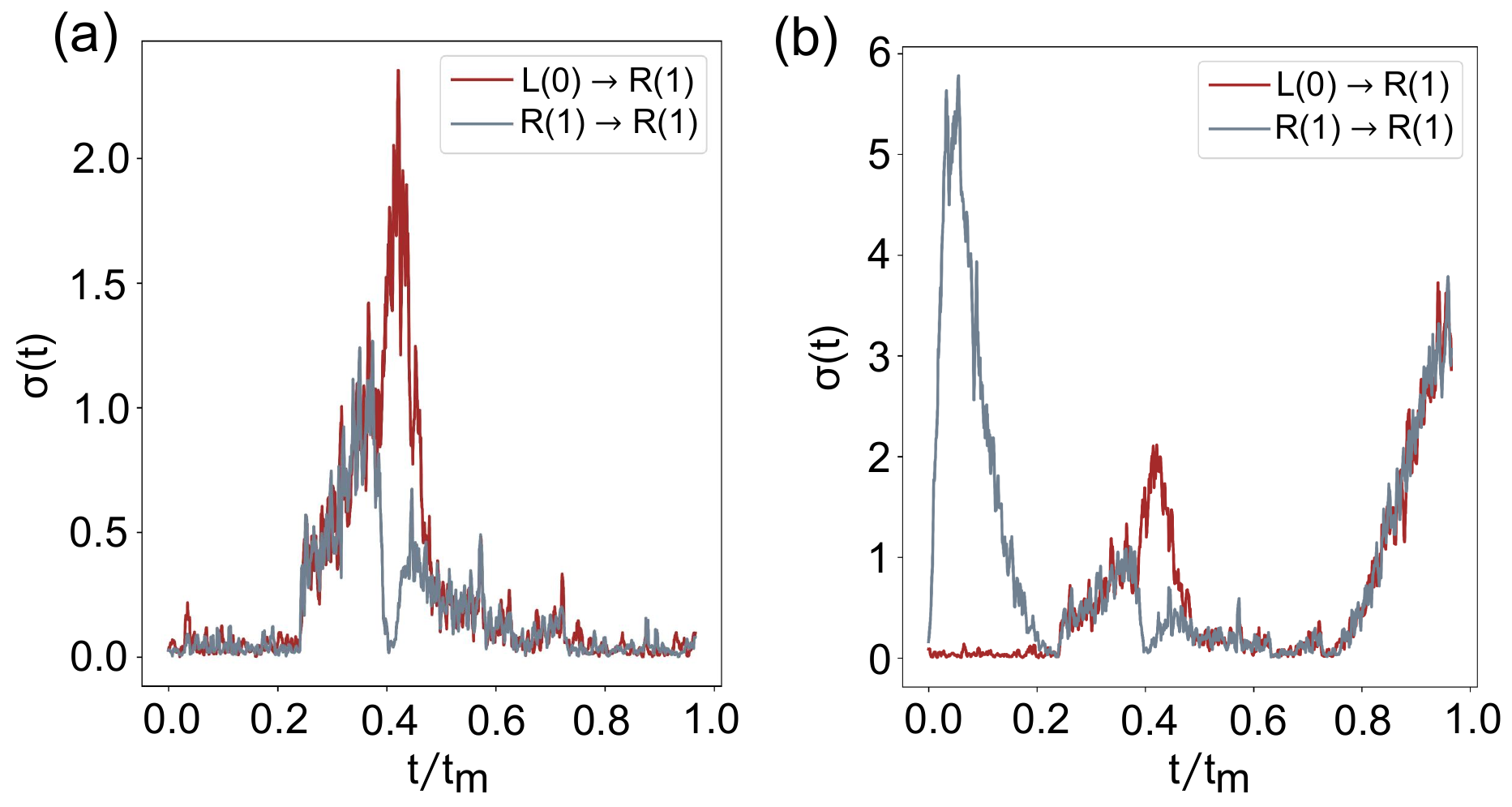}
    \caption{\textcolor{black}{\textbf{Average entropy production of bit-erasure protocol.} (a) and (b) are the plots of the time-averaged entropy production rate during the erasure protocol (with the same parameters as in Figure~\ref{fig:bit_erasure}) for the symmetric and asymmetric potential, respectively. Those curves are smoothened with moving averages over 5 points. } }
    \label{fig:bit_erasure_avg_epr}
\end{figure}

\begin{figure*}
    \centering
    \includegraphics[width=0.9\linewidth]{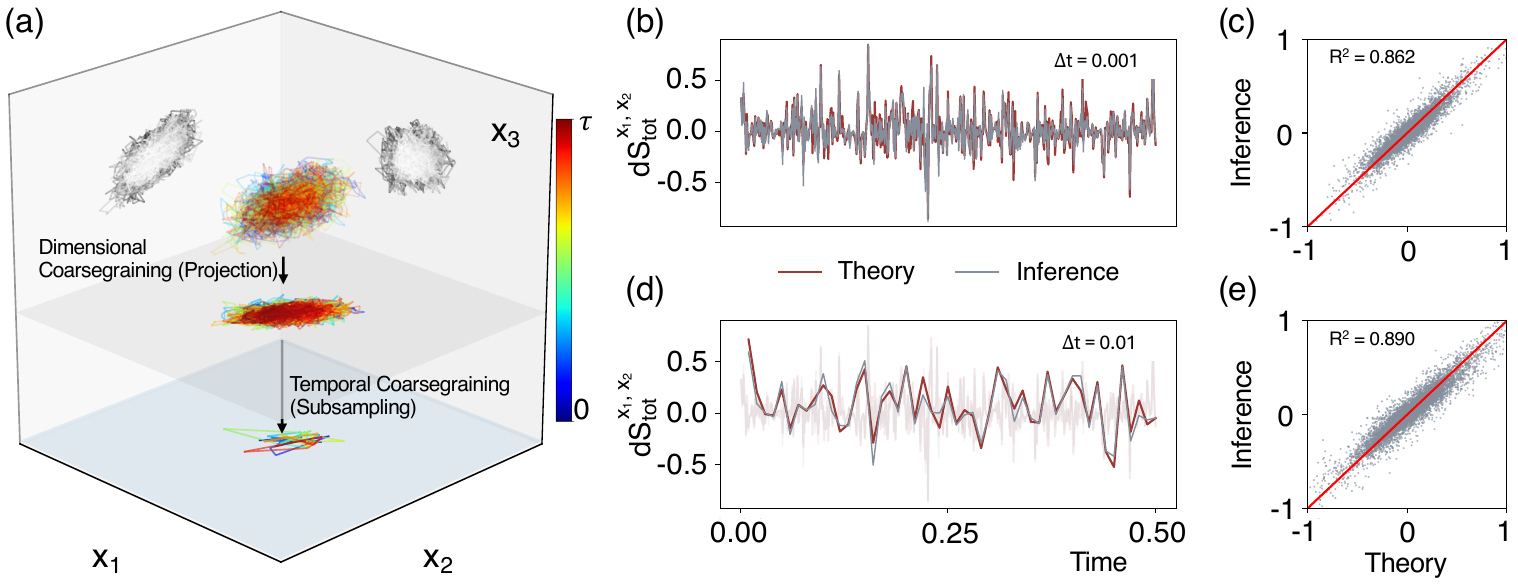}
    \caption{\textbf{Coarse-graining and hidden degrees of freedom.}  (a) Schematic of dimensional and temporal coarse-graining of the 3D harmonic gyrator model: Dimensional coarse-graining is performed by projecting the full dynamics ($x_1,x_2,x_3$) to reduced subspaces ($x_1,x_2$) while temporal coarse-graining is performed by subsampling the trajectories. (b) The fluctuating entropy production rate inferred from 
    numerical trajectories corresponding to the reduced dynamics ($(x_1,x_2)$ phase space) match well with the analytical values. (c) The inferred entropy production shows linear correlation with the theoretical values with $R^2 = 0.862$. (d) Fluctuating entropy production inferred from the subsampled trajectories ($\Delta t = 0.01$) closely follow the analytical values as well. The fine-grained entropy production from (b) is plotted in the background for a direct comparison. (e) The corresponding scatter plot of these two estimates shows a strong linear correlation with $R^2 = 0.890$.
 }
    \label{fig:coarsegraining}
\end{figure*}

 A closer inspection of both panels also reveals short temporal segments near $t = 0.4t_m$ in which bunch of few individual trajectories starting off the top well traverse the transition region (roughly identified here between $0.3 t_m < t < 0.5 t_m$ ) with nearly reversible dynamics. These segments could be associated with a momentary stationary particle as the specific erasure protocol switches the direction of the drive. During the remaining erasure time, the symmetric case maintains predominantly low entropy production, while the asymmetric case show between periods of low and high entropy production. Additionally, Figure~\ref{fig:bit_erasure_avg_epr}(a) and Figure~\ref{fig:bit_erasure_avg_epr}(b) display the time-averaged entropy production rate, which corroborates these trajectory-based observations. Note that the global estimate of entropy production rate of this process, which is traditionally computed \cite{singh2024inferring}, misses out on the distinct features that emerge for erasure from different wells or their explicit time-dependencies.

\subsection{Effect of Coarse-graining }
In many experimental and numerical settings, entropy production must be inferred from data that provide only a reduced description of the underlying dynamics. Such limitations arise both from hidden degrees of freedom, when only a subset of system variables is observed, and from finite temporal resolution, when trajectories are sampled at discrete time intervals. Both forms of coarse-graining modify the observed dynamics and can significantly alter entropy-production fluctuations.

We investigate the effect of these constraints on entropy-production inference using a three-dimensional variant of the Brownian gyrator model introduced in Sec.~\ref{sec:applications}A. As shown in Fig.~\ref{fig:coarsegraining}(a), we first perform dimensional coarse-graining by projecting the full dynamics onto a two-dimensional subspace, thereby rendering one degree of freedom unobserved. We then impose temporal coarse-graining by subsampling the resulting trajectory, retaining only every tenth data point. This setup provides a controlled framework for testing inference under reduced observational access.

Interestingly, theoretical benchmarks for both dimensional and temporal coarse-graining are already known in the literature. In the case of hidden degrees of freedom, the benchmark follows from a reduced phase-space description in which the influence of unobserved variables is absorbed into the drift term of the low-dimensional Langevin equation \cite{nicoletti2024tuning}. For temporal coarse-graining, an effective description can be constructed from the propagator of the underlying continuous-time dynamics \cite{Weiss}. Details of both constructions are provided in the \textbf{Supplementary Note 3}.

We find that the inferred entropy production in each case remains in close agreement with these theoretical benchmarks. This is shown in Fig.~\ref{fig:coarsegraining}(b,d) using the time series of the total entropy production along the trajectory and with scatter plots in Fig.~\ref{fig:coarsegraining}(c,e). In the scatter plots, we get $R^2 \simeq 0.9$. {\color{black} We emphasize that the corresponding learned force field should be interpreted as an effective coarse-grained force at the chosen resolution. Additionally, the controlled benchmark considered here is available only for linear models, where analytical reduced descriptions can be derived.}

\section{Conclusions}
\label{sec:conclusion}
In this work, we have demonstrated the applicability of the short-time inference scheme, previously proposed in Refs.~\cite{manikandan2020inferring,van2020entropy, manikandan2021quantitative, otsubo2022estimating} to localize fluctuating entropy production, as defined in \cite{seifert2005entropy}, along individual non-equilibrium trajectories. This demonstrates, for overdamped diffusive processes in a steady state, a direct link between the original thermodynamic uncertainty relation framework~\cite{barato2015thermodynamic} and the trajectory-level entropy production formalism introduced in Ref.~\cite{seifert2005entropy}. The results also show that entropy production, its fluctuations and statistical properties can vary substantially across phase space, yet showcasing predictive and testable characteristics both at the fine-grained and coarse-grained levels. 

It will be interesting to test the applicability of this method in a wide range of experimental and theoretical contexts where irreversibility and entropy production naturally arise~\cite{dabelow2019irreversibility,seara2021irreversibility,mori2023entropy,das2025irreversibility}. In biological systems, such as cellular signaling networks or metabolic pathways, the method could help quantify and spatiotemporally localize energy dissipation in a data-driven manner. This could be applied to a range of single-molecule tracking experiments to detect and quantify the underlying dissipative dynamics at molecular scales, such as the operation of molecular motors~\cite{ariga2018nonequilibrium}, driven conformational changes of proteins~\cite{stigler2011complex}, or enzymatic activity. The method could also be applied to active matter systems, such as self-propelled colloids or bacterial colonies, to resolve how localized energy dissipation drives collective behavior and self-organization~\cite{bechinger2016active,bowick2022symmetry}. Our ideas can be extended to coarse-grained processes where Markovian events can be identified as introduced in Refs~\cite{degunther2024fluctuating,degunther2024general} to consistently obtain fluctuating coarse grained entropy production along the trajectory. Additionally, extending this framework to high-dimensional systems, such as gene regulatory networks~\cite{tyson2001network} or spatially extended ecological systems~\cite{padmanabha2024landscape}, or neural systems \cite{lynn2021broken} could provide insights into the thermodynamic constraints shaping their dynamics. 

 The ability to infer entropy production locally, also opens the door to next-generation inverse-design \cite{ronellenfitsch2019inverse, chennakesavalu2024adaptive} and optimal, minimum dissipation control strategies for nonequilibrium systems \cite{aurell2011optimal, jeremie, klinger2025minimally, zhong2022limited}. Our results could enable feedback-guided schemes that identify and adaptively concentrate sampling in high-dissipation regions. They can also offer the quantitative inputs needed to construct optimal local controllers that intervene only where energetically costly dynamics occur. Similarly, for inverse-design applications, our methods could provide guiding principles for the assembly of systems, such as lattices of colloids \cite{liu2025optical}, with distinct local spatiotemporal nonequilibrium characteristics arising from tailored driving protocols. We intend to explore some of these aspects in future work.
 
\section{Acknowledgements}
BD is thankful to the Ministry Of Education of
Government of India for financial support through the Prime Minister’s Research Fellowship (PMRF) grant. SKM acknowledges the Knut and Alice Wallenberg Foundation for financial support through Grant No. KAW 2021.0328. SKM acknowledges discussions with Supriya Krishnamurthy, especially for pointing out Ref.~\cite{Weiss}. Parts of the computations and data handling were enabled by resources provided by the National Academic Infrastructure for Supercomputing in Sweden (NAISS), partially funded by the Swedish Research Council through grant agreement no. 2022-06725.

\section{Data  availability }
The datasets generated and analysed during the current study are available from the corresponding author (SKM) upon request.

\section{Code availability }
The source code for generating the data and implementing the inference algorithm for each case will be openly available in \texttt{https://doi.org/10.5281/zenodo.19057023}.  
\section{Author Contributions}
SKM conceived the idea for the project. BD and SKM designed the research framework, implemented the algorithm, and carried out the numerical studies. Both authors checked and discussed the results and wrote the manuscript. In this study, part of the work was conducted by SKM while employed at Stanford University, and the remaining part was carried out while affiliated with Gothenburg University.
\section{Competing Interests}
The authors declare no competing interests.


\begin{thebibliography}{85}%
\makeatletter
\providecommand \@ifxundefined [1]{%
 \@ifx{#1\undefined}
}%
\providecommand \@ifnum [1]{%
 \ifnum #1\expandafter \@firstoftwo
 \else \expandafter \@secondoftwo
 \fi
}%
\providecommand \@ifx [1]{%
 \ifx #1\expandafter \@firstoftwo
 \else \expandafter \@secondoftwo
 \fi
}%
\providecommand \natexlab [1]{#1}%
\providecommand \enquote  [1]{``#1''}%
\providecommand \bibnamefont  [1]{#1}%
\providecommand \bibfnamefont [1]{#1}%
\providecommand \citenamefont [1]{#1}%
\providecommand \href@noop [0]{\@secondoftwo}%
\providecommand \href [0]{\begingroup \@sanitize@url \@href}%
\providecommand \@href[1]{\@@startlink{#1}\@@href}%
\providecommand \@@href[1]{\endgroup#1\@@endlink}%
\providecommand \@sanitize@url [0]{\catcode `\\12\catcode `\$12\catcode `\&12\catcode `\#12\catcode `\^12\catcode `\_12\catcode `\%12\relax}%
\providecommand \@@startlink[1]{}%
\providecommand \@@endlink[0]{}%
\providecommand \url  [0]{\begingroup\@sanitize@url \@url }%
\providecommand \@url [1]{\endgroup\@href {#1}{\urlprefix }}%
\providecommand \urlprefix  [0]{URL }%
\providecommand \Eprint [0]{\href }%
\providecommand \doibase [0]{https://doi.org/}%
\providecommand \selectlanguage [0]{\@gobble}%
\providecommand \bibinfo  [0]{\@secondoftwo}%
\providecommand \bibfield  [0]{\@secondoftwo}%
\providecommand \translation [1]{[#1]}%
\providecommand \BibitemOpen [0]{}%
\providecommand \bibitemStop [0]{}%
\providecommand \bibitemNoStop [0]{.\EOS\space}%
\providecommand \EOS [0]{\spacefactor3000\relax}%
\providecommand \BibitemShut  [1]{\csname bibitem#1\endcsname}%
\let\auto@bib@innerbib\@empty
\bibitem [{\citenamefont {Liphardt}\ \emph {et~al.}(2005)\citenamefont {Liphardt}, \citenamefont {Ritort},\ and\ \citenamefont {Bustamante}}]{bustamante_phystoday_2005}%
  \BibitemOpen
  \bibfield  {author} {\bibinfo {author} {\bibfnamefont {J.}~\bibnamefont {Liphardt}}, \bibinfo {author} {\bibfnamefont {F.}~\bibnamefont {Ritort}},\ and\ \bibinfo {author} {\bibfnamefont {C.}~\bibnamefont {Bustamante}},\ }\bibfield  {title} {\bibinfo {title} {The nonequilibrium thermodynamics of small systems},\ }\href{https://doi.org/10.1063/1.2012462} {\bibfield  {journal} {\bibinfo  {journal} {Phys. Today}\ }\textbf {\bibinfo {volume} {58}},\ \bibinfo {pages} {43} (\bibinfo {year} {2005})}\BibitemShut {NoStop}%
\bibitem [{\citenamefont {Battle}\ \emph {et~al.}(2016)\citenamefont {Battle}, \citenamefont {Broedersz}, \citenamefont {Fakhri}, \citenamefont {Geyer}, \citenamefont {Howard}, \citenamefont {Schmidt},\ and\ \citenamefont {MacKintosh}}]{battle2016broken}%
  \BibitemOpen
  \bibfield  {author} {\bibinfo {author} {\bibfnamefont {C.}~\bibnamefont {Battle}}, \bibinfo {author} {\bibfnamefont {C.~P.}\ \bibnamefont {Broedersz}}, \bibinfo {author} {\bibfnamefont {N.}~\bibnamefont {Fakhri}}, \bibinfo {author} {\bibfnamefont {V.~F.}\ \bibnamefont {Geyer}}, \bibinfo {author} {\bibfnamefont {J.}~\bibnamefont {Howard}}, \bibinfo {author} {\bibfnamefont {C.~F.}\ \bibnamefont {Schmidt}},\ and\ \bibinfo {author} {\bibfnamefont {F.~C.}\ \bibnamefont {MacKintosh}},\ }\bibfield  {title} {\bibinfo {title} {Broken detailed balance at mesoscopic scales in active biological systems},\ }\href{https://doi.org/10.1126/science.aac8167} {\bibfield  {journal} {\bibinfo  {journal} {Science}\ }\textbf {\bibinfo {volume} {352}},\ \bibinfo {pages} {604} (\bibinfo {year} {2016})}\BibitemShut {NoStop}%
\bibitem [{\citenamefont {Gnesotto}\ \emph {et~al.}(2018)\citenamefont {Gnesotto}, \citenamefont {Mura}, \citenamefont {Gladrow},\ and\ \citenamefont {Broedersz}}]{gnesotto2018broken}%
  \BibitemOpen
  \bibfield  {author} {\bibinfo {author} {\bibfnamefont {F.~S.}\ \bibnamefont {Gnesotto}}, \bibinfo {author} {\bibfnamefont {F.}~\bibnamefont {Mura}}, \bibinfo {author} {\bibfnamefont {J.}~\bibnamefont {Gladrow}},\ and\ \bibinfo {author} {\bibfnamefont {C.~P.}\ \bibnamefont {Broedersz}},\ }\bibfield  {title} {\bibinfo {title} {Broken detailed balance and non-equilibrium dynamics in living systems: a review},\ }\href{https://doi.org/10.1088/1361-6633/aab3ed} {\bibfield  {journal} {\bibinfo  {journal} {Reports on Progress in Physics}\ }\textbf {\bibinfo {volume} {81}},\ \bibinfo {pages} {066601} (\bibinfo {year} {2018})}\BibitemShut {NoStop}%
\bibitem [{\citenamefont {Rold{\'a}n}\ and\ \citenamefont {Parrondo}(2010)}]{roldan2010estimating}%
  \BibitemOpen
  \bibfield  {author} {\bibinfo {author} {\bibfnamefont {{\'E}.}~\bibnamefont {Rold{\'a}n}}\ and\ \bibinfo {author} {\bibfnamefont {J.~M.}\ \bibnamefont {Parrondo}},\ }\bibfield  {title} {\bibinfo {title} {Estimating dissipation from single stationary trajectories},\ }\href{https://doi.org/10.1103/PhysRevLett.105.150607} {\bibfield  {journal} {\bibinfo  {journal} {Physical review letters}\ }\textbf {\bibinfo {volume} {105}},\ \bibinfo {pages} {150607} (\bibinfo {year} {2010})}\BibitemShut {NoStop}%
\bibitem [{\citenamefont {Lander}\ \emph {et~al.}(2012)\citenamefont {Lander}, \citenamefont {Mehl}, \citenamefont {Blickle}, \citenamefont {Bechinger},\ and\ \citenamefont {Seifert}}]{noninvasive}%
  \BibitemOpen
  \bibfield  {author} {\bibinfo {author} {\bibfnamefont {B.}~\bibnamefont {Lander}}, \bibinfo {author} {\bibfnamefont {J.}~\bibnamefont {Mehl}}, \bibinfo {author} {\bibfnamefont {V.}~\bibnamefont {Blickle}}, \bibinfo {author} {\bibfnamefont {C.}~\bibnamefont {Bechinger}},\ and\ \bibinfo {author} {\bibfnamefont {U.}~\bibnamefont {Seifert}},\ }\bibfield  {title} {\bibinfo {title} {Noninvasive measurement of dissipation in colloidal systems},\ }\href{https://doi.org/10.1103/PhysRevE.86.030401} {\bibfield  {journal} {\bibinfo  {journal} {Physical Review E}\ }\textbf {\bibinfo {volume} {86}},\ \bibinfo {pages} {030401} (\bibinfo {year} {2012})}\BibitemShut {NoStop}%
\bibitem [{\citenamefont {Harada}\ and\ \citenamefont {Sasa}(2005)}]{harada2005equality}%
  \BibitemOpen
  \bibfield  {author} {\bibinfo {author} {\bibfnamefont {T.}~\bibnamefont {Harada}}\ and\ \bibinfo {author} {\bibfnamefont {S.-i.}\ \bibnamefont {Sasa}},\ }\bibfield  {title} {\bibinfo {title} {Equality connecting energy dissipation with a violation of the fluctuation-response relation},\ }\href{https://doi.org/10.1103/PhysRevLett.95.130602} {\bibfield  {journal} {\bibinfo  {journal} {Physical review letters}\ }\textbf {\bibinfo {volume} {95}},\ \bibinfo {pages} {130602} (\bibinfo {year} {2005})}\BibitemShut {NoStop}%
\bibitem [{\citenamefont {Frishman}\ and\ \citenamefont {Ronceray}(2020)}]{frishman:sfi}%
  \BibitemOpen
  \bibfield  {author} {\bibinfo {author} {\bibfnamefont {A.}~\bibnamefont {Frishman}}\ and\ \bibinfo {author} {\bibfnamefont {P.}~\bibnamefont {Ronceray}},\ }\bibfield  {title} {\bibinfo {title} {Learning force fields from stochastic trajectories},\ }\href{https://doi.org/10.1103/PhysRevX.10.021009} {\bibfield  {journal} {\bibinfo  {journal} {Physical Review X}\ }\textbf {\bibinfo {volume} {10}},\ \bibinfo {pages} {021009} (\bibinfo {year} {2020})}\BibitemShut {NoStop}%
\bibitem [{\citenamefont {Gnesotto}\ \emph {et~al.}(2020)\citenamefont {Gnesotto}, \citenamefont {Gradziuk}, \citenamefont {Ronceray},\ and\ \citenamefont {Broedersz}}]{gnesotto:lnb}%
  \BibitemOpen
  \bibfield  {author} {\bibinfo {author} {\bibfnamefont {F.~S.}\ \bibnamefont {Gnesotto}}, \bibinfo {author} {\bibfnamefont {G.}~\bibnamefont {Gradziuk}}, \bibinfo {author} {\bibfnamefont {P.}~\bibnamefont {Ronceray}},\ and\ \bibinfo {author} {\bibfnamefont {C.~P.}\ \bibnamefont {Broedersz}},\ }\bibfield  {title} {\bibinfo {title} {Learning the non-equilibrium dynamics of brownian movies},\ }\href {https://doi.org/10.1038/s41467-020-18796-9} {\bibfield  {journal} {\bibinfo  {journal} {Nature communications}\ }\textbf {\bibinfo {volume} {11}},\ \bibinfo {pages} {5378} (\bibinfo {year} {2020})}\BibitemShut {NoStop}%
\bibitem [{\citenamefont {Manikandan}\ \emph {et~al.}(2020)\citenamefont {Manikandan}, \citenamefont {Gupta},\ and\ \citenamefont {Krishnamurthy}}]{manikandan2020inferring}%
  \BibitemOpen
  \bibfield  {author} {\bibinfo {author} {\bibfnamefont {S.~K.}\ \bibnamefont {Manikandan}}, \bibinfo {author} {\bibfnamefont {D.}~\bibnamefont {Gupta}},\ and\ \bibinfo {author} {\bibfnamefont {S.}~\bibnamefont {Krishnamurthy}},\ }\bibfield  {title} {\bibinfo {title} {Inferring entropy production from short experiments},\ }\href {https://doi.org/10.1103/PhysRevLett.124.120603} {\bibfield  {journal} {\bibinfo  {journal} {Physical Review Letters}\ }\textbf {\bibinfo {volume} {124}},\ \bibinfo {pages} {120603} (\bibinfo {year} {2020})}\BibitemShut {NoStop}%
\bibitem [{\citenamefont {Manikandan}\ \emph {et~al.}(2024)\citenamefont {Manikandan}, \citenamefont {Ghosh}, \citenamefont {Mandal}, \citenamefont {Biswas}, \citenamefont {Sinha},\ and\ \citenamefont {Mitra}}]{manikandan2024estimate}%
  \BibitemOpen
  \bibfield  {author} {\bibinfo {author} {\bibfnamefont {S.~K.}\ \bibnamefont {Manikandan}}, \bibinfo {author} {\bibfnamefont {T.}~\bibnamefont {Ghosh}}, \bibinfo {author} {\bibfnamefont {T.}~\bibnamefont {Mandal}}, \bibinfo {author} {\bibfnamefont {A.}~\bibnamefont {Biswas}}, \bibinfo {author} {\bibfnamefont {B.}~\bibnamefont {Sinha}},\ and\ \bibinfo {author} {\bibfnamefont {D.}~\bibnamefont {Mitra}},\ }\bibfield  {title} {\bibinfo {title} {Estimate of entropy production rate can spatiotemporally resolve the active nature of cell flickering},\ }\href {https://doi.org/10.1103/PhysRevResearch.6.023310} {\bibfield  {journal} {\bibinfo  {journal} {Phys. Rev. Res.}\ }\textbf {\bibinfo {volume} {6}},\ \bibinfo {pages} {023310} (\bibinfo {year} {2024})}\BibitemShut {NoStop}%
\bibitem [{\citenamefont {Boffi}\ and\ \citenamefont {Vanden-Eijnden}(2024{\natexlab{a}})}]{boffi2024deep}%
  \BibitemOpen
  \bibfield  {author} {\bibinfo {author} {\bibfnamefont {N.~M.}\ \bibnamefont {Boffi}}\ and\ \bibinfo {author} {\bibfnamefont {E.}~\bibnamefont {Vanden-Eijnden}},\ }\bibfield  {title} {\bibinfo {title} {Deep learning probability flows and entropy production rates in active matter},\ }\href{https://doi.org/10.1073/pnas.2318106121} {\bibfield  {journal} {\bibinfo  {journal} {Proceedings of the National Academy of Sciences}\ }\textbf {\bibinfo {volume} {121}},\ \bibinfo {pages} {e2318106121} (\bibinfo {year} {2024}{\natexlab{a}})}\BibitemShut {NoStop}%
\bibitem [{\citenamefont {Rold{\'a}n}\ \emph {et~al.}(2021)\citenamefont {Rold{\'a}n}, \citenamefont {Barral}, \citenamefont {Martin}, \citenamefont {Parrondo},\ and\ \citenamefont {J{\"u}licher}}]{roldan2021quantifying}%
  \BibitemOpen
  \bibfield  {author} {\bibinfo {author} {\bibfnamefont {{\'E}.}~\bibnamefont {Rold{\'a}n}}, \bibinfo {author} {\bibfnamefont {J.}~\bibnamefont {Barral}}, \bibinfo {author} {\bibfnamefont {P.}~\bibnamefont {Martin}}, \bibinfo {author} {\bibfnamefont {J.~M.}\ \bibnamefont {Parrondo}},\ and\ \bibinfo {author} {\bibfnamefont {F.}~\bibnamefont {J{\"u}licher}},\ }\bibfield  {title} {\bibinfo {title} {Quantifying entropy production in active fluctuations of the hair-cell bundle from time irreversibility and uncertainty relations},\ }\href {https://doi.org/10.1088/1367-2630/ac0f18} {\bibfield  {journal} {\bibinfo  {journal} {New Journal of Physics}\ }\textbf {\bibinfo {volume} {23}},\ \bibinfo {pages} {083013} (\bibinfo {year} {2021})}\BibitemShut {NoStop}%
\bibitem [{\citenamefont {Di~Terlizzi}\ \emph {et~al.}(2024)\citenamefont {Di~Terlizzi}, \citenamefont {Gironella}, \citenamefont {Herr{\'a}ez-Aguilar}, \citenamefont {Betz}, \citenamefont {Monroy}, \citenamefont {Baiesi},\ and\ \citenamefont {Ritort}}]{di2024variance}%
  \BibitemOpen
  \bibfield  {author} {\bibinfo {author} {\bibfnamefont {I.}~\bibnamefont {Di~Terlizzi}}, \bibinfo {author} {\bibfnamefont {M.}~\bibnamefont {Gironella}}, \bibinfo {author} {\bibfnamefont {D.}~\bibnamefont {Herr{\'a}ez-Aguilar}}, \bibinfo {author} {\bibfnamefont {T.}~\bibnamefont {Betz}}, \bibinfo {author} {\bibfnamefont {F.}~\bibnamefont {Monroy}}, \bibinfo {author} {\bibfnamefont {M.}~\bibnamefont {Baiesi}},\ and\ \bibinfo {author} {\bibfnamefont {F.}~\bibnamefont {Ritort}},\ }\bibfield  {title} {\bibinfo {title} {Variance sum rule for entropy production},\ }\href {https://www.science.org/doi/full/10.1126/science.adh1823} {\bibfield  {journal} {\bibinfo  {journal} {Science}\ }\textbf {\bibinfo {volume} {383}},\ \bibinfo {pages} {971} (\bibinfo {year} {2024})}\BibitemShut {NoStop}%
\bibitem [{\citenamefont {Deg\"unther}\ \emph {et~al.}(2024)\citenamefont {Deg\"unther}, \citenamefont {van~der Meer},\ and\ \citenamefont {Seifert}}]{degunther2024fluctuating}%
  \BibitemOpen
  \bibfield  {author} {\bibinfo {author} {\bibfnamefont {J.}~\bibnamefont {Deg\"unther}}, \bibinfo {author} {\bibfnamefont {J.}~\bibnamefont {van~der Meer}},\ and\ \bibinfo {author} {\bibfnamefont {U.}~\bibnamefont {Seifert}},\ }\bibfield  {title} {\bibinfo {title} {Fluctuating entropy production on the coarse-grained level: Inference and localization of irreversibility},\ }\href {https://doi.org/10.1103/PhysRevResearch.6.023175} {\bibfield  {journal} {\bibinfo  {journal} {Phys. Rev. Res.}\ }\textbf {\bibinfo {volume} {6}},\ \bibinfo {pages} {023175} (\bibinfo {year} {2024})}\BibitemShut {NoStop}%
\bibitem [{\citenamefont {Deg{\"u}nther}\ \emph {et~al.}(2024)\citenamefont {Deg{\"u}nther}, \citenamefont {van~der Meer},\ and\ \citenamefont {Seifert}}]{degunther2024general}%
  \BibitemOpen
  \bibfield  {author} {\bibinfo {author} {\bibfnamefont {J.}~\bibnamefont {Deg{\"u}nther}}, \bibinfo {author} {\bibfnamefont {J.}~\bibnamefont {van~der Meer}},\ and\ \bibinfo {author} {\bibfnamefont {U.}~\bibnamefont {Seifert}},\ }\bibfield  {title} {\bibinfo {title} {General theory for localizing the where and when of entropy production meets single-molecule experiments},\ }\href {https://doi.org/10.1073/pnas.2405371121} {\bibfield  {journal} {\bibinfo  {journal} {Proceedings of the National Academy of Sciences}\ }\textbf {\bibinfo {volume} {121}},\ \bibinfo {pages} {e2405371121} (\bibinfo {year} {2024})}\BibitemShut {NoStop}%
\bibitem [{\citenamefont {Majhi}\ \emph {et~al.}(2025)\citenamefont {Majhi}, \citenamefont {Das}, \citenamefont {Gupta}, \citenamefont {Ranjan}, \citenamefont {Mallick}, \citenamefont {Paul},\ and\ \citenamefont {Banerjee}}]{majhi2025decoding}%
  \BibitemOpen
  \bibfield  {author} {\bibinfo {author} {\bibfnamefont {A.}~\bibnamefont {Majhi}}, \bibinfo {author} {\bibfnamefont {B.}~\bibnamefont {Das}}, \bibinfo {author} {\bibfnamefont {S.}~\bibnamefont {Gupta}}, \bibinfo {author} {\bibfnamefont {A.~D.}\ \bibnamefont {Ranjan}}, \bibinfo {author} {\bibfnamefont {A.~I.}\ \bibnamefont {Mallick}}, \bibinfo {author} {\bibfnamefont {S.}~\bibnamefont {Paul}},\ and\ \bibinfo {author} {\bibfnamefont {A.}~\bibnamefont {Banerjee}},\ }\bibfield  {title} {\bibinfo {title} {Decoding active force fluctuations from spatial trajectories of active systems},\ }\href {https://doi.org/10.1103/pqrl-splf} {\bibfield  {journal} {\bibinfo  {journal} {Physical Review E}\ }\textbf {\bibinfo {volume} {111}},\ \bibinfo {pages} {065411} (\bibinfo {year} {2025})}\BibitemShut {NoStop}%
\bibitem [{\citenamefont {Seifert}(2008)}]{seifert2008stochastic}%
  \BibitemOpen
  \bibfield  {author} {\bibinfo {author} {\bibfnamefont {U.}~\bibnamefont {Seifert}},\ }\bibfield  {title} {\bibinfo {title} {Stochastic thermodynamics: principles and perspectives},\ }\href{https://doi.org/10.1140/epjb/e2008-00001-9} {\bibfield  {journal} {\bibinfo  {journal} {The European Physical Journal B}\ }\textbf {\bibinfo {volume} {64}},\ \bibinfo {pages} {423} (\bibinfo {year} {2008})}\BibitemShut {NoStop}%
\bibitem [{\citenamefont {Seifert}(2019)}]{seifert2019stochastic}%
  \BibitemOpen
  \bibfield  {author} {\bibinfo {author} {\bibfnamefont {U.}~\bibnamefont {Seifert}},\ }\bibfield  {title} {\bibinfo {title} {From stochastic thermodynamics to thermodynamic inference},\ }\href {https://www.annualreviews.org/doi/abs/10.1146/annurev-conmatphys-031218-013554} {\bibfield  {journal} {\bibinfo  {journal} {Annual Review of Condensed Matter Physics}\ }\textbf {\bibinfo {volume} {10}},\ \bibinfo {pages} {171} (\bibinfo {year} {2019})}\BibitemShut {NoStop}%
\bibitem [{\citenamefont {Ciliberto}(2017)}]{Ciliberto:2017est}%
  \BibitemOpen
  \bibfield  {author} {\bibinfo {author} {\bibfnamefont {S.}~\bibnamefont {Ciliberto}},\ }\bibfield  {title} {\bibinfo {title} {Experiments in stochastic thermodynamics: Short history and perspectives},\ }\href{https://doi.org/10.1103/PhysRevX.7.021051} {\bibfield  {journal} {\bibinfo  {journal} {Phys. Rev. X}\ }\textbf {\bibinfo {volume} {7}},\ \bibinfo {pages} {021051} (\bibinfo {year} {2017})}\BibitemShut {NoStop}%
\bibitem [{\citenamefont {Li}\ \emph {et~al.}(2019)\citenamefont {Li}, \citenamefont {Horowitz}, \citenamefont {Gingrich},\ and\ \citenamefont {Fakhri}}]{li2019quantifying}%
  \BibitemOpen
  \bibfield  {author} {\bibinfo {author} {\bibfnamefont {J.}~\bibnamefont {Li}}, \bibinfo {author} {\bibfnamefont {J.~M.}\ \bibnamefont {Horowitz}}, \bibinfo {author} {\bibfnamefont {T.~R.}\ \bibnamefont {Gingrich}},\ and\ \bibinfo {author} {\bibfnamefont {N.}~\bibnamefont {Fakhri}},\ }\bibfield  {title} {\bibinfo {title} {Quantifying dissipation using fluctuating currents},\ }\href {https://doi.org/10.1038/s41467-019-09631-x} {\bibfield  {journal} {\bibinfo  {journal} {Nature Communications}\ }\textbf {\bibinfo {volume} {10}},\ \bibinfo {pages} {1666} (\bibinfo {year} {2019})}\BibitemShut {NoStop}%
\bibitem [{\citenamefont {Otsubo}\ \emph {et~al.}(2022)\citenamefont {Otsubo}, \citenamefont {Manikandan}, \citenamefont {Sagawa},\ and\ \citenamefont {Krishnamurthy}}]{otsubo2022estimating}%
  \BibitemOpen
  \bibfield  {author} {\bibinfo {author} {\bibfnamefont {S.}~\bibnamefont {Otsubo}}, \bibinfo {author} {\bibfnamefont {S.~K.}\ \bibnamefont {Manikandan}}, \bibinfo {author} {\bibfnamefont {T.}~\bibnamefont {Sagawa}},\ and\ \bibinfo {author} {\bibfnamefont {S.}~\bibnamefont {Krishnamurthy}},\ }\bibfield  {title} {\bibinfo {title} {Estimating time-dependent entropy production from non-equilibrium trajectories},\ }\href {https://doi.org/10.1038/s42005-021-00787-x} {\bibfield  {journal} {\bibinfo  {journal} {Communications Physics}\ }\textbf {\bibinfo {volume} {5}},\ \bibinfo {pages} {11} (\bibinfo {year} {2022})}\BibitemShut {NoStop}%
\bibitem [{\citenamefont {Manikandan}\ \emph {et~al.}(2021)\citenamefont {Manikandan}, \citenamefont {Ghosh}, \citenamefont {Kundu}, \citenamefont {Das}, \citenamefont {Agrawal}, \citenamefont {Mitra}, \citenamefont {Banerjee},\ and\ \citenamefont {Krishnamurthy}}]{manikandan2021quantitative}%
  \BibitemOpen
  \bibfield  {author} {\bibinfo {author} {\bibfnamefont {S.~K.}\ \bibnamefont {Manikandan}}, \bibinfo {author} {\bibfnamefont {S.}~\bibnamefont {Ghosh}}, \bibinfo {author} {\bibfnamefont {A.}~\bibnamefont {Kundu}}, \bibinfo {author} {\bibfnamefont {B.}~\bibnamefont {Das}}, \bibinfo {author} {\bibfnamefont {V.}~\bibnamefont {Agrawal}}, \bibinfo {author} {\bibfnamefont {D.}~\bibnamefont {Mitra}}, \bibinfo {author} {\bibfnamefont {A.}~\bibnamefont {Banerjee}},\ and\ \bibinfo {author} {\bibfnamefont {S.}~\bibnamefont {Krishnamurthy}},\ }\bibfield  {title} {\bibinfo {title} {Quantitative analysis of non-equilibrium systems from short-time experimental data},\ }\href {https://www.nature.com/articles/s42005-021-00766-2} {\bibfield  {journal} {\bibinfo  {journal} {Communications Physics}\ }\textbf {\bibinfo {volume} {4}},\ \bibinfo {pages} {258} (\bibinfo {year} {2021})}\BibitemShut {NoStop}%
\bibitem [{\citenamefont {Kim}\ \emph {et~al.}(2020)\citenamefont {Kim}, \citenamefont {Bae}, \citenamefont {Lee},\ and\ \citenamefont {Jeong}}]{kim2020learning}%
  \BibitemOpen
  \bibfield  {author} {\bibinfo {author} {\bibfnamefont {D.-K.}\ \bibnamefont {Kim}}, \bibinfo {author} {\bibfnamefont {Y.}~\bibnamefont {Bae}}, \bibinfo {author} {\bibfnamefont {S.}~\bibnamefont {Lee}},\ and\ \bibinfo {author} {\bibfnamefont {H.}~\bibnamefont {Jeong}},\ }\bibfield  {title} {\bibinfo {title} {Learning entropy production via neural networks},\ }\href {https://doi.org/10.1103/PhysRevLett.125.140604} {\bibfield  {journal} {\bibinfo  {journal} {Phys. Rev. Lett.}\ }\textbf {\bibinfo {volume} {125}},\ \bibinfo {pages} {140604} (\bibinfo {year} {2020})}\BibitemShut {NoStop}%
\bibitem [{\citenamefont {Fodor}\ \emph {et~al.}(2016)\citenamefont {Fodor}, \citenamefont {Ahmed}, \citenamefont {Almonacid}, \citenamefont {Bussonnier}, \citenamefont {Gov}, \citenamefont {Verlhac}, \citenamefont {Betz}, \citenamefont {Visco},\ and\ \citenamefont {van Wijland}}]{oocites}%
  \BibitemOpen
  \bibfield  {author} {\bibinfo {author} {\bibfnamefont {{\'E}.}~\bibnamefont {Fodor}}, \bibinfo {author} {\bibfnamefont {W.~W.}\ \bibnamefont {Ahmed}}, \bibinfo {author} {\bibfnamefont {M.}~\bibnamefont {Almonacid}}, \bibinfo {author} {\bibfnamefont {M.}~\bibnamefont {Bussonnier}}, \bibinfo {author} {\bibfnamefont {N.~S.}\ \bibnamefont {Gov}}, \bibinfo {author} {\bibfnamefont {M.-H.}\ \bibnamefont {Verlhac}}, \bibinfo {author} {\bibfnamefont {T.}~\bibnamefont {Betz}}, \bibinfo {author} {\bibfnamefont {P.}~\bibnamefont {Visco}},\ and\ \bibinfo {author} {\bibfnamefont {F.}~\bibnamefont {van Wijland}},\ }\bibfield  {title} {\bibinfo {title} {Nonequilibrium dissipation in living oocytes},\ }\href{https://doi.org/10.1209/0295-5075/116/30008} {\bibfield  {journal} {\bibinfo  {journal} {EPL (Europhysics Letters)}\ }\textbf {\bibinfo {volume} {116}},\ \bibinfo {pages} {30008} (\bibinfo {year} {2016})}\BibitemShut {NoStop}%
\bibitem [{\citenamefont {Ariga}\ \emph {et~al.}(2018)\citenamefont {Ariga}, \citenamefont {Tomishige},\ and\ \citenamefont {Mizuno}}]{ariga2018nonequilibrium}%
  \BibitemOpen
  \bibfield  {author} {\bibinfo {author} {\bibfnamefont {T.}~\bibnamefont {Ariga}}, \bibinfo {author} {\bibfnamefont {M.}~\bibnamefont {Tomishige}},\ and\ \bibinfo {author} {\bibfnamefont {D.}~\bibnamefont {Mizuno}},\ }\bibfield  {title} {\bibinfo {title} {Nonequilibrium energetics of molecular motor kinesin},\ }\href{https://doi.org/10.1103/PhysRevLett.121.218101} {\bibfield  {journal} {\bibinfo  {journal} {Physical review letters}\ }\textbf {\bibinfo {volume} {121}},\ \bibinfo {pages} {218101} (\bibinfo {year} {2018})}\BibitemShut {NoStop}%
\bibitem [{\citenamefont {Seifert}(2005{\natexlab{a}})}]{seifert2005entropy}%
  \BibitemOpen
  \bibfield  {author} {\bibinfo {author} {\bibfnamefont {U.}~\bibnamefont {Seifert}},\ }\bibfield  {title} {\bibinfo {title} {Entropy production along a stochastic trajectory and an integral fluctuation theorem},\ }\href {https://doi.org/10.1103/PhysRevLett.95.040602} {\bibfield  {journal} {\bibinfo  {journal} {Physical Review Letters}\ }\textbf {\bibinfo {volume} {95}},\ \bibinfo {pages} {040602} (\bibinfo {year} {2005}{\natexlab{a}})}\BibitemShut {NoStop}%
\bibitem [{\citenamefont {Sekimoto}(1997)}]{sekimoto1997kinetic}%
  \BibitemOpen
  \bibfield  {author} {\bibinfo {author} {\bibfnamefont {K.}~\bibnamefont {Sekimoto}},\ }\bibfield  {title} {\bibinfo {title} {Kinetic characterization of heat bath and the energetics of thermal ratchet models},\ }\href{https://doi.org/10.1143/JPSJ.66.1234} {\bibfield  {journal} {\bibinfo  {journal} {Journal of the physical society of Japan}\ }\textbf {\bibinfo {volume} {66}},\ \bibinfo {pages} {1234} (\bibinfo {year} {1997})}\BibitemShut {NoStop}%
\bibitem [{\citenamefont {Sekimoto}(1998)}]{sekimoto1998langevin}%
  \BibitemOpen
  \bibfield  {author} {\bibinfo {author} {\bibfnamefont {K.}~\bibnamefont {Sekimoto}},\ }\bibfield  {title} {\bibinfo {title} {Langevin equation and thermodynamics},\ }\href{https://doi.org/10.1143/PTPS.130.17} {\bibfield  {journal} {\bibinfo  {journal} {Progress of Theoretical Physics Supplement}\ }\textbf {\bibinfo {volume} {130}},\ \bibinfo {pages} {17} (\bibinfo {year} {1998})}\BibitemShut {NoStop}%
\bibitem [{\citenamefont {Seifert}(2012)}]{Seifert:2012stf}%
  \BibitemOpen
  \bibfield  {author} {\bibinfo {author} {\bibfnamefont {U.}~\bibnamefont {Seifert}},\ }\bibfield  {title} {\bibinfo {title} {Stochastic thermodynamics, fluctuation theorems and molecular machines},\ }\href {http://stacks.iop.org/0034-4885/75/i=12/a=126001} {\bibfield  {journal} {\bibinfo  {journal} {Rep. Prog. Phys.}\ }\textbf {\bibinfo {volume} {75}},\ \bibinfo {pages} {126001} (\bibinfo {year} {2012})}\BibitemShut {NoStop}%
\bibitem [{\citenamefont {Maes}\ and\ \citenamefont {Neto{\v{c}}n{\`y}}(2003)}]{maes2003time}%
  \BibitemOpen
  \bibfield  {author} {\bibinfo {author} {\bibfnamefont {C.}~\bibnamefont {Maes}}\ and\ \bibinfo {author} {\bibfnamefont {K.}~\bibnamefont {Neto{\v{c}}n{\`y}}},\ }\bibfield  {title} {\bibinfo {title} {Time-reversal and entropy},\ }\href{https://doi.org/10.1023/A:1021026930129} {\bibfield  {journal} {\bibinfo  {journal} {Journal of statistical physics}\ }\textbf {\bibinfo {volume} {110}},\ \bibinfo {pages} {269} (\bibinfo {year} {2003})}\BibitemShut {NoStop}%
\bibitem [{\citenamefont {Gaspard}(2004)}]{gaspard2004time}%
  \BibitemOpen
  \bibfield  {author} {\bibinfo {author} {\bibfnamefont {P.}~\bibnamefont {Gaspard}},\ }\bibfield  {title} {\bibinfo {title} {Time-reversed dynamical entropy and irreversibility in markovian random processes},\ }\href{https://doi.org/10.1007/s10955-004-3455-1} {\bibfield  {journal} {\bibinfo  {journal} {Journal of statistical physics}\ }\textbf {\bibinfo {volume} {117}},\ \bibinfo {pages} {599} (\bibinfo {year} {2004})}\BibitemShut {NoStop}%
\bibitem [{\citenamefont {Andrieux}\ \emph {et~al.}(2008)\citenamefont {Andrieux}, \citenamefont {Gaspard}, \citenamefont {Ciliberto}, \citenamefont {Garnier}, \citenamefont {Joubaud},\ and\ \citenamefont {Petrosyan}}]{andrieux2008thermodynamic}%
  \BibitemOpen
  \bibfield  {author} {\bibinfo {author} {\bibfnamefont {D.}~\bibnamefont {Andrieux}}, \bibinfo {author} {\bibfnamefont {P.}~\bibnamefont {Gaspard}}, \bibinfo {author} {\bibfnamefont {S.}~\bibnamefont {Ciliberto}}, \bibinfo {author} {\bibfnamefont {N.}~\bibnamefont {Garnier}}, \bibinfo {author} {\bibfnamefont {S.}~\bibnamefont {Joubaud}},\ and\ \bibinfo {author} {\bibfnamefont {A.}~\bibnamefont {Petrosyan}},\ }\bibfield  {title} {\bibinfo {title} {Thermodynamic time asymmetry in non-equilibrium fluctuations},\ }\href{https://doi.org/10.1088/1742-5468/2008/01/P01002} {\bibfield  {journal} {\bibinfo  {journal} {Journal of Statistical Mechanics: Theory and Experiment}\ }\textbf {\bibinfo {volume} {2008}},\ \bibinfo {pages} {P01002} (\bibinfo {year} {2008})}\BibitemShut {NoStop}%
\bibitem [{\citenamefont {Andrieux}\ \emph {et~al.}(2007)\citenamefont {Andrieux}, \citenamefont {Gaspard}, \citenamefont {Ciliberto}, \citenamefont {Garnier}, \citenamefont {Joubaud},\ and\ \citenamefont {Petrosyan}}]{EpTA}%
  \BibitemOpen
  \bibfield  {author} {\bibinfo {author} {\bibfnamefont {D.}~\bibnamefont {Andrieux}}, \bibinfo {author} {\bibfnamefont {P.}~\bibnamefont {Gaspard}}, \bibinfo {author} {\bibfnamefont {S.}~\bibnamefont {Ciliberto}}, \bibinfo {author} {\bibfnamefont {N.}~\bibnamefont {Garnier}}, \bibinfo {author} {\bibfnamefont {S.}~\bibnamefont {Joubaud}},\ and\ \bibinfo {author} {\bibfnamefont {A.}~\bibnamefont {Petrosyan}},\ }\bibfield  {title} {\bibinfo {title} {Entropy production and time asymmetry in nonequilibrium fluctuations},\ }\href {https://doi.org/10.1103/PhysRevLett.98.150601} {\bibfield  {journal} {\bibinfo  {journal} {Phys. Rev. Lett.}\ }\textbf {\bibinfo {volume} {98}},\ \bibinfo {pages} {150601} (\bibinfo {year} {2007})}\BibitemShut {NoStop}%
\bibitem [{\citenamefont {Di~Terlizzi}(2025)}]{diterlizzi2025forcefree}%
  \BibitemOpen
  \bibfield  {author} {\bibinfo {author} {\bibfnamefont {I.}~\bibnamefont {Di~Terlizzi}},\ }\bibfield  {title} {\bibinfo {title} {Force-free kinetic inference of entropy production},\ }\href {https://doi.org/10.1103/fsph-437v} {\bibfield  {journal} {\bibinfo  {journal} {Phys. Rev. Lett.}\ }\textbf {\bibinfo {volume} {135}},\ \bibinfo {pages} {237101} (\bibinfo {year} {2025})}\BibitemShut {NoStop}%
\bibitem [{\citenamefont {Barato}\ and\ \citenamefont {Seifert}(2015)}]{barato2015thermodynamic}%
  \BibitemOpen
  \bibfield  {author} {\bibinfo {author} {\bibfnamefont {A.~C.}\ \bibnamefont {Barato}}\ and\ \bibinfo {author} {\bibfnamefont {U.}~\bibnamefont {Seifert}},\ }\bibfield  {title} {\bibinfo {title} {Thermodynamic uncertainty relation for biomolecular processes},\ }\href {https://doi.org/10.1103/PhysRevLett.114.158101} {\bibfield  {journal} {\bibinfo  {journal} {Physical Review Letters}\ }\textbf {\bibinfo {volume} {114}},\ \bibinfo {pages} {158101} (\bibinfo {year} {2015})}\BibitemShut {NoStop}%
\bibitem [{\citenamefont {Gingrich}\ \emph {et~al.}(2017)\citenamefont {Gingrich}, \citenamefont {Rotskoff},\ and\ \citenamefont {Horowitz}}]{gingrich2017}%
  \BibitemOpen
  \bibfield  {author} {\bibinfo {author} {\bibfnamefont {T.~R.}\ \bibnamefont {Gingrich}}, \bibinfo {author} {\bibfnamefont {G.~M.}\ \bibnamefont {Rotskoff}},\ and\ \bibinfo {author} {\bibfnamefont {J.~M.}\ \bibnamefont {Horowitz}},\ }\bibfield  {title} {\bibinfo {title} {Inferring dissipation from current fluctuations},\ }\href {https://iopscience.iop.org/article/10.1088/1751-8121/aa672f/meta} {\bibfield  {journal} {\bibinfo  {journal} {Journal of Physics A: Mathematical and Theoretical}\ }\textbf {\bibinfo {volume} {50}},\ \bibinfo {pages} {184004} (\bibinfo {year} {2017})}\BibitemShut {NoStop}%
\bibitem [{\citenamefont {Gingrich}\ \emph {et~al.}(2016)\citenamefont {Gingrich}, \citenamefont {Horowitz}, \citenamefont {Perunov},\ and\ \citenamefont {England}}]{Gingrich2016}%
  \BibitemOpen
  \bibfield  {author} {\bibinfo {author} {\bibfnamefont {T.~R.}\ \bibnamefont {Gingrich}}, \bibinfo {author} {\bibfnamefont {J.~M.}\ \bibnamefont {Horowitz}}, \bibinfo {author} {\bibfnamefont {N.}~\bibnamefont {Perunov}},\ and\ \bibinfo {author} {\bibfnamefont {J.~L.}\ \bibnamefont {England}},\ }\bibfield  {title} {\bibinfo {title} {Dissipation bounds all steady-state current fluctuations},\ }\href {https://doi.org/10.1103/PhysRevLett.116.120601} {\bibfield  {journal} {\bibinfo  {journal} {Phys. Rev. Lett.}\ }\textbf {\bibinfo {volume} {116}},\ \bibinfo {pages} {120601} (\bibinfo {year} {2016})}\BibitemShut {NoStop}%
\bibitem [{\citenamefont {Otsubo}\ \emph {et~al.}(2020)\citenamefont {Otsubo}, \citenamefont {Ito}, \citenamefont {Dechant},\ and\ \citenamefont {Sagawa}}]{Shun:eem}%
  \BibitemOpen
  \bibfield  {author} {\bibinfo {author} {\bibfnamefont {S.}~\bibnamefont {Otsubo}}, \bibinfo {author} {\bibfnamefont {S.}~\bibnamefont {Ito}}, \bibinfo {author} {\bibfnamefont {A.}~\bibnamefont {Dechant}},\ and\ \bibinfo {author} {\bibfnamefont {T.}~\bibnamefont {Sagawa}},\ }\bibfield  {title} {\bibinfo {title} {Estimating entropy production by machine learning of short-time fluctuating currents},\ }\href{https://doi.org/10.1103/PhysRevE.101.062106} {\bibfield  {journal} {\bibinfo  {journal} {Physical Review E}\ }\textbf {\bibinfo {volume} {101}},\ \bibinfo {pages} {062106} (\bibinfo {year} {2020})}\BibitemShut {NoStop}%
\bibitem [{\citenamefont {Das}\ \emph {et~al.}(2022)\citenamefont {Das}, \citenamefont {Manikandan},\ and\ \citenamefont {Banerjee}}]{das2022inferring}%
  \BibitemOpen
  \bibfield  {author} {\bibinfo {author} {\bibfnamefont {B.}~\bibnamefont {Das}}, \bibinfo {author} {\bibfnamefont {S.~K.}\ \bibnamefont {Manikandan}},\ and\ \bibinfo {author} {\bibfnamefont {A.}~\bibnamefont {Banerjee}},\ }\bibfield  {title} {\bibinfo {title} {Inferring entropy production in anharmonic brownian gyrators},\ }\href {https://doi.org/10.1103/PhysRevResearch.4.043080} {\bibfield  {journal} {\bibinfo  {journal} {Physical Review Research}\ }\textbf {\bibinfo {volume} {4}},\ \bibinfo {pages} {043080} (\bibinfo {year} {2022})}\BibitemShut {NoStop}%
\bibitem [{\citenamefont {Boffi}\ and\ \citenamefont {Vanden-Eijnden}(2023)}]{boffi2023probability}%
  \BibitemOpen
  \bibfield  {author} {\bibinfo {author} {\bibfnamefont {N.~M.}\ \bibnamefont {Boffi}}\ and\ \bibinfo {author} {\bibfnamefont {E.}~\bibnamefont {Vanden-Eijnden}},\ }\bibfield  {title} {\bibinfo {title} {Probability flow solution of the fokker--planck equation},\ }\href {https://doi.org/10.1088/2632-2153/ace2aa} {\bibfield  {journal} {\bibinfo  {journal} {Machine Learning: Science and Technology}\ }\textbf {\bibinfo {volume} {4}},\ \bibinfo {pages} {035012} (\bibinfo {year} {2023})}\BibitemShut {NoStop}%
\bibitem [{\citenamefont {Boffi}\ and\ \citenamefont {Vanden-Eijnden}(2024{\natexlab{b}})}]{boffi2024modelfree}%
  \BibitemOpen
  \bibfield  {author} {\bibinfo {author} {\bibfnamefont {N.~M.}\ \bibnamefont {Boffi}}\ and\ \bibinfo {author} {\bibfnamefont {E.}~\bibnamefont {Vanden-Eijnden}},\ }\href {https://arxiv.org/abs/2411.14317} {\bibinfo {title} {Model-free learning of probability flows: Elucidating the nonequilibrium dynamics of flocking}} (\bibinfo {year} {2024}{\natexlab{b}}),\ \Eprint {https://arxiv.org/abs/2411.14317} {arXiv:2411.14317 [cond-mat.stat-mech]} \BibitemShut {NoStop}%
\bibitem [{\citenamefont {Spinney}\ and\ \citenamefont {Ford}(2012)}]{Spinney2012}%
  \BibitemOpen
  \bibfield  {author} {\bibinfo {author} {\bibfnamefont {R.~E.}\ \bibnamefont {Spinney}}\ and\ \bibinfo {author} {\bibfnamefont {I.~J.}\ \bibnamefont {Ford}},\ }\bibfield  {title} {\bibinfo {title} {Entropy production in full phase space for continuous stochastic dynamics},\ }\href {https://doi.org/10.1103/PhysRevE.85.051113} {\bibfield  {journal} {\bibinfo  {journal} {Phys. Rev. E}\ }\textbf {\bibinfo {volume} {85}},\ \bibinfo {pages} {051113} (\bibinfo {year} {2012})}\BibitemShut {NoStop}%
\bibitem [{\citenamefont {Van~Vu}\ \emph {et~al.}(2020{\natexlab{b}})\citenamefont {Van~Vu}, \citenamefont {Hasegawa} \emph {et~al.}}]{van2020entropy}%
  \BibitemOpen
  \bibfield  {author} {\bibinfo {author} {\bibfnamefont {T.}~\bibnamefont {Van~Vu}}, \bibinfo {author} {\bibfnamefont {Y.}~\bibnamefont {Hasegawa}}, \emph {et~al.},\ }\bibfield  {title} {\bibinfo {title} {Entropy production estimation with optimal current},\ }\href {https://doi.org/10.1103/PhysRevE.101.042138} {\bibfield  {journal} {\bibinfo  {journal} {Physical Review E}\ }\textbf {\bibinfo {volume} {101}},\ \bibinfo {pages} {042138} (\bibinfo {year} {2020}{\natexlab{b}})}\BibitemShut {NoStop}%
\bibitem [{\citenamefont {Manikandan}\ and\ \citenamefont {Krishnamurthy}(2018)}]{Manikandan:2018erf}%
  \BibitemOpen
  \bibfield  {author} {\bibinfo {author} {\bibfnamefont {S.~K.}\ \bibnamefont {Manikandan}}\ and\ \bibinfo {author} {\bibfnamefont {S.}~\bibnamefont {Krishnamurthy}},\ }\bibfield  {title} {\bibinfo {title} {Exact results for the finite time thermodynamic uncertainty relation},\ }\href{https://doi.org/10.1088/1751-8121/aaaa54} {\bibfield  {journal} {\bibinfo  {journal} {J. Phys. A: Math. Theor.}\ }\textbf {\bibinfo {volume} {51}},\ \bibinfo {pages} {11LT01} (\bibinfo {year} {2018})}\BibitemShut {NoStop}%
\bibitem [{\citenamefont {Rotskoff}\ and\ \citenamefont {Vanden-Eijnden}(2018)}]{rotskoff2018advances}%
  \BibitemOpen
  \bibfield  {author} {\bibinfo {author} {\bibfnamefont {G.}~\bibnamefont {Rotskoff}}\ and\ \bibinfo {author} {\bibfnamefont {E.}~\bibnamefont {Vanden-Eijnden}},\ }in\ \href{https://doi.org/10.1002/cpa.22074} {\emph {\bibinfo {booktitle} {Advances in Neural Information Processing Systems 31}}},\ \bibinfo {editor} {edited by\ \bibinfo {editor} {\bibfnamefont {S.}~\bibnamefont {Bengio}}, \bibinfo {editor} {\bibfnamefont {H.}~\bibnamefont {Wallach}}, \bibinfo {editor} {\bibfnamefont {H.}~\bibnamefont {Larochelle}}, \bibinfo {editor} {\bibfnamefont {K.}~\bibnamefont {Grauman}}, \bibinfo {editor} {\bibfnamefont {N.}~\bibnamefont {Cesa-Bianchi}},\ and\ \bibinfo {editor} {\bibfnamefont {R.}~\bibnamefont {Garnett}}}\ (\bibinfo  {publisher} {Curran Associates},\ \bibinfo {address} {Red Hook, NY},\ \bibinfo {year} {2018})\ pp.\ \bibinfo {pages} {7146--7155}\BibitemShut {NoStop}%
\bibitem [{\citenamefont {Chizat}\ and\ \citenamefont {Bach}(2018)}]{chizat2018advances}%
  \BibitemOpen
  \bibfield  {author} {\bibinfo {author} {\bibfnamefont {L.}~\bibnamefont {Chizat}}\ and\ \bibinfo {author} {\bibfnamefont {F.}~\bibnamefont {Bach}},\ }in\ \href{https://doi.org/10.48550/arXiv.1805.09545} {\emph {\bibinfo {booktitle} {Advances in Neural Information Processing Systems 31}}},\ \bibinfo {editor} {edited by\ \bibinfo {editor} {\bibfnamefont {S.}~\bibnamefont {Bengio}}, \bibinfo {editor} {\bibfnamefont {H.}~\bibnamefont {Wallach}}, \bibinfo {editor} {\bibfnamefont {H.}~\bibnamefont {Larochelle}}, \bibinfo {editor} {\bibfnamefont {K.}~\bibnamefont {Grauman}}, \bibinfo {editor} {\bibfnamefont {N.}~\bibnamefont {Cesa-Bianchi}},\ and\ \bibinfo {editor} {\bibfnamefont {R.}~\bibnamefont {Garnett}}}\ (\bibinfo  {publisher} {Curran Associates},\ \bibinfo {address} {Red Hook, NY},\ \bibinfo {year} {2018})\ pp.\ \bibinfo {pages} {3036--3046}\BibitemShut {NoStop}%
\bibitem [{\citenamefont {Mei}\ \emph {et~al.}(2018)\citenamefont {Mei}, \citenamefont {Montanari},\ and\ \citenamefont {Nguyen}}]{mei2018mean}%
  \BibitemOpen
  \bibfield  {author} {\bibinfo {author} {\bibfnamefont {S.}~\bibnamefont {Mei}}, \bibinfo {author} {\bibfnamefont {A.}~\bibnamefont {Montanari}},\ and\ \bibinfo {author} {\bibfnamefont {P.-M.}\ \bibnamefont {Nguyen}},\ }\bibfield  {title} {\bibinfo {title} {A mean field view of the landscape of two-layer neural networks},\ }\href {https://doi.org/10.1073/pnas.1806579115} {\bibfield  {journal} {\bibinfo  {journal} {Proceedings of the National Academy of Sciences}\ }\textbf {\bibinfo {volume} {115}},\ \bibinfo {pages} {E7665} (\bibinfo {year} {2018})}\BibitemShut {NoStop}%
\bibitem [{\citenamefont {Sirignano}\ and\ \citenamefont {Spiliopoulos}(2020)}]{sirignano2020mean}%
  \BibitemOpen
  \bibfield  {author} {\bibinfo {author} {\bibfnamefont {J.}~\bibnamefont {Sirignano}}\ and\ \bibinfo {author} {\bibfnamefont {K.}~\bibnamefont {Spiliopoulos}},\ }\bibfield  {title} {\bibinfo {title} {Mean field analysis of neural networks: A law of large numbers},\ }\href {https://doi.org/10.1137/18M1192184} {\bibfield  {journal} {\bibinfo  {journal} {SIAM Journal on Applied Mathematics}\ }\textbf {\bibinfo {volume} {80}},\ \bibinfo {pages} {725} (\bibinfo {year} {2020})}\BibitemShut {NoStop}%
\bibitem [{\citenamefont {Barron}(1993)}]{barron1993universal}%
  \BibitemOpen
  \bibfield  {author} {\bibinfo {author} {\bibfnamefont {A.~R.}\ \bibnamefont {Barron}},\ }\bibfield  {title} {\bibinfo {title} {Universal approximation bounds for superpositions of a sigmoidal function},\ }\href {https://doi.org/10.1109/18.256500} {\bibfield  {journal} {\bibinfo  {journal} {IEEE Transactions on Information theory}\ }\textbf {\bibinfo {volume} {39}},\ \bibinfo {pages} {930} (\bibinfo {year} {1993})}\BibitemShut {NoStop}%
\bibitem [{\citenamefont {Cybenko}(1989)}]{cybenko1989approximation}%
  \BibitemOpen
  \bibfield  {author} {\bibinfo {author} {\bibfnamefont {G.}~\bibnamefont {Cybenko}},\ }\bibfield  {title} {\bibinfo {title} {Approximation by superpositions of a sigmoidal function},\ }\href {https://doi.org/10.1007/BF02551274} {\bibfield  {journal} {\bibinfo  {journal} {Mathematics of control, signals and systems}\ }\textbf {\bibinfo {volume} {2}},\ \bibinfo {pages} {303} (\bibinfo {year} {1989})}\BibitemShut {NoStop}%
\bibitem [{git()}]{git_repo}%
  \BibitemOpen
  \href@noop {} {}\bibinfo {note} {\url{https://github.com/sreekmnoneq/Inference-short}}\BibitemShut {NoStop}%
\bibitem [{\citenamefont {Yu}\ \emph {et~al.}(2018)\citenamefont {Yu} \emph {et~al.}}]{yu2018deep}%
  \BibitemOpen
  \bibfield  {author} {\bibinfo {author} {\bibfnamefont {B.}~\bibnamefont {Yu}} \emph {et~al.},\ }\bibfield  {title} {\bibinfo {title} {The deep ritz method: a deep learning-based numerical algorithm for solving variational problems},\ }\href {https://doi.org/10.1007/s40304-018-0127-z} {\bibfield  {journal} {\bibinfo  {journal} {Communications in Mathematics and Statistics}\ }\textbf {\bibinfo {volume} {6}},\ \bibinfo {pages} {1} (\bibinfo {year} {2018})}\BibitemShut {NoStop}%
\bibitem [{\citenamefont {Yan}\ \emph {et~al.}(2022)\citenamefont {Yan}, \citenamefont {Touchette},\ and\ \citenamefont {Rotskoff}}]{yan2022learning}%
  \BibitemOpen
  \bibfield  {author} {\bibinfo {author} {\bibfnamefont {J.}~\bibnamefont {Yan}}, \bibinfo {author} {\bibfnamefont {H.}~\bibnamefont {Touchette}},\ and\ \bibinfo {author} {\bibfnamefont {G.~M.}\ \bibnamefont {Rotskoff}},\ }\bibfield  {title} {\bibinfo {title} {Learning nonequilibrium control forces to characterize dynamical phase transitions},\ }\href {https://doi.org/10.1103/PhysRevE.105.024115} {\bibfield  {journal} {\bibinfo  {journal} {Physical Review E}\ }\textbf {\bibinfo {volume} {105}},\ \bibinfo {pages} {024115} (\bibinfo {year} {2022})}\BibitemShut {NoStop}%
\bibitem [{\citenamefont {Filliger}\ and\ \citenamefont {Reimann}(2007)}]{filliger2007brownian}%
  \BibitemOpen
  \bibfield  {author} {\bibinfo {author} {\bibfnamefont {R.}~\bibnamefont {Filliger}}\ and\ \bibinfo {author} {\bibfnamefont {P.}~\bibnamefont {Reimann}},\ }\bibfield  {title} {\bibinfo {title} {Brownian gyrator: A minimal heat engine on the nanoscale},\ }\href {https://doi.org/10.1103/PhysRevLett.99.230602} {\bibfield  {journal} {\bibinfo  {journal} {Physical Review Letters}\ }\textbf {\bibinfo {volume} {99}},\ \bibinfo {pages} {230602} (\bibinfo {year} {2007})}\BibitemShut {NoStop}%
\bibitem [{\citenamefont {Argun}\ \emph {et~al.}(2017)\citenamefont {Argun}, \citenamefont {Soni}, \citenamefont {Dabelow}, \citenamefont {Bo}, \citenamefont {Pesce}, \citenamefont {Eichhorn},\ and\ \citenamefont {Volpe}}]{argun2017experimental}%
  \BibitemOpen
  \bibfield  {author} {\bibinfo {author} {\bibfnamefont {A.}~\bibnamefont {Argun}}, \bibinfo {author} {\bibfnamefont {J.}~\bibnamefont {Soni}}, \bibinfo {author} {\bibfnamefont {L.}~\bibnamefont {Dabelow}}, \bibinfo {author} {\bibfnamefont {S.}~\bibnamefont {Bo}}, \bibinfo {author} {\bibfnamefont {G.}~\bibnamefont {Pesce}}, \bibinfo {author} {\bibfnamefont {R.}~\bibnamefont {Eichhorn}},\ and\ \bibinfo {author} {\bibfnamefont {G.}~\bibnamefont {Volpe}},\ }\bibfield  {title} {\bibinfo {title} {Experimental realization of a minimal microscopic heat engine},\ }\href {https://doi.org/10.1103/PhysRevE.96.052106} {\bibfield  {journal} {\bibinfo  {journal} {Physical Review E}\ }\textbf {\bibinfo {volume} {96}},\ \bibinfo {pages} {052106} (\bibinfo {year} {2017})}\BibitemShut {NoStop}%
\bibitem [{\citenamefont {Chang}\ \emph {et~al.}(2021)\citenamefont {Chang}, \citenamefont {Lee}, \citenamefont {Lai},\ and\ \citenamefont {Chen}}]{chang2021autonomous}%
  \BibitemOpen
  \bibfield  {author} {\bibinfo {author} {\bibfnamefont {H.}~\bibnamefont {Chang}}, \bibinfo {author} {\bibfnamefont {C.-L.}\ \bibnamefont {Lee}}, \bibinfo {author} {\bibfnamefont {P.-Y.}\ \bibnamefont {Lai}},\ and\ \bibinfo {author} {\bibfnamefont {Y.-F.}\ \bibnamefont {Chen}},\ }\bibfield  {title} {\bibinfo {title} {Autonomous brownian gyrators: A study on gyrating characteristics},\ }\href {https://doi.org/10.1103/PhysRevE.103.022128} {\bibfield  {journal} {\bibinfo  {journal} {Physical Review E}\ }\textbf {\bibinfo {volume} {103}},\ \bibinfo {pages} {022128} (\bibinfo {year} {2021})}\BibitemShut {NoStop}%
\bibitem [{\citenamefont {Manikandan}\ \emph {et~al.}(2022)\citenamefont {Manikandan}, \citenamefont {Das}, \citenamefont {Kundu}, \citenamefont {Dey}, \citenamefont {Banerjee},\ and\ \citenamefont {Krishnamurthy}}]{manikandan2022nonmonotonic}%
  \BibitemOpen
  \bibfield  {author} {\bibinfo {author} {\bibfnamefont {S.~K.}\ \bibnamefont {Manikandan}}, \bibinfo {author} {\bibfnamefont {B.}~\bibnamefont {Das}}, \bibinfo {author} {\bibfnamefont {A.}~\bibnamefont {Kundu}}, \bibinfo {author} {\bibfnamefont {R.}~\bibnamefont {Dey}}, \bibinfo {author} {\bibfnamefont {A.}~\bibnamefont {Banerjee}},\ and\ \bibinfo {author} {\bibfnamefont {S.}~\bibnamefont {Krishnamurthy}},\ }\bibfield  {title} {\bibinfo {title} {Nonmonotonic skewness of currents in nonequilibrium steady states},\ }\href {https://doi.org/10.1103/PhysRevResearch.4.043067} {\bibfield  {journal} {\bibinfo  {journal} {Phys. Rev. Res.}\ }\textbf {\bibinfo {volume} {4}},\ \bibinfo {pages} {043067} (\bibinfo {year} {2022})}\BibitemShut {NoStop}%
\bibitem [{\citenamefont {Delaunay}(1934)}]{delaunay1934}%
  \BibitemOpen
  \bibfield  {author} {\bibinfo {author} {\bibfnamefont {B.}~\bibnamefont {Delaunay}},\ }\bibfield  {title} {\bibinfo {title} {Sur la sphère vide},\ }\href{http://mi.mathnet.ru/eng/im4937} {\bibfield  {journal} {\bibinfo  {journal} {Izvestia Akademii Nauk SSSR, Otdelenie Matematicheskikh i Estestvennykh Nauk}\ }\textbf {\bibinfo {volume} {7}},\ \bibinfo {pages} {793} (\bibinfo {year} {1934})}\BibitemShut {NoStop}%
\bibitem [{\citenamefont {Community}()}]{SciPy}%
  \BibitemOpen
  \bibfield  {author} {\bibinfo {author} {\bibfnamefont {S.}~\bibnamefont {Community}},\ }\href{} {\bibinfo {title} {{SciPy}: Python-based ecosystem of open-source software for mathematics, science, and engineering}},\ \bibinfo {howpublished} {\url{https://scipy.org/}}\BibitemShut {NoStop}%
\bibitem [{\citenamefont {Virtanen}\ \emph {et~al.}(2020)\citenamefont {Virtanen} \emph {et~al.}}]{2020SciPy-NMeth}%
  \BibitemOpen
  \bibfield  {author} {\bibinfo {author} {\bibfnamefont {P.}~\bibnamefont {Virtanen}} \emph {et~al.},\ }\bibfield  {title} {\bibinfo {title} {{SciPy} 1.0: Fundamental algorithms for scientific computing in python},\ }\href {https://doi.org/10.1038/s41592-019-0686-2} {\bibfield  {journal} {\bibinfo  {journal} {Nature Methods}\ }\textbf {\bibinfo {volume} {17}},\ \bibinfo {pages} {261} (\bibinfo {year} {2020})}\BibitemShut {NoStop}%
\bibitem [{\citenamefont {Nadrowski}\ \emph {et~al.}(2004)\citenamefont {Nadrowski}, \citenamefont {Martin},\ and\ \citenamefont {J{\"u}licher}}]{nadrowski2004active}%
  \BibitemOpen
  \bibfield  {author} {\bibinfo {author} {\bibfnamefont {B.}~\bibnamefont {Nadrowski}}, \bibinfo {author} {\bibfnamefont {P.}~\bibnamefont {Martin}},\ and\ \bibinfo {author} {\bibfnamefont {F.}~\bibnamefont {J{\"u}licher}},\ }\bibfield  {title} {\bibinfo {title} {Active hair-bundle motility harnesses noise to operate near an optimum of mechanosensitivity},\ }\href {https://www.pnas.org/doi/full/10.1073/pnas.0403020101} {\bibfield  {journal} {\bibinfo  {journal} {Proceedings of the National Academy of Sciences}\ }\textbf {\bibinfo {volume} {101}},\ \bibinfo {pages} {12195} (\bibinfo {year} {2004})}\BibitemShut {NoStop}%
\bibitem [{\citenamefont {Tucci}\ \emph {et~al.}(2022)\citenamefont {Tucci}, \citenamefont {Rold\'an}, \citenamefont {Gambassi}, \citenamefont {Belousov}, \citenamefont {Berger}, \citenamefont {Alonso},\ and\ \citenamefont {Hudspeth}}]{tucci2022modelling}%
  \BibitemOpen
  \bibfield  {author} {\bibinfo {author} {\bibfnamefont {G.}~\bibnamefont {Tucci}}, \bibinfo {author} {\bibfnamefont {E.}~\bibnamefont {Rold\'an}}, \bibinfo {author} {\bibfnamefont {A.}~\bibnamefont {Gambassi}}, \bibinfo {author} {\bibfnamefont {R.}~\bibnamefont {Belousov}}, \bibinfo {author} {\bibfnamefont {F.}~\bibnamefont {Berger}}, \bibinfo {author} {\bibfnamefont {R.~G.}\ \bibnamefont {Alonso}},\ and\ \bibinfo {author} {\bibfnamefont {A.~J.}\ \bibnamefont {Hudspeth}},\ }\bibfield  {title} {\bibinfo {title} {Modeling active non-markovian oscillations},\ }\href {https://doi.org/10.1103/PhysRevLett.129.030603} {\bibfield  {journal} {\bibinfo  {journal} {Phys. Rev. Lett.}\ }\textbf {\bibinfo {volume} {129}},\ \bibinfo {pages} {030603} (\bibinfo {year} {2022})}\BibitemShut {NoStop}%
\bibitem [{\citenamefont {Singh}\ and\ \citenamefont {Proesmans}(2024)}]{singh2024inferring}%
  \BibitemOpen
  \bibfield  {author} {\bibinfo {author} {\bibfnamefont {P.}~\bibnamefont {Singh}}\ and\ \bibinfo {author} {\bibfnamefont {K.}~\bibnamefont {Proesmans}},\ }\bibfield  {title} {\bibinfo {title} {Inferring entropy production from time-dependent moments},\ }\href {https://doi.org/10.1038/s42005-024-01725-3} {\bibfield  {journal} {\bibinfo  {journal} {Communications Physics}\ }\textbf {\bibinfo {volume} {7}},\ \bibinfo {pages} {231} (\bibinfo {year} {2024})}\BibitemShut {NoStop}%
\bibitem [{\citenamefont {B{\'e}rut}\ \emph {et~al.}(2012)\citenamefont {B{\'e}rut}, \citenamefont {Arakelyan}, \citenamefont {Petrosyan}, \citenamefont {Ciliberto}, \citenamefont {Dillenschneider},\ and\ \citenamefont {Lutz}}]{berut2012experimental}%
  \BibitemOpen
  \bibfield  {author} {\bibinfo {author} {\bibfnamefont {A.}~\bibnamefont {B{\'e}rut}}, \bibinfo {author} {\bibfnamefont {A.}~\bibnamefont {Arakelyan}}, \bibinfo {author} {\bibfnamefont {A.}~\bibnamefont {Petrosyan}}, \bibinfo {author} {\bibfnamefont {S.}~\bibnamefont {Ciliberto}}, \bibinfo {author} {\bibfnamefont {R.}~\bibnamefont {Dillenschneider}},\ and\ \bibinfo {author} {\bibfnamefont {E.}~\bibnamefont {Lutz}},\ }\bibfield  {title} {\bibinfo {title} {Experimental verification of landauer’s principle linking information and thermodynamics},\ }\href {https://doi.org/10.1038/nature10872} {\bibfield  {journal} {\bibinfo  {journal} {Nature}\ }\textbf {\bibinfo {volume} {483}},\ \bibinfo {pages} {187} (\bibinfo {year} {2012})}\BibitemShut {NoStop}%
\bibitem [{\citenamefont {Jun}\ \emph {et~al.}(2014)\citenamefont {Jun}, \citenamefont {Gavrilov},\ and\ \citenamefont {Bechhoefer}}]{jun2014high}%
  \BibitemOpen
  \bibfield  {author} {\bibinfo {author} {\bibfnamefont {Y.}~\bibnamefont {Jun}}, \bibinfo {author} {\bibfnamefont {M.~c.~v.}\ \bibnamefont {Gavrilov}},\ and\ \bibinfo {author} {\bibfnamefont {J.}~\bibnamefont {Bechhoefer}},\ }\bibfield  {title} {\bibinfo {title} {High-precision test of landauer's principle in a feedback trap},\ }\href {https://doi.org/10.1103/PhysRevLett.113.190601} {\bibfield  {journal} {\bibinfo  {journal} {Physical Review Letters}\ }\textbf {\bibinfo {volume} {113}},\ \bibinfo {pages} {190601} (\bibinfo {year} {2014})}\BibitemShut {NoStop}%
\bibitem [{\citenamefont {Gavrilov}\ and\ \citenamefont {Bechhoefer}(2016)}]{gavrilov2016erasure}%
  \BibitemOpen
  \bibfield  {author} {\bibinfo {author} {\bibfnamefont {M.~c.~v.}\ \bibnamefont {Gavrilov}}\ and\ \bibinfo {author} {\bibfnamefont {J.}~\bibnamefont {Bechhoefer}},\ }\bibfield  {title} {\bibinfo {title} {Erasure without work in an asymmetric double-well potential},\ }\href {https://doi.org/10.1103/PhysRevLett.117.200601} {\bibfield  {journal} {\bibinfo  {journal} {Phys. Rev. Lett.}\ }\textbf {\bibinfo {volume} {117}},\ \bibinfo {pages} {200601} (\bibinfo {year} {2016})}\BibitemShut {NoStop}%
\bibitem [{\citenamefont {Nicoletti}\ and\ \citenamefont {Busiello}(2024)}]{nicoletti2024tuning}%
  \BibitemOpen
  \bibfield  {author} {\bibinfo {author} {\bibfnamefont {G.}~\bibnamefont {Nicoletti}}\ and\ \bibinfo {author} {\bibfnamefont {D.~M.}\ \bibnamefont {Busiello}},\ }\bibfield  {title} {\bibinfo {title} {Tuning transduction from hidden observables to optimize information harvesting},\ }\href {https://doi.org/10.1103/PhysRevLett.133.158401} {\bibfield  {journal} {\bibinfo  {journal} {Phys. Rev. Lett.}\ }\textbf {\bibinfo {volume} {133}},\ \bibinfo {pages} {158401} (\bibinfo {year} {2024})}\BibitemShut {NoStop}%
\bibitem [{\citenamefont {Weiss}(2007)}]{Weiss}%
  \BibitemOpen
  \bibfield  {author} {\bibinfo {author} {\bibfnamefont {J.~B.}\ \bibnamefont {Weiss}},\ }\bibfield  {title} {\bibinfo {title} {Fluctuation properties of steady-state langevin systems},\ }\href {https://doi.org/10.1103/PhysRevE.76.061128} {\bibfield  {journal} {\bibinfo  {journal} {Phys. Rev. E}\ }\textbf {\bibinfo {volume} {76}},\ \bibinfo {pages} {061128} (\bibinfo {year} {2007})}\BibitemShut {NoStop}%
\bibitem [{\citenamefont {Dabelow}\ \emph {et~al.}(2019)\citenamefont {Dabelow}, \citenamefont {Bo},\ and\ \citenamefont {Eichhorn}}]{dabelow2019irreversibility}%
  \BibitemOpen
  \bibfield  {author} {\bibinfo {author} {\bibfnamefont {L.}~\bibnamefont {Dabelow}}, \bibinfo {author} {\bibfnamefont {S.}~\bibnamefont {Bo}},\ and\ \bibinfo {author} {\bibfnamefont {R.}~\bibnamefont {Eichhorn}},\ }\bibfield  {title} {\bibinfo {title} {Irreversibility in active matter systems: Fluctuation theorem and mutual information},\ }\href {https://doi.org/10.1103/PhysRevX.9.021009} {\bibfield  {journal} {\bibinfo  {journal} {Physical Review X}\ }\textbf {\bibinfo {volume} {9}},\ \bibinfo {pages} {021009} (\bibinfo {year} {2019})}\BibitemShut {NoStop}%
\bibitem [{\citenamefont {Seara}\ \emph {et~al.}(2021)\citenamefont {Seara}, \citenamefont {Machta},\ and\ \citenamefont {Murrell}}]{seara2021irreversibility}%
  \BibitemOpen
  \bibfield  {author} {\bibinfo {author} {\bibfnamefont {D.~S.}\ \bibnamefont {Seara}}, \bibinfo {author} {\bibfnamefont {B.~B.}\ \bibnamefont {Machta}},\ and\ \bibinfo {author} {\bibfnamefont {M.~P.}\ \bibnamefont {Murrell}},\ }\bibfield  {title} {\bibinfo {title} {Irreversibility in dynamical phases and transitions},\ }\href {https://doi.org/10.1038/s41467-020-20281-2} {\bibfield  {journal} {\bibinfo  {journal} {Nature Communications}\ }\textbf {\bibinfo {volume} {12}},\ \bibinfo {pages} {392} (\bibinfo {year} {2021})}\BibitemShut {NoStop}%
\bibitem [{\citenamefont {Mori}\ \emph {et~al.}(2023)\citenamefont {Mori}, \citenamefont {Olsen},\ and\ \citenamefont {Krishnamurthy}}]{mori2023entropy}%
  \BibitemOpen
  \bibfield  {author} {\bibinfo {author} {\bibfnamefont {F.}~\bibnamefont {Mori}}, \bibinfo {author} {\bibfnamefont {K.~S.}\ \bibnamefont {Olsen}},\ and\ \bibinfo {author} {\bibfnamefont {S.}~\bibnamefont {Krishnamurthy}},\ }\bibfield  {title} {\bibinfo {title} {Entropy production of resetting processes},\ }\href {https://doi.org/10.1103/PhysRevResearch.5.023103} {\bibfield  {journal} {\bibinfo  {journal} {Phys. Rev. Res.}\ }\textbf {\bibinfo {volume} {5}},\ \bibinfo {pages} {023103} (\bibinfo {year} {2023})}\BibitemShut {NoStop}%
\bibitem [{\citenamefont {Das}\ \emph {et~al.}(2025)\citenamefont {Das}, \citenamefont {Manikandan}, \citenamefont {Paul}, \citenamefont {Kundu}, \citenamefont {Krishnamurthy},\ and\ \citenamefont {Banerjee}}]{das2025irreversibility}%
  \BibitemOpen
  \bibfield  {author} {\bibinfo {author} {\bibfnamefont {B.}~\bibnamefont {Das}}, \bibinfo {author} {\bibfnamefont {S.~K.}\ \bibnamefont {Manikandan}}, \bibinfo {author} {\bibfnamefont {S.}~\bibnamefont {Paul}}, \bibinfo {author} {\bibfnamefont {A.}~\bibnamefont {Kundu}}, \bibinfo {author} {\bibfnamefont {S.}~\bibnamefont {Krishnamurthy}},\ and\ \bibinfo {author} {\bibfnamefont {A.}~\bibnamefont {Banerjee}},\ }\bibfield  {title} {\bibinfo {title} {Irreversibility of mesoscopic processes with hydrodynamic interactions},\ }\href {https://doi.org/10.1103/1f5n-6s92} {\bibfield  {journal} {\bibinfo  {journal} {Phys. Rev. E}\ }\textbf {\bibinfo {volume} {112}},\ \bibinfo {pages} {L023401} (\bibinfo {year} {2025})}\BibitemShut {NoStop}%
\bibitem [{\citenamefont {Stigler}\ \emph {et~al.}(2011)\citenamefont {Stigler}, \citenamefont {Ziegler}, \citenamefont {Gieseke}, \citenamefont {Gebhardt},\ and\ \citenamefont {Rief}}]{stigler2011complex}%
  \BibitemOpen
  \bibfield  {author} {\bibinfo {author} {\bibfnamefont {J.}~\bibnamefont {Stigler}}, \bibinfo {author} {\bibfnamefont {F.}~\bibnamefont {Ziegler}}, \bibinfo {author} {\bibfnamefont {A.}~\bibnamefont {Gieseke}}, \bibinfo {author} {\bibfnamefont {J.~C.~M.}\ \bibnamefont {Gebhardt}},\ and\ \bibinfo {author} {\bibfnamefont {M.}~\bibnamefont {Rief}},\ }\bibfield  {title} {\bibinfo {title} {The complex folding network of single calmodulin molecules},\ }\href {https://doi.org/10.1126/science.1207598} {\bibfield  {journal} {\bibinfo  {journal} {Science}\ }\textbf {\bibinfo {volume} {334}},\ \bibinfo {pages} {512} (\bibinfo {year} {2011})}\BibitemShut {NoStop}%
\bibitem [{\citenamefont {Bechinger}\ \emph {et~al.}(2016)\citenamefont {Bechinger}, \citenamefont {Di~Leonardo}, \citenamefont {L{\"o}wen}, \citenamefont {Reichhardt}, \citenamefont {Volpe},\ and\ \citenamefont {Volpe}}]{bechinger2016active}%
  \BibitemOpen
  \bibfield  {author} {\bibinfo {author} {\bibfnamefont {C.}~\bibnamefont {Bechinger}}, \bibinfo {author} {\bibfnamefont {R.}~\bibnamefont {Di~Leonardo}}, \bibinfo {author} {\bibfnamefont {H.}~\bibnamefont {L{\"o}wen}}, \bibinfo {author} {\bibfnamefont {C.}~\bibnamefont {Reichhardt}}, \bibinfo {author} {\bibfnamefont {G.}~\bibnamefont {Volpe}},\ and\ \bibinfo {author} {\bibfnamefont {G.}~\bibnamefont {Volpe}},\ }\bibfield  {title} {\bibinfo {title} {Active particles in complex and crowded environments},\ }\href {https://doi.org/10.1103/RevModPhys.88.045006} {\bibfield  {journal} {\bibinfo  {journal} {Reviews of Modern Physics}\ }\textbf {\bibinfo {volume} {88}},\ \bibinfo {pages} {045006} (\bibinfo {year} {2016})}\BibitemShut {NoStop}%
\bibitem [{\citenamefont {Bowick}\ \emph {et~al.}(2022)\citenamefont {Bowick}, \citenamefont {Fakhri}, \citenamefont {Marchetti},\ and\ \citenamefont {Ramaswamy}}]{bowick2022symmetry}%
  \BibitemOpen
  \bibfield  {author} {\bibinfo {author} {\bibfnamefont {M.~J.}\ \bibnamefont {Bowick}}, \bibinfo {author} {\bibfnamefont {N.}~\bibnamefont {Fakhri}}, \bibinfo {author} {\bibfnamefont {M.~C.}\ \bibnamefont {Marchetti}},\ and\ \bibinfo {author} {\bibfnamefont {S.}~\bibnamefont {Ramaswamy}},\ }\bibfield  {title} {\bibinfo {title} {Symmetry, thermodynamics, and topology in active matter},\ }\href {https://doi.org/10.1103/PhysRevX.12.010501} {\bibfield  {journal} {\bibinfo  {journal} {Phys. Rev. X}\ }\textbf {\bibinfo {volume} {12}},\ \bibinfo {pages} {010501} (\bibinfo {year} {2022})}\BibitemShut {NoStop}%
\bibitem [{\citenamefont {Tyson}\ \emph {et~al.}(2001)\citenamefont {Tyson}, \citenamefont {Chen},\ and\ \citenamefont {Novak}}]{tyson2001network}%
  \BibitemOpen
  \bibfield  {author} {\bibinfo {author} {\bibfnamefont {J.~J.}\ \bibnamefont {Tyson}}, \bibinfo {author} {\bibfnamefont {K.}~\bibnamefont {Chen}},\ and\ \bibinfo {author} {\bibfnamefont {B.}~\bibnamefont {Novak}},\ }\bibfield  {title} {\bibinfo {title} {Network dynamics and cell physiology},\ }\href {https://doi.org/10.1038/35103078} {\bibfield  {journal} {\bibinfo  {journal} {Nature reviews Molecular cell biology}\ }\textbf {\bibinfo {volume} {2}},\ \bibinfo {pages} {908} (\bibinfo {year} {2001})}\BibitemShut {NoStop}%
\bibitem [{\citenamefont {Padmanabha}\ \emph {et~al.}(2024)\citenamefont {Padmanabha}, \citenamefont {Nicoletti}, \citenamefont {Bernardi}, \citenamefont {Suweis}, \citenamefont {Azaele}, \citenamefont {Rinaldo},\ and\ \citenamefont {Maritan}}]{padmanabha2024landscape}%
  \BibitemOpen
  \bibfield  {author} {\bibinfo {author} {\bibfnamefont {P.}~\bibnamefont {Padmanabha}}, \bibinfo {author} {\bibfnamefont {G.}~\bibnamefont {Nicoletti}}, \bibinfo {author} {\bibfnamefont {D.}~\bibnamefont {Bernardi}}, \bibinfo {author} {\bibfnamefont {S.}~\bibnamefont {Suweis}}, \bibinfo {author} {\bibfnamefont {S.}~\bibnamefont {Azaele}}, \bibinfo {author} {\bibfnamefont {A.}~\bibnamefont {Rinaldo}},\ and\ \bibinfo {author} {\bibfnamefont {A.}~\bibnamefont {Maritan}},\ }\bibfield  {title} {\bibinfo {title} {Landscape and environmental heterogeneity support coexistence in competitive metacommunities},\ }\href {https://doi.org/10.1073/pnas.2410932121} {\bibfield  {journal} {\bibinfo  {journal} {Proceedings of the National Academy of Sciences}\ }\textbf {\bibinfo {volume} {121}},\ \bibinfo {pages} {e2410932121} (\bibinfo {year} {2024})}\BibitemShut {NoStop}%
\bibitem [{\citenamefont {Lynn}\ \emph {et~al.}(2021)\citenamefont {Lynn}, \citenamefont {Cornblath}, \citenamefont {Papadopoulos}, \citenamefont {Bertolero},\ and\ \citenamefont {Bassett}}]{lynn2021broken}%
  \BibitemOpen
  \bibfield  {author} {\bibinfo {author} {\bibfnamefont {C.~W.}\ \bibnamefont {Lynn}}, \bibinfo {author} {\bibfnamefont {E.~J.}\ \bibnamefont {Cornblath}}, \bibinfo {author} {\bibfnamefont {L.}~\bibnamefont {Papadopoulos}}, \bibinfo {author} {\bibfnamefont {M.~A.}\ \bibnamefont {Bertolero}},\ and\ \bibinfo {author} {\bibfnamefont {D.~S.}\ \bibnamefont {Bassett}},\ }\bibfield  {title} {\bibinfo {title} {Broken detailed balance and entropy production in the human brain},\ }\href{https://doi.org/10.1073/pnas.2109889118} {\bibfield  {journal} {\bibinfo  {journal} {Proceedings of the National Academy of Sciences}\ }\textbf {\bibinfo {volume} {118}},\ \bibinfo {pages} {e2109889118} (\bibinfo {year} {2021})}\BibitemShut {NoStop}%
\bibitem [{\citenamefont {Ronellenfitsch}\ \emph {et~al.}(2019)\citenamefont {Ronellenfitsch}, \citenamefont {Stoop}, \citenamefont {Yu}, \citenamefont {Forrow},\ and\ \citenamefont {Dunkel}}]{ronellenfitsch2019inverse}%
  \BibitemOpen
  \bibfield  {author} {\bibinfo {author} {\bibfnamefont {H.}~\bibnamefont {Ronellenfitsch}}, \bibinfo {author} {\bibfnamefont {N.}~\bibnamefont {Stoop}}, \bibinfo {author} {\bibfnamefont {J.}~\bibnamefont {Yu}}, \bibinfo {author} {\bibfnamefont {A.}~\bibnamefont {Forrow}},\ and\ \bibinfo {author} {\bibfnamefont {J.}~\bibnamefont {Dunkel}},\ }\bibfield  {title} {\bibinfo {title} {Inverse design of discrete mechanical metamaterials},\ }\href{https://doi.org/10.1103/PhysRevMaterials.3.095201} {\bibfield  {journal} {\bibinfo  {journal} {Physical Review Materials}\ }\textbf {\bibinfo {volume} {3}},\ \bibinfo {pages} {095201} (\bibinfo {year} {2019})}\BibitemShut {NoStop}%
\bibitem [{\citenamefont {Chennakesavalu}\ \emph {et~al.}(2024)\citenamefont {Chennakesavalu}, \citenamefont {Manikandan}, \citenamefont {Hu},\ and\ \citenamefont {Rotskoff}}]{chennakesavalu2024adaptive}%
  \BibitemOpen
  \bibfield  {author} {\bibinfo {author} {\bibfnamefont {S.}~\bibnamefont {Chennakesavalu}}, \bibinfo {author} {\bibfnamefont {S.~K.}\ \bibnamefont {Manikandan}}, \bibinfo {author} {\bibfnamefont {F.}~\bibnamefont {Hu}},\ and\ \bibinfo {author} {\bibfnamefont {G.~M.}\ \bibnamefont {Rotskoff}},\ }\bibfield  {title} {\bibinfo {title} {Adaptive nonequilibrium design of actin-based metamaterials: Fundamental and practical limits of control},\ }\href{https://doi.org/10.1073/pnas.2310238121} {\bibfield  {journal} {\bibinfo  {journal} {Proceedings of the National Academy of Sciences}\ }\textbf {\bibinfo {volume} {121}},\ \bibinfo {pages} {e2310238121} (\bibinfo {year} {2024})}\BibitemShut {NoStop}%
\bibitem [{\citenamefont {Aurell}\ \emph {et~al.}(2011)\citenamefont {Aurell}, \citenamefont {Mej{\'\i}a-Monasterio},\ and\ \citenamefont {Muratore-Ginanneschi}}]{aurell2011optimal}%
  \BibitemOpen
  \bibfield  {author} {\bibinfo {author} {\bibfnamefont {E.}~\bibnamefont {Aurell}}, \bibinfo {author} {\bibfnamefont {C.}~\bibnamefont {Mej{\'\i}a-Monasterio}},\ and\ \bibinfo {author} {\bibfnamefont {P.}~\bibnamefont {Muratore-Ginanneschi}},\ }\bibfield  {title} {\bibinfo {title} {Optimal protocols and optimal transport in stochastic thermodynamics},\ }\href{https://doi.org/10.1103/PhysRevLett.106.250601} {\bibfield  {journal} {\bibinfo  {journal} {Physical review letters}\ }\textbf {\bibinfo {volume} {106}},\ \bibinfo {pages} {250601} (\bibinfo {year} {2011})}\BibitemShut {NoStop}%
\bibitem [{\citenamefont {Klinger}\ and\ \citenamefont {Rotskoff}(2025{\natexlab{a}})}]{jeremie}%
  \BibitemOpen
  \bibfield  {author} {\bibinfo {author} {\bibfnamefont {J.}~\bibnamefont {Klinger}}\ and\ \bibinfo {author} {\bibfnamefont {G.~M.}\ \bibnamefont {Rotskoff}},\ }\bibfield  {title} {\bibinfo {title} {Universal energy-speed-accuracy trade-offs in driven nonequilibrium systems},\ }\href {https://doi.org/10.1103/PhysRevE.111.014114} {\bibfield  {journal} {\bibinfo  {journal} {Phys. Rev. E}\ }\textbf {\bibinfo {volume} {111}},\ \bibinfo {pages} {014114} (\bibinfo {year} {2025}{\natexlab{a}})}\BibitemShut {NoStop}%
\bibitem [{\citenamefont {Klinger}\ and\ \citenamefont {Rotskoff}(2025{\natexlab{b}})}]{klinger2025minimally}%
  \BibitemOpen
  \bibfield  {author} {\bibinfo {author} {\bibfnamefont {J.}~\bibnamefont {Klinger}}\ and\ \bibinfo {author} {\bibfnamefont {G.~M.}\ \bibnamefont {Rotskoff}},\ }\bibfield  {title} {\bibinfo {title} {Minimally dissipative multi-bit logical operations},\ }\href{https://doi.org/10.48550/arXiv.2506.24021} {\bibfield  {journal} {\bibinfo  {journal} {arXiv preprint arXiv:2506.24021}\ } (\bibinfo {year} {2025}{\natexlab{b}})}\BibitemShut {NoStop}%
\bibitem [{\citenamefont {Zhong}\ and\ \citenamefont {DeWeese}(2022)}]{zhong2022limited}%
  \BibitemOpen
  \bibfield  {author} {\bibinfo {author} {\bibfnamefont {A.}~\bibnamefont {Zhong}}\ and\ \bibinfo {author} {\bibfnamefont {M.~R.}\ \bibnamefont {DeWeese}},\ }\bibfield  {title} {\bibinfo {title} {Limited-control optimal protocols arbitrarily far from equilibrium},\ }\href{https://doi.org/10.1103/PhysRevE.106.044135} {\bibfield  {journal} {\bibinfo  {journal} {Physical Review E}\ }\textbf {\bibinfo {volume} {106}},\ \bibinfo {pages} {044135} (\bibinfo {year} {2022})}\BibitemShut {NoStop}%
\bibitem [{\citenamefont {Liu}\ \emph {et~al.}(2025)\citenamefont {Liu}, \citenamefont {Nguyen}, \citenamefont {Lin}, \citenamefont {Sun},\ and\ \citenamefont {Zheng}}]{liu2025optical}%
  \BibitemOpen
  \bibfield  {author} {\bibinfo {author} {\bibfnamefont {S.-F.}\ \bibnamefont {Liu}}, \bibinfo {author} {\bibfnamefont {K.}~\bibnamefont {Nguyen}}, \bibinfo {author} {\bibfnamefont {L.}~\bibnamefont {Lin}}, \bibinfo {author} {\bibfnamefont {H.-B.}\ \bibnamefont {Sun}},\ and\ \bibinfo {author} {\bibfnamefont {Y.}~\bibnamefont {Zheng}},\ }\bibfield  {title} {\bibinfo {title} {Optical colloidal assembly},\ }\href{https://doi.org/10.1021/acs.chemrev.5c00644} {\bibfield  {journal} {\bibinfo  {journal} {Chemical Reviews}\ }\textbf {\bibinfo {volume} {126}},\ \bibinfo {pages} {448} (\bibinfo {year} {2025})}\BibitemShut {NoStop}%
\end{thebibliography}
%

\clearpage
\onecolumngrid

\section*{Supplemental Material: Localising entropy production along non-equilibrium trajectories}

\setcounter{section}{0}
\renewcommand{\thesection}{S\arabic{section}}
\renewcommand{\thefigure}{S\arabic{figure}}
\setcounter{figure}{0}
\renewcommand{\theequation}{S\arabic{equation}}
\setcounter{equation}{0}

\section{Supplementary Note 1: Performance test of learning algorithm}

\label{supp_note1}
\begin{figure}[h]
    \centering
    \includegraphics[width=0.55\linewidth]{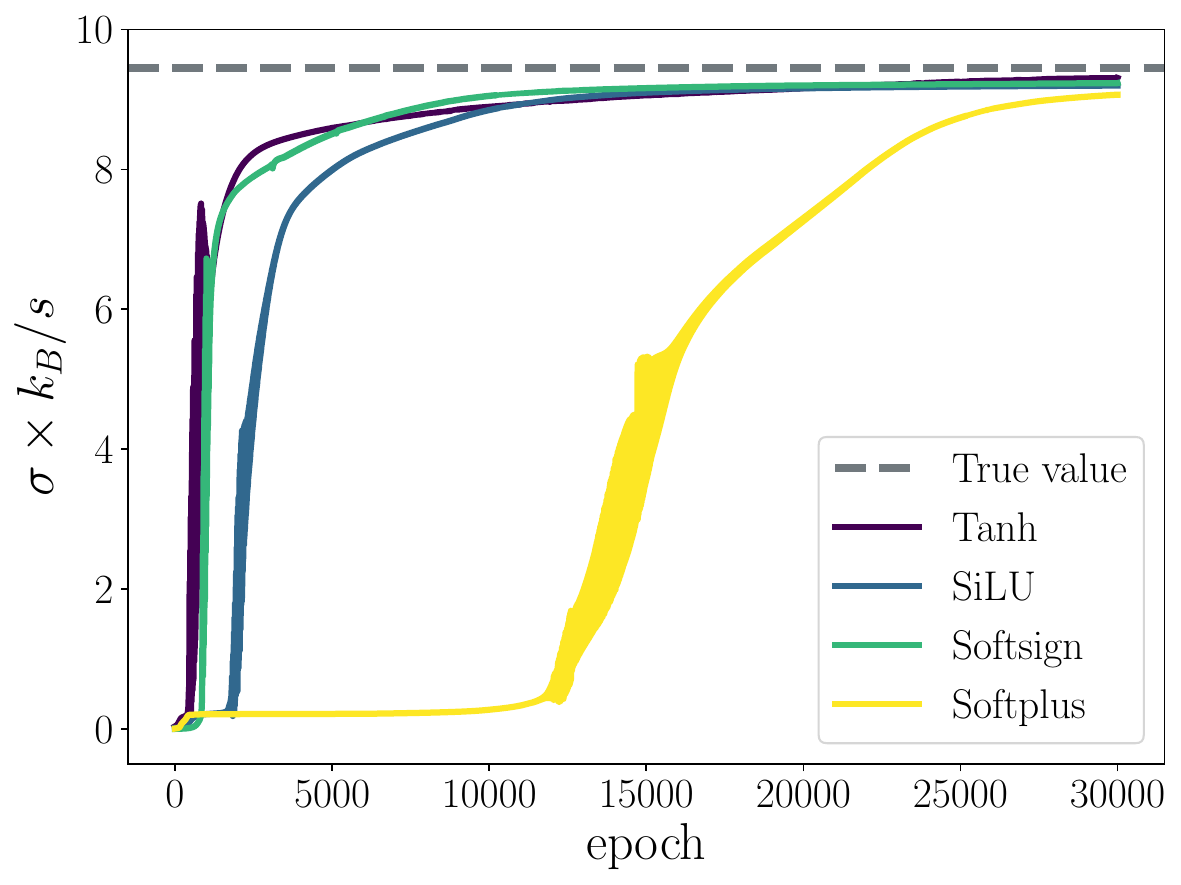}
    \caption{Learning rate of the algorithm with different activation functions (shown as different colored lines) for the bistable Brownian gyrator system. The True value (shown as the dashed horizontal line) is obtained as the medium entropy production rate~\cite{seifert2005entropy}. }
    
    \label{fig:learning_curve}
\end{figure}

As noted in the main text, we use the $\tanh(\cdot)$ activation in our DeepRitz-based neural network algorithm because it is a smooth nonlinearity with outputs naturally confined to $(-1,1)$. In Fig.~\ref{fig:learning_curve}, we compare the performance of the algorithm using several different activation functions. The results clearly show that both $\tanh$ and Softsign lead to faster convergence than other activation functions of similar kinds.

\section{Supplementary Note 2: $R^2$ test of learning algorithm}
\label{supp_note2}
\begin{figure*}
    \centering
    \includegraphics[width=\linewidth]{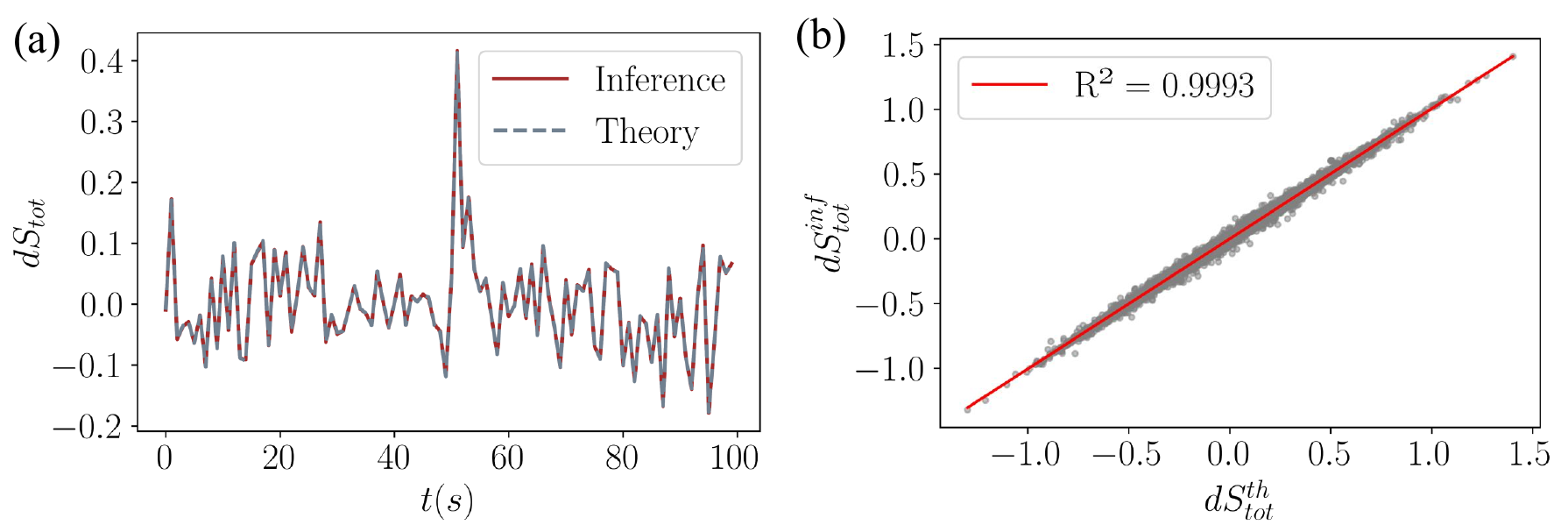}
    \caption{(a) Inferred fluctuating entropy production of the 2D harmonic Brownian gyrator overlaps with the analytical values. (b) $R^2$ test of the neural network-based inference for the same. The scatter points represent individual data samples, while the red solid line corresponds to the linear fit between theoretically computed and numerically inferred entropy production. }
    \label{fig:harmonic_rsquare}
\end{figure*}

For the higher-dimensional system, the fluctuating entropy production obtained from theory and inference shows reasonable agreement, with the data distributed around a linear fit yielding $R^2 = 0.7576$ (Figure~\ref{fig:nBgr}(d) of the main text). In contrast, the same neural network performs significantly better for the low-dimensional case, as illustrated in Figure~\ref{fig:harmonic_rsquare}(a) for the 2D harmonic Brownian gyrator model. In this case, the inferred local entropy production closely overlaps with the analytical prediction obtained from Eq.~\eqref{eq:th_localepr} of the main text, which is based on the force field in Eq.~\eqref{eq:th_ffield} of the main text. The corresponding scatter plot in Figure~\ref{fig:harmonic_rsquare}(b) shows an almost perfect linear relation between the two estimates, reflected in the high $R^2$ value of 0.9993. 

\section{Supplementary Note 3: Local entropy production for coarse-grained dynamics}
\label{ap:local_epr_coarsegraining}

We consider nonequilibrium dynamics described by linear Langevin equations
\begin{equation}
\dot{\mathbf{x}}=\mathbf{A}\mathbf{x}+\boldsymbol{\xi}(t),
\label{eq:langevin_general}
\end{equation}
where $\boldsymbol{\xi}(t)$ is Gaussian white noise with $\langle \xi_i(t) \xi_j(t^\prime)\rangle = 2\mathbf{D} \delta(t-t')$. In the spirit of the N-dimensional Brownian gyrator model, we consider a diagonal diffusion matrix $\mathbf{D}$ with distinct entries, while the elements of the drift matrix $\mathbf{A}$ are defined according to Eq.~\eqref{eq:AMatrix} of the main text. 
The steady state is Gaussian with covariance matrix $\mathbf{C}$ satisfying
\begin{equation}
\mathbf{A}\mathbf{C}+\mathbf{C}\mathbf{A}^{T}+2\mathbf{D}=0.
\end{equation}
Within this framework, entropy production can be characterized both under
\emph{spatial coarse graining}, induced by projection onto a reduced phase space,
and under \emph{temporal coarse graining}, induced by finite-time sampling.
We discuss these two cases in turn.

\noindent
\subsection{Spatial coarse graining and reduced local entropy production}

The local entropy production rate in a reduced phase space obtained via a
projection operation can be estimated analogously to the full description using
Eq.~\eqref{eq:th_localepr} of the main text, with the force field given by
Eq.~\eqref{eq:th_ffield} of the main text. Under coarse graining, however, the drift, diffusion,
and covariance matrices must be appropriately renormalized, as discussed in
Ref.~\cite{nicoletti2024tuning}.

For the reduced phase space $(x_1,x_2)$ of a three-dimensional Brownian gyrator
with full state vector $(x_1,x_2,x_3)$, the effective drift matrix is
\begin{equation}
\mathbf{A}^{\mathrm{red}}_{x_1 x_2}
=
\mathbf{A}_{x_1 x_2}
+
\mathbf{M}\,\mathbf{C}_{x_1 x_2}^{-1},
\qquad
(\mathbf{M})_{ij}
=
(\mathbf{A})_{i x_3}(\mathbf{C})_{j x_3},
\label{ap:eq_reduced_F}
\end{equation}
where $i,j\in\{x_1,x_2\}$ and $\mathbf{C}_{x_1 x_2}$ denotes the corresponding
submatrix of the steady-state covariance $\mathbf{C}$.
The reduced covariance matrix $\mathbf{C}_{x_1 x_2}$ must be used consistently in
Eq.~\eqref{eq:th_ffield} of the main text, while the diffusion matrix entering the reduced
description is given by the submatrix $\mathbf{D}_{x_1 x_2}$.
Together, these renormalized quantities determine the local entropy production
rate in the reduced phase space.

\noindent
\subsection{Temporal coarse graining and trajectory-level entropy production}

A closely related notion of coarse-grained entropy production arises when the
same Langevin dynamics~\eqref{eq:langevin_general} is observed only at discrete
times separated by a finite interval $\Delta t$, as discussed in Ref.\ \cite{Weiss}. Over this interval, the transition probability
$p(\mathbf{x}_1|\mathbf{x}_0)$ is Gaussian with covariance
$\mathbf{C}_{\Delta t}
=
\mathbf{C}
-
e^{\mathbf{A}\Delta t}\,
\mathbf{C}\,
e^{\mathbf{A}^{T}\Delta t}$.
The joint probability of a trajectory segment specified by its endpoints
$(\mathbf{x}_0,\mathbf{x}_1)$ can therefore be written as
\begin{equation}
p(\mathbf{x}_0,\mathbf{x}_1)
\propto
\exp\!\left[-\tfrac12\,\mathbf{z}^{T}\mathbf{R}_{01}\mathbf{z}\right],
\qquad
\mathbf{z}=(\mathbf{x}_0,\mathbf{x}_1)^{T},
\end{equation}
with trajectory concentration matrix
\begin{equation}
\mathbf{R}_{01}=
\begin{pmatrix}
e^{\mathbf{A}^{T}\Delta t}\mathbf{Q}_{\Delta t}e^{\mathbf{A}\Delta t}+\mathbf{Q}
&
-\,e^{\mathbf{A}^{T}\Delta t}\mathbf{Q}_{\Delta t}
\\[4pt]
-\,\mathbf{Q}_{\Delta t}e^{\mathbf{A}\Delta t}
&
\mathbf{Q}_{\Delta t}
\end{pmatrix},
\end{equation}
where $\mathbf{Q}=\mathbf{C}^{-1}$ and
$\mathbf{Q}_{\Delta t}=\mathbf{C}_{\Delta t}^{-1}$.
The probability of the time-reversed trajectory segment is obtained by exchanging
initial and final states,
\begin{equation}
\mathbf{R}_{10}=\mathbf{J}\mathbf{R}_{01}\mathbf{J},
\qquad
\mathbf{J}=
\begin{pmatrix}
0 & \mathbf{I} \\
\mathbf{I} & 0
\end{pmatrix}.
\end{equation}
The irreversibility of a trajectory segment is then defined as
\begin{equation}
r
=
\ln\frac{p(\mathbf{x}_0,\mathbf{x}_1)}{p(\mathbf{x}_1,\mathbf{x}_0)}
=
\tfrac12\,\mathbf{z}^{T}
\bigl(\mathbf{R}_{10}-\mathbf{R}_{01}\bigr)
\mathbf{z}.
\end{equation}
Since all fluctuations within the interval $\Delta t$ have been integrated out,
$r$ naturally represents the entropy production associated with
\emph{time--coarse-grained dynamics}.
This coarse-grained entropy production vanishes under detailed balance and
quantifies the breaking of time-reversal symmetry in nonequilibrium steady state
of process described by Eq.\ \eqref{eq:langevin_general}, sampled at a finite time interval $\Delta t$. It is this quantity $r$ that we use to benchmark the inferred time-coarse-grained local entropy production in Fig.\ \ref{fig:coarsegraining} of the main text. 
\begin{figure*}[!t]
    \centering
\includegraphics[width=1\linewidth]{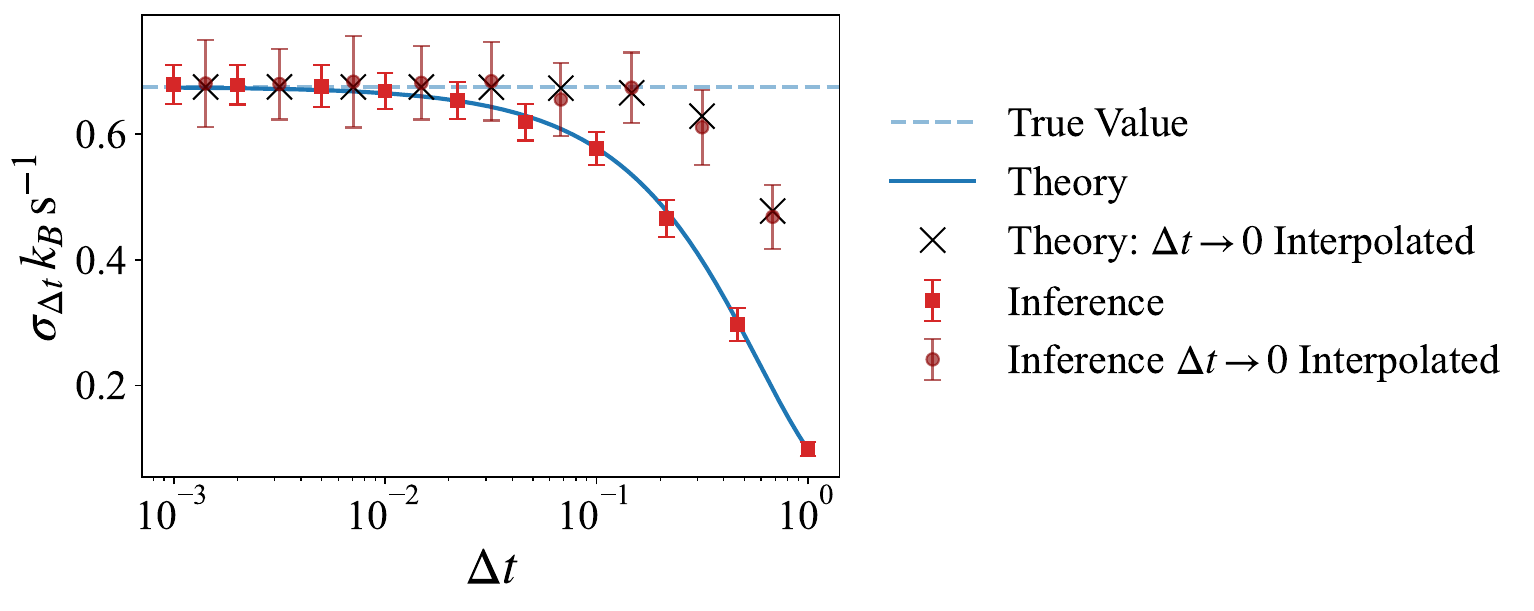}
\caption{
Analytical and inferred time--coarse-grained entropy production rate $\sigma_{\Delta t}$ for the two-dimensional harmonic Brownian gyrator as a function of the sampling interval $\Delta t$.
The solid blue curve shows the analytical prediction for $\sigma_{\Delta t}$, while red square markers denote values inferred from trajectory data, with error bars indicating statistical uncertainty.
Dark red circles show two-point interpolated estimates of the $\Delta t\!\to\!0$ limit obtained from adjacent sampling intervals, and black crosses indicate the corresponding interpolated values computed from the analytical curve.
The horizontal dashed line marks the true entropy production rate. The shortest relevant time scale in this case is $\tau = 0.5s$.
}
\label{fig:avsigm}
\end{figure*}
As shown in the maintext, for both spatial and temporal coarse graining, we find that the entropy
production computed within the corresponding reduced descriptions
theoretically agrees well with inference results obtained from the
respective coarse-grained data, with nearly 90\% accuracy. Furthermore, the average time--coarse-grained entropy production rate can be expressed as
a function of $\Delta t$ as
\begin{equation}
\sigma_{\Delta t} = \langle r \rangle
= \frac{1}{2\Delta t}\,\mathrm{Tr}\!\left(
\mathbf{R}_{01}^{-1}\left(\mathbf{R}_{10}-\mathbf{R}_{01}\right)
\right).
\end{equation}
Figure~\ref{fig:avsigm} compares this analytical prediction with estimates inferred directly from trajectory data for the two-dimensional linear Brownian gyrator. The solid blue curve shows the theoretical prediction $\sigma_{\Delta t}$ evaluated as a continuous function of $\Delta t$, while the red square markers denote the entropy production rate inferred from simulated time series at discrete sampling intervals, with error bars indicating statistical uncertainty. We observe excellent agreement between theory and inference across the entire range of $\Delta t$ shown, confirming that the finite--$\Delta t$ dependence is accurately captured.

In addition, we illustrate a simple two-point interpolation procedure to improve estimates of the continuum-time limit $\Delta t \to 0$. Using pairs of adjacent sampling intervals $(\Delta t_1,\Delta t_2)$, we linearly interpolate in $\Delta t$ to remove the leading-order discretization bias, yielding an improved estimate
\begin{equation}
\sigma_0 \approx \frac{\Delta t_2\,\sigma_{\Delta t_1} - \Delta t_1\,\sigma_{\Delta t_2}}{\Delta t_2 - \Delta t_1}.
\end{equation}
The resulting interpolated estimates obtained from the inferred data are shown as dark red circles, while the corresponding interpolated values computed from the theoretical curve are shown as black crosses. For visualization, these extrapolated values are plotted at $\Delta t=\sqrt{\Delta t_1\Delta t_2}$, although they represent estimates of the $\Delta t\to0$ limit. The horizontal dashed line indicates the true entropy production rate. We find that the extrapolated estimates lie very close to the true value even when the individual points lie significantly below the true value. We note that, in principle, such an interpolation can also be applied to obtain first order $\Delta t$ corrections to the thermodynamic force field.

\end{document}